\documentclass[final,1p,times]{elsarticle}

\usepackage{amsmath}
\usepackage{amsthm}
\usepackage{amsmath,bm}
\usepackage{amsfonts}
\usepackage{tikz}
\usepackage{algorithm}
\usepackage{algpseudocode}
\usepackage{graphics}
\usepackage{subfigure}
\usepackage{epsfig}
\usepackage{epstopdf}
\usepackage{hyperref}
\usepackage{natbib} \setcitestyle{comma,sort&compress,numbers,square}
\hypersetup{hypertex=true,
            colorlinks=true,
            linkcolor=blue,
            anchorcolor=blue,
            citecolor=blue}

\DeclareMathOperator*{\argmin}{argmin}
\usepackage{geometry}
\begin{document}
\begin{frontmatter}
\title{Data-driven nonintrusive reduced order modeling for dynamical systems with moving boundaries using Gaussian process regression}
\date{}
\author[WISC]{Zhan Ma}
\author[WISC]{Wenxiao Pan\corref{cor}}
\ead{wpan9@wisc.edu}
\cortext[cor]{Corresponding author}
\address[WISC]{Department of Mechanical Engineering, University of Wisconsin-Madison, Madison, WI 53706, USA}

\begin{abstract}
We present a data-driven nonintrusive model order reduction method for dynamical systems with moving boundaries. The proposed method draws on the proper orthogonal decomposition, Gaussian process regression, and moving least squares interpolation. It combines several attributes that are not simultaneously satisfied in the existing model order reduction methods for dynamical systems with moving boundaries. Specifically, the method requires only snapshot data of state variables at discrete time instances and the parameters that characterize the boundaries, but not further knowledge of the full-order model and the underlying governing equations. The dynamical systems can be generally nonlinear. The movements of boundaries are not limited to prescribed or periodic motions but can be free motions. In addition, we numerically investigate the ability of the reduced order model constructed by the proposed method to forecast the full-order solutions for future times beyond the range of snapshot data. The error analysis for the proposed reduced order modeling and the criteria to determine the furthest forecast time are also provided. Through numerical experiments, we assess the accuracy and efficiency of the proposed method in several benchmark problems. The snapshot data used to construct and validate the reduced order model are from analytical/numerical solutions and  experimental measurements. 
\end{abstract}


\end{frontmatter}
%===========================================

\section{Introduction}\label{sec:intro}

%1) What is ROM and why?
Many natural and engineering systems can exhibit complex dynamics with a wide range of temporal and spatial features. In the analysis of these systems, a first step is to seek and extract dominant features or modes \cite{ROM_FluidFlow_Review_Taira2017,PEHERSTORFER2016196,ROM_GPR_Wan2017,ROM_LSTM_Vlachas2018}. This step typically starts with a modal decomposition of a data set of state variables of interest attained at discrete time instances from experiments or computations. The proper orthogonal decomposition (POD) is one of the most widely used techniques for accomplishing modal decomposition, which can decompose the set of data into spatially dependent POD bases and temporally dependent coefficients. The dominant POD modes can effectively capture the major dynamical evolution of the non-stationary state variables and thereby provide a means to describe a complex dynamical system in a low-dimensional form, the so-called reduced-order model (ROM). 
%By adding more modes, one can reconstruct the original variable field more accurately, but their contributions are much smaller in comparison to the most dominant modes. 
%Thus, a low-order dynamical system is obtained by projecting the original full-order system in the space of  a smaller number of POD modes. , as the basis for the development of reduced-order modeling (ROM)
%As computational and experimental techniques are advancing their ability in providing  high-fidelity data, the compression of a vast amount of field data to a low-dimensional form becomes more important in studying various systems and developing models for understanding and modeling their dynamical behavior. 

%2) how to construct the temporal coefficients: intrusive vs. non-intrusive
%Our approach utilizes a training process from full-order scale direct numerical simulation data projected on proper orthogonal decomposition (POD) modes
% in a POD basis spanned space.
To determine the temporal coefficient for each POD mode retained, both intrusive and non-intrusive approaches have been developed. The Galerkin projection projects the full model (usually partial differential equations (PDEs)) onto the space of truncated POD modes. Through the orthonormality and energy-optimality attributes of the POD bases, a truncated set of coupled ordinary differential equations (ODEs) is obtained, which constitutes a low-order dynamical system and governs the evolution of the temporal coefficients. By solving the ODEs, usually numerically, the temporal coefficients can be determined. Along with the spatial POD basis functions, such constructed ROM can then be used to reproduce the full-order solutions. Since the Galerkin projection requires the prior knowledge of the governing PDEs, this approach is referred to as intrusive model reduction. The quadratic nonlinearity and triadic interactions in the ODEs derived from Galerkin projection call for a computational load in the order of $\mathcal{O}(R^3)$, where $R$ is the number of POD modes retained in the ROM.
Alternative to the Galerkin projection, 
%the fully data-driven, nonintrusive ROM (NIROM). NIROM is a family of methods that solely access available datasets to extract and mimic the system’s dynamics, with little-to-no knowledge of the governing equations.
nonintrusive approaches can be employed to determine the temporal modes.
For example, replace the set of ODEs with a set of hyper surfaces, which is constructed using the interpolation methods such as the Smolyak sparse grid \cite{NIROM_Smolyak_XIAO2015522} and radial basis function (RBF) \cite{ROM_PODRBF_WALTON2013,NIROM_RBF_XIAO2015}. The work in \cite{PEHERSTORFER2016196} proposed to infer the operators of the ODEs via the least squares optimization \cite{PEHERSTORFER2016196}. Wan et al. employed the Gaussian Processes to model the temporal modes \cite{ROM_GPR_Wan2017}. In addition, several efforts adopted deep learning techniques and trained a neural network model for the temporal modes in the form of the  artificial \cite{ROM_Flow_ANN_Omer2019} or deep feedforward neural network \cite{ROM_Turbulent_DNN_Lui2019}, the long short-term memory (LSTM) recurrent neural network \cite{ROM_PODLSTM_Wang2018,ROM_LSTM_Vlachas2018},
or the temporal convolutional neural network \cite{ROM_Flow_Data_WU2020}. These nonintrusive approaches can effectively capture the generally nonlinear time evolution of the temporal coefficients without performing the Galerkin projection. Among them, some are limited to predictions within the database range \cite{PEHERSTORFER2016196}; and some can predict both inside and outside the database range \cite{ROM_Turbulent_DNN_Lui2019,ROM_GPR_Wan2017,ROM_LSTM_Vlachas2018}. 
The non-intrusive approaches do not necessarily require the exact form of the full-order equations and hence are applicable to  experimental data where the governing equations are often not well established or the associated parameters have considerable uncertainties.
%For reduced-order data-driven forecast of high-dimensional chaotic dynamical systems, researchers also proposed to employ the Gaussain Process regression \cite{ROM_GPR_Wan2017} or long short-term memory (LSTM) networks \cite{ROM_LSTM_Vlachas2018} to construct a model for predicting the evolution of the time coefficients. 

%3) literature review on ROM related to moving boundaries or FSI and for prediction in time not in parameter space
% The model is based on Galerkin projection of governing equation onto space spanned by modes obtained from high-fidelity computations. The motion of the boundary defined  results in additional convective term in Galerkin system.
The above-mentioned model order reduction methods are applicable to fixed-domain problems. For many systems with moving objects/boundaries, e.g., in fluid-solid interactions, additional care is needed to derive the ROM. The efforts reported in literature also fall into either of two categories: intrusive or nonintrusive. And those efforts mainly focus on fluid-solid interaction problems. In the intrusive category, the Navier-Stokes (NS) equations that govern the fluid flow are extended to the solid domain. The POD modes are computed for the combined fluid-solid domain. The extended NS equations are then projected via Galerkin projection onto the POD modes to yield the low-order dynamical system with extra terms related to the solid motion. Liberge and Hamdouni proposed a multiphase method similar to the fictitious domain method to extend the NS equations to the solid domain by using a penalization method and a Lagrangian multiplier  \cite{ROM_OscillatingCylinder_Liberge2010}. Gao and Wei proposed to add extra body-force terms to the NS equations, similar to the immersed boundary method \cite{ROM_MovingBoundary_Gao2016,ROM_MovingBoundary_GaoPHDthesis2018}. For generally unprescribed solid motion, the additional terms contributed by the unsteady solid motion must be recomputed at each time step. The significantly increased computational cost by these terms can overshadow the benefit of reduced order modeling. To alleviate this issue, the most expensive nonlinear term was neglected in practice \cite{ROM_MovingBoundary_Gao2016}. Those intrusive approaches were validated by the tests on a rigid cylinder or sphere oscillating in a fluid \cite{ROM_OscillatingCylinder_Liberge2010,ROM_MovingBoundary_Gao2016,ROM_MovingBoundary_GaoPHDthesis2018}, and only prescribed motion was considered in \cite{ROM_MovingBoundary_Gao2016,ROM_MovingBoundary_GaoPHDthesis2018}. In the non-intrusive category, Xiao et al. \cite{NIROM_FSI_XIAO201635} proposed to employ the POD and RBF multi-dimensional interpolation to construct the ROM for fluid-solid interaction problems with a free-moving solid body in the fluid. In this work, the predictions of the ROM were examined for the time instances inside the database range.

\begin{figure}[htbp]
\centering
\includegraphics[width=12cm]{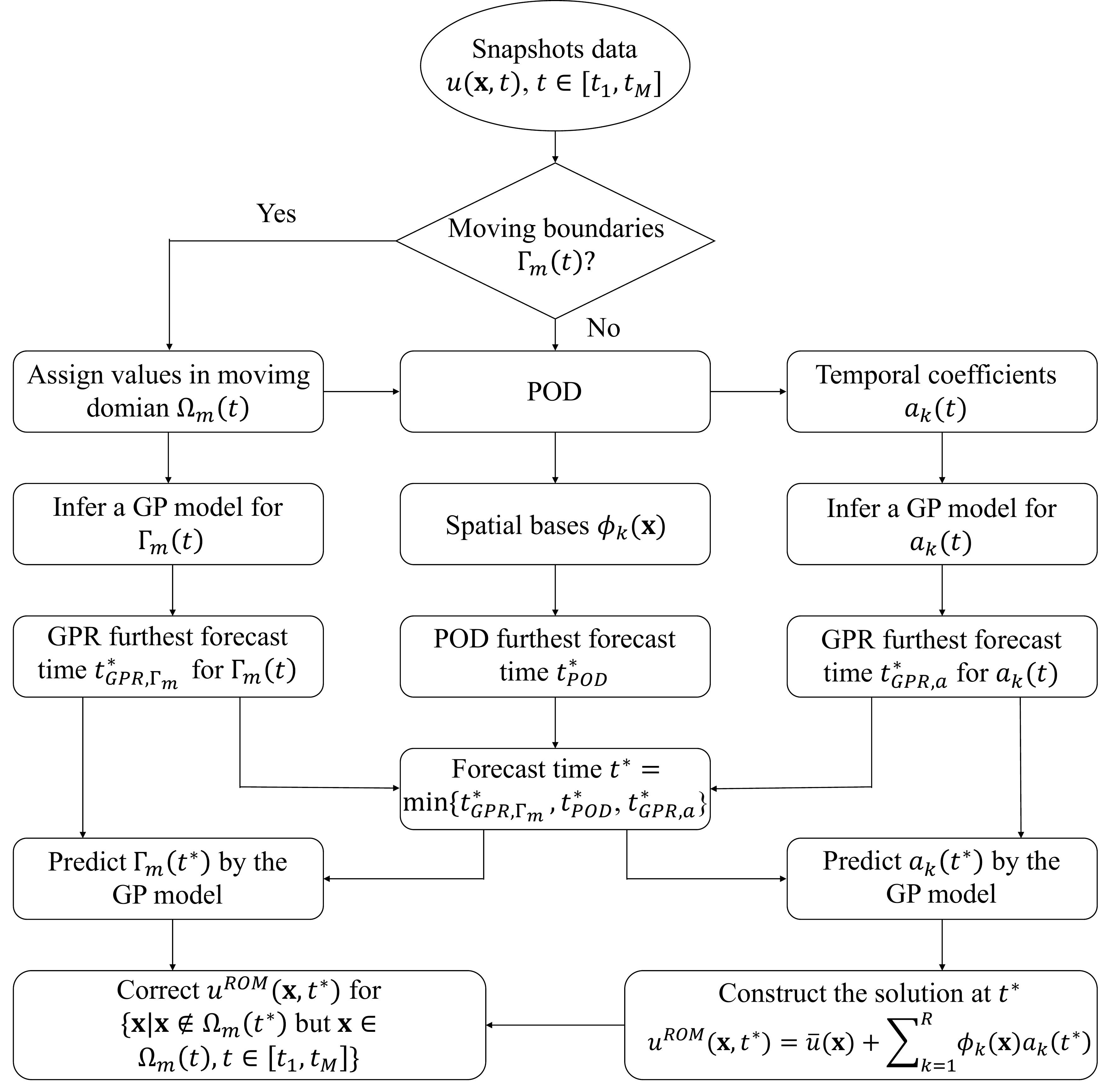}
\caption{Schematic of the proposed data-driven nonintrusive model order reduction method for dynamical systems with moving boundaries.}
\label{fig:schematic_of_method}
\end{figure}

%4) what we propose in this work with problem statement 
In this paper, we present a new data-driven nonintrusive model order reduction method for dynamical systems involving moving boundaries. The dynamical systems can be generally nonlinear. The proposed method is based on the POD, Gaussian process regression (GPR) and moving least squares (MLS) interpolation. It combines three attributes that are not simultaneously satisfied in the existing reduced order modeling methods for dynamical systems with moving boundaries    \cite{ROM_OscillatingCylinder_Liberge2010,ROM_MovingBoundary_Gao2016,NIROM_FSI_XIAO201635}. First, our method only needs snapshots of state variables at discrete time instances and the parameters that characterize the boundaries. Otherwise, it does not require any prior knowledge of the full-order model or the underlying governing equations for construction of the ROM. The snapshot data can be from simulations or experiments. Second, the moving boundaries are not limited to prescribed or periodic motions. Third, the ability of the ROM constructed to forecast the full-order solutions for future times beyond the range of snapshot data is studied. We provide the criteria to determine the furthest future time that the ROM constructed from a given set of data can predict. Figure \ref{fig:schematic_of_method} summarizes the main components of the proposed method.
%%%%%%%%%%%%%%%%%%%%%
%%%%%%%%%%%%%%%%%%%%%

The paper is organized as follows. Section \ref{sec:method} explains in detail each component of the proposed method and provides algorithms for practical implementation. In Section \ref{sec:results}, we outline the results of numerical experiments, where the proposed method is assessed in several benchmark problems. We first present the benchmarks with fixed domains only and then proceed to the cases with moving boundaries. The data used to construct and validate the ROMs are from analytical/numerical solutions and experiments. Finally, we conclude in Section \ref{sec:conclu} and discuss the limitation and possible extensions of the present work.

\section{Methodology}\label{sec:method}
\begin{figure}[htbp]
\centering
\includegraphics[width=5cm]{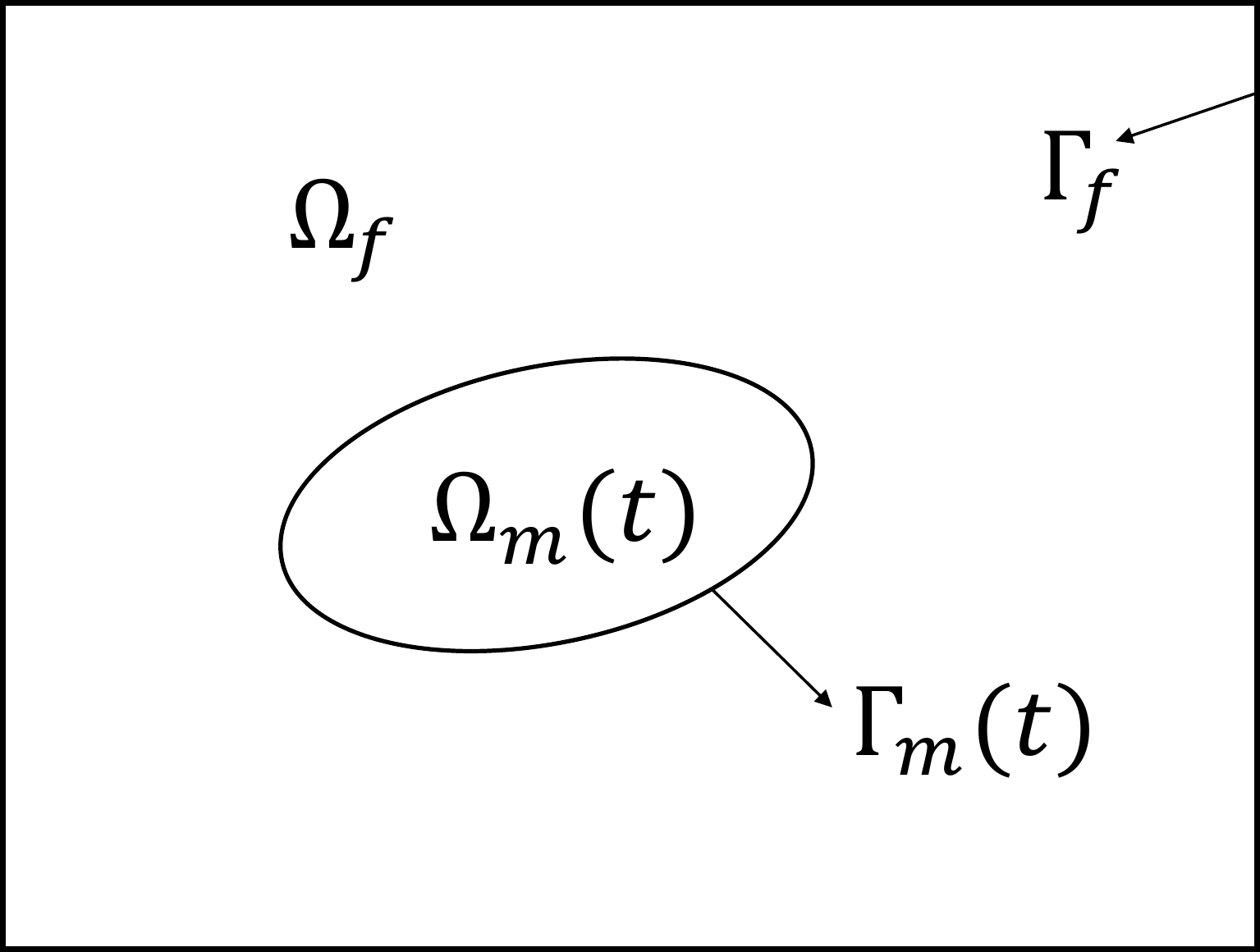}
\caption{Schematic of the setup of spatial domain and boundaries.}
\label{fig:domain}
\end{figure}

Suppose $u({{\mathbf{x}}},t)$ is the full-order solution for a dynamical system. The spatial domain and boundaries for this dynamical system generally consist of the fixed domain $\Omega_f$, the moving domain $\Omega_m$ occupied by other phase(s) (e.g., a cavity or a moving solid body), the fixed boundary $\Gamma_f$, and the moving boundary $\Gamma_m$, as illustrated in Figure \ref{fig:domain}.
%%%%%%%%%%%%%%%%%%%
%%%%%%%%%%%%%%%%%%%

\subsection{POD}\label{subsubsec:POD}
POD starts from the snapshot data of the full-order solutions of the dynamical system, which can be obtained from either numerical simulations or experimental measurements. The snapshot data correspond to $u({{\mathbf x}},t_i)$ at $M$ time instances with $t_i \in \{t_1, t_2, \dots, t_M\}$. Any $u({{\mathbf x}},t_i)$ may be decomposed into
\begin{equation}
u({{\mathbf x}},t_i) = \bar u ({{\mathbf x}}) + \hat u({{\mathbf x}},t_i),\ \ \bar u ({{\mathbf x}}) = \frac{1}{M} \sum\limits_{i = 1}^M u({{\mathbf x}},t_i) \;,
\end{equation}
where $\bar u$ is the temporal mean of snapshot data and $\hat u$ is the fluctuating part. The task is to find a set of spatial bases $\phi_k({\mathbf x})$ and temporal coefficients $a_k(t)$ such that the fluctuating part $\hat u$ can be reproduced as: 
\begin{equation}
    \hat u({{\mathbf x}},t)=\sum\limits_{k = 1}^M a_k(t)\phi_k({{\mathbf x}}).
\label{equ:hatu_decomp}
\end{equation}
To this end, the correlation matrix $\textbf{A}\in \mathbb{R}^{M\times M}$ of the fluctuating parts of snapshot data is constructed by:
\begin{equation}
{A_{ij}} = \int_\Omega  {{{\hat u}}({{\mathbf x}},t_i){{\hat u}}({{\mathbf x}},t_j)d{{\mathbf x}}},
\end{equation}
where $i$ and $j$ refer to the $i$th and $j$th snapshots, respectively. By definition, $\textbf{A}$ is symmetric and semi-positive definite.
% Then for simplicity, we define the inner-product of any two fields $u_1$ and $u_2$ on domain $\Omega$ as 
% \begin{equation}
%     (u_1,u_2)=\int_\Omega  {u_1({{\mathbf x}})u_2({{\mathbf x}})d{{\mathbf x}}}.
% \end{equation}
Thus, the set of spatial bases $\phi_k({\mathbf x})$ (i.e., POD basis function) can be obtained via eigendecomposition of $\textbf{A}$: 
% \begin{equation}
%     CW = C\Lambda,
% \end{equation}
% $\Lambda=diag[\lambda_1, \lambda_2, ..., \lambda_M]$ and $W=[{\mathbf{w}}^1,{\mathbf{w}}^2,..., \mathbf{w}^M]$. We let eigenvalues in descending order.
\begin{equation}\label{equ:phi_k}
    {\phi _k}({{\mathbf x}}) = \frac{1}{{\sqrt {{\lambda _k}} }}\sum\limits_{i = 1}^M {w_i^k\hat u({{\mathbf x}},{t_i})} \;,
\end{equation}
where $\{\lambda_1, \lambda_2, ..., \lambda_M\}$ are eigenvalues in descending order and $\mathbf{w}^k$ is the eigenvector corresponding to $\lambda_k$. The POD basis functions satisfy the condition of orthogonality:
\begin{equation}
    ({\phi _k},{\phi _l}) = \left\{ \begin{array}{l}
1,\ \ k = l\\
0,\ \ k \ne l
\end{array} \right.,
\end{equation}
where $({\phi _k},{\phi _l})=\int_\Omega  {\phi_k({{\mathbf x}})\phi_l({{\mathbf x}})d{{\mathbf x}}}$ defines the inner-product of two functions.

For model order reduction, only the first $R\ll M$ largest eigenvectors (or POD modes) are retained. As a result, a snapshot $\hat u({{\mathbf x}},{t_i})$ can be approximated as: \begin{equation}\label{equ:POD_ti}
    \hat u({{\mathbf x}},{t_i}) \approx \hat u^{POD}({{\mathbf x}},{t_i}) = \sum\limits_{k = 1}^R \sqrt {{\lambda _k}}{w_k^i}{\phi_k}({{\mathbf x}}) 
\end{equation}
with the following truncation error: 
\begin{equation}\label{equ:POD_truncation_error}
    \epsilon^\text{POD} = \frac{1}{M}\sum\limits_{i=1}^M \left \|\hat u({{\mathbf x}},{t_i})-\hat u^{POD}({{\mathbf x}},{t_i}) \right \|_{L_2} = \frac{1}{M}\sum\limits_{i=1}^M \left \| \sum\limits_{k=R+1}^{M} {\sqrt {{\lambda_k}}{w_k^i}{\phi _k}({{\mathbf x}})} \right \|_{L_2} = \sqrt{\sum\limits_{k=R+1}^M \lambda_k} \; .
\end{equation}
By increasing $R$, i.e., by keeping more POD modes, the truncation error of POD ($\epsilon^\text{POD}$) can be reduced. Given Eq. \eqref{equ:POD_ti}, Eq. \eqref{equ:hatu_decomp} can be truncated to $R$ dominant POD modes as:  
\begin{equation}
    \hat u({{\mathbf x}},t) \approx \sum\limits_{k = 1}^R a_k(t)\phi_k({{\mathbf x}})\; .
\end{equation}
Define the relative root mean square (RRMS) truncation error as: 
\begin{equation}
    RRMS \ \ Error = \sqrt{ \frac{\sum\limits_{k=R+1}^M \lambda _k}{\sum\limits_{k=1}^M \lambda_k} } \;.
\end{equation}
By requiring $ RRMS \ \ Error < \alpha^\text{POD} $ with $\alpha^\text{POD}$ the preset threshold, we can determine how many dominant POD modes ($R$) to retain. 

Once the spatial POD bases $\phi_k({\mathbf x})$ are determined from Eq. \eqref{equ:phi_k}, the next task is to construct the temporal coefficients $a_k(t)$ such that the full-order solution can be effectively reproduced at a given time $t'$ either within or beyond the database range:
%%%%%%%%%%%%%%%%%%%%%
\begin{equation}\label{equ:ROM_POD}
u({{\mathbf x}},{t'}) = \bar u ({{\mathbf x}}) + \hat u({{\mathbf x}},t') \approx u^\text{ROM}({{\mathbf x}},t')= \bar u ({{\mathbf x}}) + \sum\limits_{k = 1}^R a_k(t')\phi_k({{\mathbf x}}) ~~~~\text {for}~ {\mathbf x} \in \Omega_f~~ \text{and}~~ t' \in [t_1,t_M]~~ \text{or}~~ t'>t_M \; .
\end{equation}
%%%%%%%%%%%%%%%%%%
Eq. \eqref{equ:ROM_POD} is hence the ROM constructed via POD.
% Consider a black-box full order trajectories $u({{\mathbf x}},t)$, using proper orthogonal decomposition introduced in section 2, we can get the basis function $\phi_k({\mathbf x})$ over ${{\mathbf x}} \in \Omega$ and $a_k(t)$ over $t \in [T_1,T_2]$, where $k=1,2,...,R$. The time domain $[T_1,T_2]$ is discretized into $M \in \mathbb{N}$ equidistant time steps, $T_1=t_0, t_1,..., t_M=T_2$ and $dt=\frac{T_2-T_1}{M}$. The problem is to predict the reduced order solution $\hat u({{\mathbf x}},t^*)$ at $t^*=T_2+dt,T_2+2dt,...,T_2+M_1dt $, where $M_1$ is the assumed prediction length in time domain. 
%Supposing the space basis $\phi_k({{\mathbf x}})$ would not change when the assumed prediction length $M_1$ is small, 

\subsection{Classical intrusive model order reduction based on Galerkin Projection}\label{subsec:Galerkin}
In this section, we review the classical, intrusive approach to determine the temporal coefficients $a_k(t)$ based on the Galerkin projection. Assume the dynamical system of interest can be described by the following full-order model:
\begin{equation}
    \dot{u}(\textbf{x},t) = \mathcal{F}(u)\; ,
\label{equ:full_order}
\end{equation}
where $\mathcal{F}$ denotes  general operators, e.g.,  spatial differential operators. 
%The idea of Galerkin projection is to project the original full-order model Eq. \eqref{equ:full_order} into the POD manifold: $\dot{u} = P_s \mathcal{o}(u)$, where $u(t) \in S$ and $P_s$ is the projection map. Given the reduced order space $S$ of $H$, Galerkin projection can give an approximate dynamic system evolving on $S$, which is denoted as following:
% \begin{equation}
%     a_k(t) = (\hat u({{\mathbf x}},t),\phi_k({{\mathbf x}})) \; 
% \end{equation}
To determine $a_k(t)$, the idea of Galerkin projection is to project the original full-order model (Eq. \eqref{equ:full_order}) onto the POD manifold, by which and noting the orthogonality of POD basis functions a set of ordinary differential equations (ODEs) of $a_k(t)$ can be derived as: 
\begin{equation}\label{equ:Galerkin_proj}
    \dot{a}_k(t) = (\mathcal{F}(u), \phi_k), \ \ \ \ k=1,2,...,R \;.
\end{equation}
To close the above ODEs, the initial conditions can be given using the following projection:
\begin{equation}
    {a_k}(t = 0) = (u(\mathbf x,t = 0) - \bar u(x),{\phi _k}).
\end{equation}
Hence, to solve for $a_k(t)$, one needs to evaluate the inner product in the right hand side of Eq. \eqref{equ:Galerkin_proj}, which requires the knowledge of $\mathcal{F}$. If $\mathcal{F}$ is not attainable for ``black-box" problems, e.g., for experimental data, the Galerkin projection is not feasible. For problems with moving boundaries, even if $\mathcal{F}$ is known, it needs to include the contributions from the moving boundaries, and hence evaluating the inner product $(\mathcal{F}(u), \phi_k)$ demands expensive computation. For example, in fluid-solid interactions, the NS equations that govern the fluid flow are augmented with extra body-force terms to account for the coupling of fluid flow and solid motion \cite{ROM_MovingBoundary_Gao2016,ROM_MovingBoundary_GaoPHDthesis2018}. For generally unsteady solid motion, the contributions of these extra terms in the inner product $(\mathcal{F}(u), \phi_k)$ must be recomputed at each time step, which significantly increases the computational cost of the ROM. In addition, the ODEs in Eq.  \eqref{equ:Galerkin_proj} usually must be numerically solved. Thus, the temporal integrator and time step size must be chosen properly for the desired accuracy and stability. 

\subsection{Gaussian process-enabled nonintrusive model order reduction}\label{subsec:Gaussian_ROM}

%\textcolor{blue}{I use $t'$ to represent the future time to predict and $t^*$ to represent the further forecast time.}

In this paper, we propose a Gaussian process-enabled nonintrusive model order reduction method. Different from the Galerkin projection discussed in \S \ref{subsec:Galerkin}, the proposed method does not rely on any prior knowledge of the full-model operator and is well suited for black-box problems or the problems requiring nontrivial evaluation and solution of Eq. \eqref{equ:Galerkin_proj}. For each temporal coefficient $a_k(t)$, we infer a Gaussian process model from the dataset $a_k(t_i)$ for $i=1,\dots, M$. The obtained Gaussian process model is then employed to predict $a_k(t’)$ at a given time $t'$, where $t'>t_1$ can be within ($t'<t_M$) or beyond ($t'>t_M$) the dataset range. We provide the criteria to determine the furthest time extrapolation permitted in this method. Once $a_k(t’)$ is predicted, the full-order solution can be reconstructed from Eq. \eqref{equ:ROM_POD}.

\subsubsection{Inference of temporal coefficients via Gaussian process}\label{subsubsec:Gaussian}
Consider the data set $\mathbf{t} =[t_1,...,t_M]^T$ as the training inputs and $\mathbf{t}'=[t'_1,t'_2,...,t'_N]^T$ as the inputs for prediction. For each $k=1,2,...,R$, the training outputs are $\mathbf{y} =[a_k(t_1),a_k(t_2),...,a_k(t_M)]^T$, and the predicted outputs are $\mathbf{y}'=[a_k(t'_1),a_k(t'_2),...,a_k(t'_{N})]^T$. The Gaussian process model is given by: 
$\mathbf{y} (\mathbf{t} ) \sim  \mathcal{GP}(\boldsymbol{\mu}(\mathbf{t} ),\mathbf{C} (\mathbf{t} ,\mathbf{t} '))$
, where $\boldsymbol{\mu}(\mathbf{t})$ is the mean function and $\mathbf{C} (\mathbf{t} ,\mathbf{t}')$ is the covariance function. To predict $\mathbf{y}'$, the key is to determine the posterior distribution $p(\mathbf{y'}|\mathbf{y} )$. Since the joint distribution of the training outputs $\mathbf{y} $ and the predicted outputs $\mathbf{y}'$ satisfy:
\begin{equation}
\begin{bmatrix}
    \mathbf{y} \\
    \mathbf{y}'
\end{bmatrix}
\sim \mathcal{N}
\begin{pmatrix}
\begin{bmatrix}
    \boldsymbol{\mu}(\mathbf{t} )\\
    \boldsymbol{\mu}(\mathbf{t}')
\end{bmatrix},
\begin{bmatrix}
    \mathbf{C} (\mathbf{t} ,\mathbf{t} )+\sigma^2 \mathbf{I}  &\mathbf{C} (\mathbf{t}',\mathbf{t} )^T \\
    \mathbf{C} (\mathbf{t}',\mathbf{t} ) & \mathbf{C} (\mathbf{t}',\mathbf{t}')
\end{bmatrix}
\end{pmatrix} \; ,
\end{equation}
%where $\mathbf{\mu}({\mathbf x})=\mathbf{\mu}$, $\mathbf{\mu}(\mathbf{Z})=[\mu(z_1),\mu(z_2),...,\mu(z_m)]^T $, $\mathbf{K}({\mathbf x},{\mathbf x})=\mathbf{K}$, $\mathbf{K}(\mathbf{Z},{\mathbf x})$ is an $m \times n$ matrix of which the $(i,j)$-th element $[\mathbf{K}(\mathbf{Z},{\mathbf x})]_{ij}=k(z_i,x_j)$, and $\mathbf{K}(\mathbf{Z},\mathbf{Z})$ is an $m \times m$ matrix of which the $(i,j)$-th element $[\mathbf{K}(\mathbf{Z},\mathbf{Z})]_{ij}=k(z_i,z_j)$. 
where $\sigma^2$ is the variance of identically independent normally distributed noise (with zero mean) assumed in the Gaussian process model,
the posterior distribution can be determined as: 
\begin{equation}
    p(\mathbf{y'}|\mathbf{y} )=\mathcal{N}(\hat{\boldsymbol{\mu}},\hat{\mathbf{C} }),
\end{equation}
where
\begin{equation}\label{equ:posterior_muC}
\begin{split}
    \hat{\boldsymbol{\mu}}&=\mathbf{C} (\mathbf{t'},\mathbf{t} )\left [\mathbf{C} (\mathbf{t} ,\mathbf{t}) +\sigma^2\mathbf{I}\right]^{-1}\left [\mathbf{y} -\boldsymbol{\mu}(\mathbf{t} )\right ]+\boldsymbol{\mu}(\mathbf{t'}),\\
    \hat{\mathbf{C} }&=\mathbf{C} (\mathbf{t'},\mathbf{t'})-\mathbf{C} (\mathbf{t'},\mathbf{t} )\left [\mathbf{C} (\mathbf{t} ,\mathbf{t} )+\sigma^2\mathbf{I}\right]^{-1}\mathbf{C} (\mathbf{t'},\mathbf{t} )^T.
\end{split}
\end{equation}
In this work, the covariance function $\mathbf{C} (\mathbf{t} ,\mathbf{t} ')$ is assumed a squared exponential form, i.e.,
\begin{equation}
    C_{ij}(t_i,t'_j;\boldsymbol{\theta})=\theta_f^2 \exp \left [-\frac{1}{2} \theta_l^2(t_i-t'_j)^2 \right ],
\end{equation}
where $\boldsymbol{\theta} = (\theta_f, \theta_l)$ denotes the hyper-parameters. To determine the hyper-parameters $\boldsymbol{\theta}$ as well as $\sigma^2$ in Eq. \eqref{equ:posterior_muC}, we minimize the negative log marginal likelihood \cite{GPML_2006}:
\begin{equation}\label{equ:marginal_likelihood}
    -\log p(\mathbf{y} |\boldsymbol{\theta}, \sigma^2)=\frac{1}{2}\mathbf{y} ^T\mathbf{\mathcal{C}}^{-1} \mathbf{y} +\frac{1}{2} \log |\mathbf{\mathcal{C}} | + \frac{M}{2} \log (2\pi) \; ,
\end{equation}
where $\mathbf{\mathcal{C}} = \mathbf{C} (\mathbf{t} ,\mathbf{t})+\sigma^2\mathbf{I}$, via the Quasi-Newton optimizer L-BFGS \cite{BFGS_1989Liu}. Here, $M$ is the total number of training data; and $|\mathbf{\mathcal{C}}|$ is the determinant of matrix $\mathbf{\mathcal{C}}$. The marginal likelihood as in Eq. \eqref{equ:marginal_likelihood} is chosen because it entails a trade-off between data-fit and model complexity: while the term $ \frac{1}{2}\mathbf{y} ^T\mathbf{\mathcal{C}} \mathbf{y}  $ targets better fitting the training data, the term $\log |\mathbf{\mathcal{C}}|$ penalizes the model complexity. A key advantage of GPR is that uncertainty bounds can be analytically derived from the hyper-parameters. 

The method proposed herein is different from that in \cite{ROM_GPR_Wan2017}, where the time derivative $\dot{a_k}=f(a_1,\dots,a_R)$ is modeled as a Gaussian process, instead of $a_k(t)$ as in this work. To obtain an accurate model for the time derivative, it requires to approximate the time derivative by a numerical integrator of high-order accuracy in the training process. To predict $a_k(t'>t_M)$ from the trained model of $\dot{a_k}$, it needs to march from $a_k(t_M)$ by the increment of $\delta t$ (time step size) each time and finally reaches $t'$. Also, compared with training a model for $a_k(t)$, it needs more training data. In addition, if the data are noisy, e.g., from experimental measurements, training an accurate model for the time derivative can be challenging. Due to these considerations, we choose to model $a_k(t)$ itself as a Gaussian process.

\subsubsection{Criteria for furthest forecast time}\label{subsubsec:criteria_t*}

In this section, we propose two criteria to determine the furthest time extrapolation  that the ROM can achieve. Since the ROM is constructed via POD and GPR, each of them poses a constraint on the time extrapolation.  

To forecast the solution at $t'>t_M$, the same POD subspace is assumed, which is constructed by the dominant spatial bases extracted from the snapshot data. The error arising from this assumption constrains the time extrapolation of the ROM. While in some cases, the dominant POD bases vary slowly in time, e.g., flow dynamics at low Reynolds numbers described by the Burgers equation; in other cases, they can change rapidly, e.g., shock wave and advection problems. Theoretically, the decay rate of  Kolmogorov $n$-width \cite{nwidth_Kolmogoroff1936,nwidth_Pinkus1986} can measure how fast the POD spatial bases change in time for a dynamical system.  The Kolmogorov $n$-width is defined as \cite{decay_of_Kol_2019}:
%%%%%%%%%%%%
\begin{equation}
\label{equ:Kol_width}
    d_n(\mathcal{M})= \inf_{\mathcal{S}_n} \sup_{u \in \mathcal{M}} \min_{v \in \mathcal{S}_n} || u-v  || \; ,
\end{equation}
%%%%%%%%%%%%
where $\mathcal{M}$ is the manifold of the full-order solutions over the entire time considered, including the snapshot data;  $\mathcal{S}_n$ denotes $n$-dimensional linear subspaces constructed from the snapshot data; $u$ is a full-order solution in $\mathcal{M}$; and $v$ represents a reduced-order solution in $\mathcal{S}_n$. In the context of this paper, $\mathcal{S}_n$ refers to the subspaces formed by the first $R$ POD spatial bases. The Kolmogorov $n$-width as in Eq. \eqref{equ:Kol_width} provides the worst-case error resulting from a projection onto the best-possible linear subspace of a given dimension $n$ \cite{decay_of_Kol_2019}. 
%In other words, Kolmogorov $n$-width represents a barrier on system's linear reducibility. 
If the Kolmogorov $n$-width decays faster with respect to $n$, the reduced-order solution in the subspace constructed by the first $R$ POD bases can be an accurate approximation of the full-order solution for longer time beyond the range of snapshot data. 
%While the decay of Kolmogorov $n$-width is slow, the Kolmogorov barrier arises. To address this issue, the author of \cite{POD_time_decomposition2019} decomposes the time domain into many subdomains and employs POD into every subdomain to find the optimal subspace. Through this method, Kolmogorov $n$-width decays rapidly in every subdomain and thus overcomes the Kolmogorov barrier. The time domain needs to be decomposed into more parts when the decay of Kolmogorov $n$-width is slower, which indicates that the dominant spatial bases change rapidly. Besides, 
%\textcolor{red}{It has been been  proven  that  for  certain  linear,  coercive  parameterized  problems, the  Kolmogorov $n$-width decays exponentially fast} \cite{n-width_exp2012}, i.e.,
% \begin{equation}
% \label{equ:n-width_exp}
%     d_n(\mathcal{M}) \le C e^{\beta n}
% \end{equation}
% with some constants $C < \infty$ and $\beta >0$. 
In practice, a direct evaluation of the Kolmogorov $n$-width and its decay rate is difficult. In this paper, we instead estimate the decay rate of dominant eigenvalues, i.e., $\frac{\ln {\lambda_2}-\ln{\lambda_{R+2}}}{R}$. Note that $\lambda_2$ dominates the POD truncation error if only the first POD mode is retained in the ROM; $\lambda_{R+2}$ dominates the POD truncation error if the first $R+1$ modes are retained in the ROM. We employ this decay rate of dominant eigenvalues to characterize how fast the POD spatial bases change in time for a dynamical system. Based on that, 
we propose the following constraint posed by POD on the furthest forecast time of the ROM:
%%%%%%%%%%%%%%
\begin{equation}
\label{equ:Predict_POD_limit}
    \frac{\Delta t_{\text{POD}}^*}{\Delta t_{\text{Snapshots}}} \le \beta^{\text{POD}} \frac{\ln {\lambda_2}-\ln{\lambda_{R+2}}}{R},
\end{equation}
%%%%%%%%%%%%%%
% \begin{equation}
% \label{equ:Predict_POD_limit_1}
%     \frac{\Delta t_{\text{POD}}^*}{\Delta t_{\text{Snapshots}}^*} \le \beta^{\text{POD}} \frac{\ln \left[ {\epsilon^\text{POD}(1)}\right]-\ln\left[{\epsilon^\text{POD}(R+1)}\right]}{R} \; ,
% \end{equation}
% %%%%%%%%%%%%%%
where $\Delta t_{\text{snapshots}}=t_M-t_1$ denotes the time span of snapshot data; $\Delta t^*_{\text{POD}} = t_{\text{POD}}^*-t_M$ defines how long the furthest forecast time $t_{\text{POD}}^*$ is beyond $t_M$ (the latest time of snapshot data); and $\beta^{\text{POD}} < 1$ is a non-dimensional constant related to the tolerance for the error induced by the change of POD spatial bases. %Throughout the numerical experiments conducted in this paper, it is set as $\beta^{\text{POD}} = 0.3-0.8$. 
Eq. \eqref{equ:Predict_POD_limit} provides the criterion to determine the furthest forecast time $t^*_\text{POD}$ permitted in time extrapolation using the dominant POD spatial bases extracted from the $M$ snapshot data.
%, and based on that we propose a criterion to determine how long we can assume the same dominant POD spatial bases within acceptable error tolerance.
Note that although the eigenvalues can be over/under-estimated by picking different snapshots in a fixed range, the decay rate of eigenvalues has little change if we pick different snapshots in the same range. Thus, the criterion in Eq. \eqref{equ:Predict_POD_limit} does not require snapshots to be uniformly sampled.

The second criterion is from the uncertainty level of GPR for predicting each $a_k$ at a given time $t'$. The standard deviation (or uncertainty level) $\hat \sigma_k (t')$ grows when the forecast time $t'$ is further from the snapshot data, and hence it poses a constraint on the furthest forecast time $t_{\text{GPR},a}^*$ permitted by GPR. Considering each temporal mode $a_k$ contributes differently in the total energy of the dynamical system, the weighted average of standard deviation is defined to measure the uncertainty level:
\begin{equation}\label{equ:weighted_sd}
    \hat \sigma (t') = \frac{\sum_{k=1}^{R}\lambda_k \hat \sigma_k(t')}{\sum_{k=1}^{M}\lambda_k}.
\end{equation}
And based on that, we propose the following constraint posed by GPR on the furthest forecast time of the ROM: 
%%%%%%%%%%%%%%%%%
\begin{equation} \label{equ:Gaussian_presetToL}
    \frac{\sum_{k=1}^{R}\lambda_k \hat \sigma_k(t_{\text{GPR},a}^*) }{\sum_{k=1}^{R} \lambda_k |\hat {\mu}_k(t_{\text{GPR},a}^*) |} \le \beta^{\text{GPR},a}
\end{equation}
%%%%%%%%%%%%%%%%%
where $\beta^{\text{GPR},a}$ denotes the preset tolerance. Larger $\beta^{\text{GPR},a}$ means larger tolerance for the uncertainty associated in GPR. From Eq. \eqref{equ:Gaussian_presetToL}, the furthest forecast time permitted in time extrapolation by GPR, $t_{\text{GPR},a}^*$, can be adaptively determined according to the preset tolerance. %\textcolor{red}{Throughout the numerical experiments in this paper,  $\beta^{\text{GPR,a}}=0.01-0.1$.}

\subsubsection{Algorithm for Gaussian process-enabled nonintrusive model order reduction}\label{subsubsec:algorithm_ROMGaussian}
The procedure of the proposed nonintrusive model order reduction method based on the POD and GPR is summarized in Algorithm 1. 
%%%%%%%%%%%%%%%%%%
\begin{algorithm}
\caption{Gaussian process-enabled nonintrusive model order reduction}
\label{algorithm1}
\begin{algorithmic}[1]
\Require
$M$ snapshot data of  $u({{\mathbf x}},t_i)$, $i=1,2,...,M$
\Ensure
 $u({{\mathbf x}},t^*)$ predicted at $t^* > t_M $
\State Determine $R$ dominant POD spatial bases $\phi_k({{\mathbf x}})$, $k=1,2,...,R$, with $RRMS Error < \alpha^{\text{POD}}$
\State Project $u({{\mathbf x}},t_i)$ to the reduced space and obtain $a_k(t) = (\hat u({{\mathbf x}},t),\phi_k({{\mathbf x}}))$
\State Determine the furthest forecast time permitted by the POD, $t_{\text{POD}}^*$, from Eq. \eqref{equ:Predict_POD_limit}
\For {$k=1,2,...,R$}
\State Infer a Gaussian Process model for $a_k(t)$ following \S \ref{subsubsec:Gaussian}
%\State Use $[t_0,t_1,...,t_M]^T$ as training inputs, and $[a_k(t_1),a_k(t_2),...,a_k(t_M)]^T$ as training outputs to optimize the hyper-parameters and noise in Gaussian Process Regression. 
\State Determine the furthest forecast time permitted by the GPR, $t_{\mathrm{GPR},a}^*$, from Eq. \eqref{equ:Gaussian_presetToL}
\EndFor
\State Determine the final furthest forecast time as $t^* = \min \{ t_\text{POD}^*, t_{\mathrm{GPR},a}^*\}$
\State Predict each $a_k(t^*)$ by the inferred Gaussian Process models
\State Reconstruct the full-order solution $u({{\mathbf x}},t^*)$ by Eq. \eqref{equ:ROM_POD}
\State \Return $u({{\mathbf x}},t^*)$
\end{algorithmic}
\end{algorithm}
%%%%%%%%%%%%%%%%

When the numerical simulations or experimental measurements are demanding or expensive for a dynamical system, one can employ the ROM to forecast the full-order solutions at future times beyond the database range. As we have discussed in \S\ref{subsubsec:criteria_t*}, the furthest forecast time is constrained by the POD and GPR. Thus, after the furthest forecast time of the ROM is reached, the consecutive prediction would still need numerical simulations or experimental measurements. Thus, for long-time prediction of a dynamical system, the numerical simulations/experimental measurements and the ROM can be called alternatively in an automated process:
\begin{enumerate}
    \item Generate $M$ snapshot data using numerical simulations/experimental measurements. 
    \item Construct a ROM from the snapshot data and determine the furthest forecast time $t^*$ by the proposed criteria. 
    \item Call the ROM to predict the full-order solutions at any desired times until $t=t^*$. The full-order solution at $t^*$ is taken as the initial condition for the consecutive numerical simulations/experimental measurements.
    \item Repeat Steps 1-3 until the target prediction time.
\end{enumerate}
This process adaptively combines numerical simulations/experimental measurements with ROMs and hence optimizes the efficiency for long-time prediction of a dynamical system.

%More specifically, we first generate $M$ snapshot data from the DNS or experiments and then construct a ROM to forecast the full-order solution at any desired time $t'>t_M$ up to the furthest forecast time $t^*$ determined by the proposed criteria. Taking the full-order solution reconstructed by the ROM at $t^*$ as the initial condition, the DNS or experiments are restarted to generate another $M$ snapshot data, which are then used to construct a new ROM to forecast the full-order solution until the new $t^*$. According to the proposed criteria, the furthest forecast time $t^*$ for different ROMs can be different. Repeating this process, we adaptively combine the DNS/experimental measurements with the ROM so as to optimize the efficiency of long-time prediction for a dynamical system.

\subsection{Nonintrusive model order reduction for problems with moving boundaries}\label{subsec:MB_Gaussian_correction}
%In this section, we focus on the problem with moving boundary. In this type of problems, the method of body motion prediction is discussed in Section 4.1. And a correction method, which are dealing with the field of moving boundary, is presented in Section 4.2. Section 4.3 analyzes the error of this method.
In the presence of time-evolving boundaries $\Gamma_m(t)$, reproducing the full-order solution from Eq. \eqref{equ:ROM_POD} needs additional efforts. First, the snapshot data are also needed in $\Omega_m$. Thus, we need to assign the values of $\hat u({{\mathbf x}},t_i)$ for ${\mathbf x} \in \Omega_m$ and $t_i \in \{ t_1, t_2, \dots, t_M\}$. If a moving solid body is involved (in fluid-solid interaction), the values of the velocity in $\Omega_m$ coincide with the motion of the solid body; and the values of the pressure in $\Omega_m$ are assigned via interpolation of the surrounding fluid pressure in $\Omega_f$. If a cavity is involved, 
the values of the state variable in $\Omega_m$ are extrapolated by least squares from the values in $\Omega_f$ near the cavity. With the snapshot data given in both $\Omega_f$ and $\Omega_m$, the POD can be conducted for the entire domain $\Omega_f \cup \Omega_m$.

\subsubsection{Inference of the time-evolution of moving boundaries}\label{subsubsec:infer_MB}
The next task is to infer the time-evolution of the moving boundaries given a set of trajectory data. We assume the evolution of a moving boundary can be fully determined by a finite set of parameters, for example, the radius of a spherical surface (see \S \ref{subsec:buble_cavity}) or the translational and rotational displacements and velocities of the center of mass of a rigid solid body (see \S \ref{subsec:FSI}). Without loss of generality, we denote the parameters that fully characterize a time-evolving boundary $\Gamma_m(t)$ as $\boldsymbol{\gamma} (t)=[\gamma_1(t), \gamma_2(t), \dots]$. 
For each $\gamma_l(t)$, we seek a Gaussian process model using $\bm{t} =[t_1,...,t_M]^T$ as the training inputs and $[\gamma_l(t_1),\gamma_l(t_2),...,\gamma_l(t_M)]^T$ as the training outputs. By the constructed Gaussian process model, we predict $\gamma_l(\mathbf{t'})$ at a future time series $\mathbf{t'}>t_M$. We consider a Gaussian process model ${\gamma_l}(\mathbf{t} ) \sim  \mathcal{GP}(\boldsymbol{\mu}^{\gamma_l}(\mathbf{t} ),\mathbf{C} ^{\gamma_l}(\mathbf{t} ,\mathbf{t}'))$, where $\boldsymbol{\mu}^{\gamma_l}(\mathbf{t} )$ is the mean function and $\mathbf{C} ^{\gamma_l}(\mathbf{t} ,\mathbf{t}')$ is the covariance function. The procedure to construct this Gaussian process model is the same as described in \S\ref{subsubsec:Gaussian}. The furthest forecast time for $\Gamma_m(t)$ is determined by $t_\mathrm{GPR,{\Gamma_m}}^* = \min \{ t_\mathrm{GPR,{\gamma_1}}^*, t_\mathrm{GPR,{\gamma_2}}^*, \dots \}$. And each $t_\mathrm{GPR,{\gamma_l}}^*$ for $\gamma_l(t)$ is determined from:
%%%%%%%%%%%%%%%%
\begin{equation}
\label{equ:predict_boudary_limit}
   \frac{ \hat \sigma^{\gamma_l}(t_{\mathrm{GPR,\gamma_l}}^*) }{|\hat {\mu}^{\gamma_l}(t_{\mathrm{GPR,\gamma_l}}^*) |} \le \beta^{\mathrm{GPR},\Gamma_m}\; ,
\end{equation}
%%%%%%%%%%%%%%%%
where $\beta^{\mathrm{GPR},\Gamma_m}$ is the preset tolerance.

\subsubsection{Correction near the moving boundaries} \label{subsubsec:correction_MB}
We note that when reproducing the full-order solution in $\Omega_f$ at a future time $t'>t_M$, the largest errors appear in the regions near the moving boundaries on the downstream side. %, as illustrated in Figure \ref{fig:shaded_region} (shaded region). 
The large errors stem from the fact that those regions were part of $\Omega_m$ but not in $\Omega_f$ during $[t_1, t_M]$, and the gradient of the solution (e.g., velocity gradient) is discontinuous across $\Gamma_m$. Thus, we propose a correction step to recover the accuracy of the prediction for those regions. In particular, the correction is achieved via high-order interpolation from their neighbor regions using the MLS \cite{Wendland_Book2005}. 
% %%%%%%%%%%%%%%%%%%%%%%%%%
% \begin{figure}[htbp]
% \centering
% \includegraphics[width=6cm]{Figures/moving_domain.pdf}
% \caption{At $t=t^*>t_M$, the shaded region in $\Omega_f$ was previously part of $\Omega_m$ for \textcolor{red}{$t\in[t_1, t_M]$}. \textcolor{red}{Use arrow to point to the boundary}}
% \label{fig:shaded_region}
% \end{figure}
% %%%%%%%%%%%%%%%%%%%%%%

More specifically, consider a point ${\mathbf x}_p \in \Omega_m ~\text{during}~ [t_1, t_M] ~\text{but}~\in \Omega_f ~\text{at}~ t>t_M$, a kernel length $h$ is set to search the neighbor points ${\mathbf x}_q$ satisfying $\|{\mathbf x}_q-{\mathbf x}_p \| < h$ and ${\mathbf x}_q \in \Omega_f$ during $[t_1,t^*]$. Here, the furthest forecast time $t^*$ is determined by $t^* = \min \{ t_\text{POD}^*, t_{\text{GPR},a}^*, t_{\text{GPR},{\Gamma_m}}^*\}$. A polynomial $u_h({\mathbf x})=\textbf{P}^\intercal({\bf x}) \textbf{c}^*$ of order $s$ is sought to approximate $u$ at ${\mathbf x}_p$, where $\textbf{P}({\mathbf x})$ denotes the polynomial basis. The coefficient vector $\mathbf{c}^*$ is determined by minimizing the following weighted residual functional:  
%%%%%%%%%%%%%%%%%%%%%%%
\begin{equation}\label{equ:MLS_min}
\textbf{c}^*_p = \argmin \limits_{\textbf{c}_p} \sum_q \left[ u({\mathbf x}_q) - \textbf{P}^\intercal({\mathbf x}_q)\textbf{c}_p \right]^2 W_{pq} \;,
\end{equation}
%%%%%%%%%%%%%%%%%%%%%%%
where $q \in \mathbb{N}_{h,p} = \left\{ {\mathbf x}_q~ \text{s.t.}~ \|{\mathbf x}_q-{\mathbf x}_p \| <  h, {\mathbf x}_q \in \Omega_f ~\text{during}~ [t_1,t^*]\right\}$ and $W_{pq}=W(\|{\mathbf x}_q-{\mathbf x}_p \|)$ with $W$ a positive function with the compact support $h$. The choice of $h$ is determined by the polynomial order $s$ to ensure necessarily sufficient neighbor points ${\mathbf x}_q$ to solve Eq. \eqref{equ:MLS_min}. Due to its polynomial consistency, the MLS interpolation can achieve high-order accuracy by taking large $s$, e.g., $s=3$ used in this work. Following standard arguments for the minimization of a symmetric positive definite quadratic form, the solution of Eq. \eqref{equ:MLS_min} is given by:
%---
\begin{equation}
\mathbf{c}^*_p= \left(\sum_{j \in \mathbb{N}_{h,p}} \textbf{P}_p({\mathbf x}_j) W_{pj} \textbf{P}_p^\intercal({\bf x}_j)\right)^{-1} \left( \sum_{q \in \mathbb{N}_{h,p}} \textbf{P}_p({\mathbf x}_q) W_{pq} u({\mathbf x}_q) \right)\; . 
\label{equ:MLS_approx}
\end{equation}
%---
%For example, for the 3rd order two dimensional MLS interpolation, $\xi$ must be large enough to include at least 10 neighbor grids. 
The solution reproduced from the ROM (Eq. \eqref{equ:ROM_POD}) at ${\mathbf x}_p$ is then replaced with the interpolated value, i.e., $u({\mathbf x}_p)\approx u_h({\mathbf x}_p) = \textbf{P}_p^\intercal({\mathbf x}_p) \mathbf{c}^*_p$. By such, we improve the the accuracy of the ROM's predictions for the regions that fall in $\Omega_f$ at $t>t_M$ but belong to $\Omega_m$ during $[t_1, t_M]$. 

\subsubsection{Algorithm} \label{subsubsec:algorithm_ROMMB}
Algorithm \ref{algorithm2} outlines the procedure of the proposed nonintrusive model order reduction method for problems with moving boundaries, which augments Algorithm 1 with the steps to infer the time evolution of the moving boundaries and to correct the solutions for the regions in $\Omega_f$ at $t^*>t_M$ but in $\Omega_m$ during $[t_1, t_M]$.  
%%%%%%%%%%%%%%%%%%%%%%%%%%%
\begin{algorithm}[H]
\caption{Nonintrusive model order reduction for problems with moving boundaries}\label{algorithm2}
\begin{algorithmic}[1]
\Require
$M$ snapshot data of $u({{\mathbf x}},t_i)$ and  $\gamma_l(t_i)$ with  $i=1,2,\dots,M$ and $l = 1,2,\dots$
\Ensure
$u({{\mathbf x}},t^*)$ and $\gamma_l(t^*)$ predicted at $t^*$
\For {$l=1,2,\dots$}
\State Infer a Gaussian Process model for $\gamma_l(t)$ following \S \ref{subsubsec:infer_MB}
\State Determine  $t_{\mathrm{GPR,\gamma_l}}^*$ by Eq. \eqref{equ:predict_boudary_limit}
\EndFor 
\State Determine the furthest forecast time permitted for $\Gamma_m$ as $t_\mathrm{GPR,{\Gamma_m}}^* = \min \{ t_\mathrm{GPR,{\gamma_1}}^*, t_\mathrm{GPR,{\gamma_2}}^*,\dots \}$
\State Generate snapshot data for $\mathbf{x} \in \Omega_m(t)$, $t \in [t_1, t_M]$
\State Determine $R$ dominant POD spatial bases $\phi_k({{\mathbf x}})$, $k=1,2,...,R$, with $RRMS Error < \alpha^{\text{POD}}$
\State Project $u({{\mathbf x}},t_i)$ to the reduced space and obtain $a_k(t) = (\hat u({{\mathbf x}},t),\phi_k({{\mathbf x}}))$
\State Determine the furthest forecast time permitted by the POD, $t_{\text{POD}}^*$, from Eq. \eqref{equ:Predict_POD_limit}
\For {$k=1,2,...,R$}
\State Infer a Gaussian Process model for $a_k(t)$ following \S \ref{subsubsec:Gaussian}
\State Determine the furthest forecast time permitted by the GPR, $t_{\mathrm{GPR,a}}^*$, from Eq. \eqref{equ:Gaussian_presetToL}
\EndFor
\State Determine the final furthest forecast time as $t^* = \min \{ t_\text{POD}^*, t_{\mathrm{GPR},a}^*,t_\mathrm{GPR,\Gamma_m}^*\}$
\State Predict $\gamma_l(t^*)$ by the inferred Gaussian Process models for $\gamma_l(t)$
\State Predict each $a_k(t^*)$ by the inferred Gaussian Process models for $a_k(t)$
\State Reconstruct the full-order solution $u({{\mathbf x}},t^*)$ by Eq. \eqref{equ:ROM_POD}
\State Correct $u({{\mathbf x}},t^*)$ for $\{\mathbf{x} | \mathbf x \in \Omega_f ~\text{at}~t^* ~\text{but}~ \mathbf{x} \in \Omega_m \text{ during } [t_1, t_M] \}$
\State \Return $u({{\mathbf x}},t^*)$ and $\gamma_l(t^*)$
\end{algorithmic}
\end{algorithm}
%%%%%%%%%%%%%%%%%%%%%%%%%

\subsection{Error analysis}
The error of the proposed nonintrusive model order reduction is originated from three resources: the truncation error in POD, the error caused by GPR, and the interpolation error introduced in \S \ref{subsubsec:correction_MB}, i.e., 
\begin{equation}\label{equ:error_ROM}
\epsilon^\text{ROM} = \left \|u({{\mathbf x}},t)- u^\text{ROM}({{\mathbf x}},t) \right\|_{L_2} = \epsilon^\text{POD} + \epsilon^\text{GPR} + \epsilon^\text{MLS} \;.
\end{equation}
Here, $\epsilon^\text{POD}$ is the truncation error of POD given by Eq. \eqref{equ:POD_truncation_error}. $\epsilon^\text{GPR}$ is the error caused by GPR and can be indicated by the uncertainty level $\hat \sigma(t)$:
\begin{equation}\label{eq:err_GPR}
    \epsilon^\text{GPR}=  \left \|u^\text{POD}({\bf{x}},t)- u^\text{ROM}({\bf{x}},t) \right\|_{L_2} \approx C_1 \hat \sigma(t)\; ,
\end{equation}
where $u^\text{POD}({\bf{x}},t)$ is the projection of the full-order solution $u({\bf{x}},t)$ onto the reduced space formed by $R$ POD bases;  $C_1>0$ is a constant; and $\hat \sigma(t)$ is given by Eq. \eqref{equ:weighted_sd}. $\epsilon^\text{MLS}$ denotes the error of the MLS interpolation and can be estimated by \cite{MLSDiffuse_Mirzaei2012}:
%%%%%%%%%%%%%%%%%%%%%%%
\begin{equation}
\epsilon^\text{MLS}=\left \|u({{\mathbf x}},t)- u_h({{\mathbf x}},t) \right\|_{L_2} \le C_2 h^{s+1}|u|_{C^{s+1}(\Omega_f)} \;,
\end{equation}
%%%%%%%%%%%%%%%%%%%%%%%
where $C_2>0$ is a constant; and $|u|_{C^{s+1}(\Omega_f)}:=\max\limits_{\zeta \le s+1}\|D^\zeta u\|_{L_\infty}$ with $D^\zeta u$ the $\zeta$-th order spatial derivative of $u$ for ${\mathbf x} \in \Omega_f$. The relative error is then defined as:
\begin{equation}\label{equ:relative_error}
   \epsilon_r^\text{ROM}= \frac{\epsilon^\text{ROM}}{\left \|u({\bf{x}},t)\right\|_{L_2}} = \frac{\left \|u({\bf{x}},t)- u^\text{ROM}({\bf{x}},t) \right\|_{L_2}}{\left \|u({\bf{x}},t)\right\|_{L_2}} \; .
\end{equation}

\section{Numerical experiments}\label{sec:results}
We assessed the accuracy and efficiency of the proposed data-driven nonintrusive model order reduction method in several benchmark dynamical systems. We started with three benchmarks without moving boundaries and then moved to problems with moving boundaries. The data used to construct and validate the ROM are either from analytical/numerical solutions or experimental data. The values of variables or parameters are all non-dimensional.%, unless otherwise stated.  

\subsection{Burgers equation} %: nonlinear wave propagation(NWP) problem}
First, we constructed the ROM to predict the solution of the Burgers equation, which is a typical benchmark used in literature for validating model order reduction methods (e.g., in \cite{POD_Burgers_2017, POD_Burgers_1999, POD_Burgers_2014}). 
%which is generally considered a low-dimensional equivalent of the full Navier-Stokes equation due to its advective and diffusive behavior. 
The Burgers equation considered herein is:
\begin{equation}\label{equ:Burgers}
\begin{split}
&\frac{{\partial u}}{{\partial t}} + u\frac{{\partial u}}{{\partial x}} = \frac{1}{{{\mathop{Re}\nolimits} }}\frac{{{\partial ^2}u}}{{\partial {x^2}}}, \ \ x \in (0,1) \\
&u(0,t) = 0,\ \ u(1,t) = 0, \ \ t \ge 0  \\
&u(x,0) =  \frac{x}{1+\sqrt{\frac{1}{t_0}}\exp({Re \frac{x^2}{4}})} ,\ \ x \in (0,1) 
\end{split}
\end{equation}
where $Re$ is the Reynolds number and $t_0=\exp({\frac{Re}{8}})$. % we consider a challenging convective system in our numerical test since it characterizes the structure of localized flow  structures such as shock waves. 
The analytical solution of Eq. \eqref{equ:Burgers} is given by:
\begin{equation}\label{equ:Burgers_analytical}
    u(x,t) = \frac{\frac{x}{t+1}}{1+\sqrt{\frac{t+1}{t_0}}\exp({Re \frac{x^2}{4t+4}})} \; .
\end{equation}

\begin{figure}[htbp]
\centering
\subfigure[$Re=1$, $R =1$]{
\includegraphics[width=6cm]{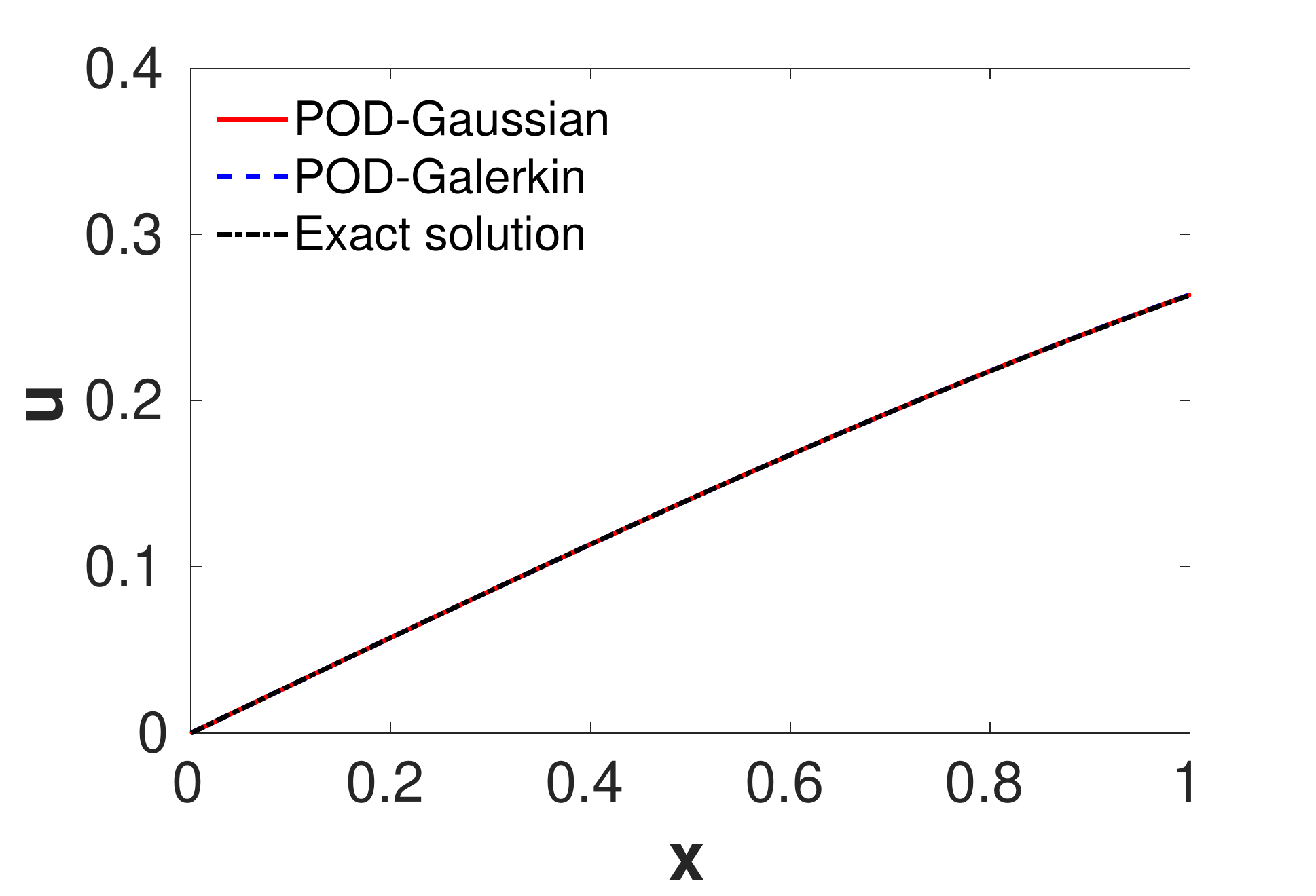}
}
\quad
\subfigure[$Re=100$, $R=2$]{
\includegraphics[width=6cm]{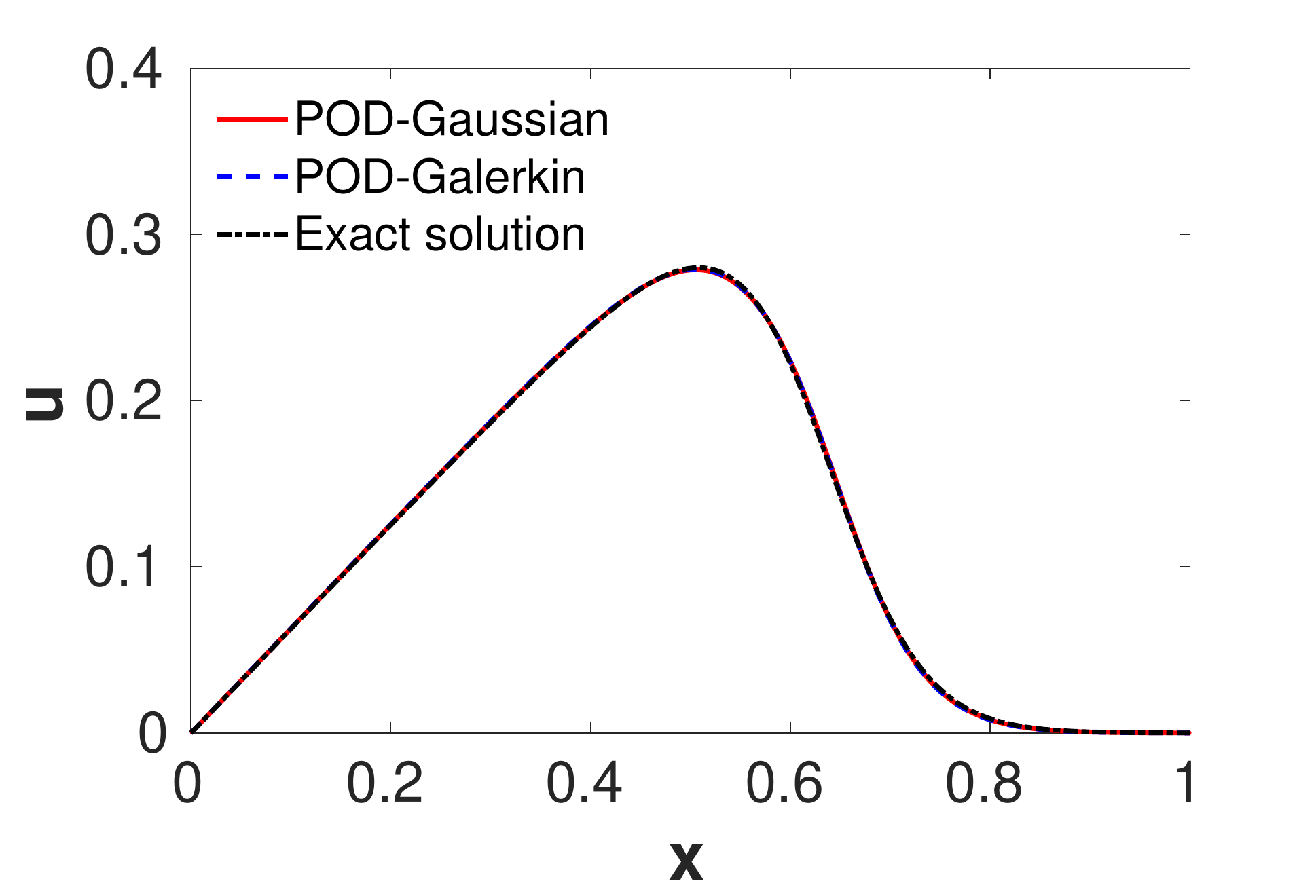}
}
\quad
\subfigure[$Re=300$, $R=4$]{
\includegraphics[width=6cm]{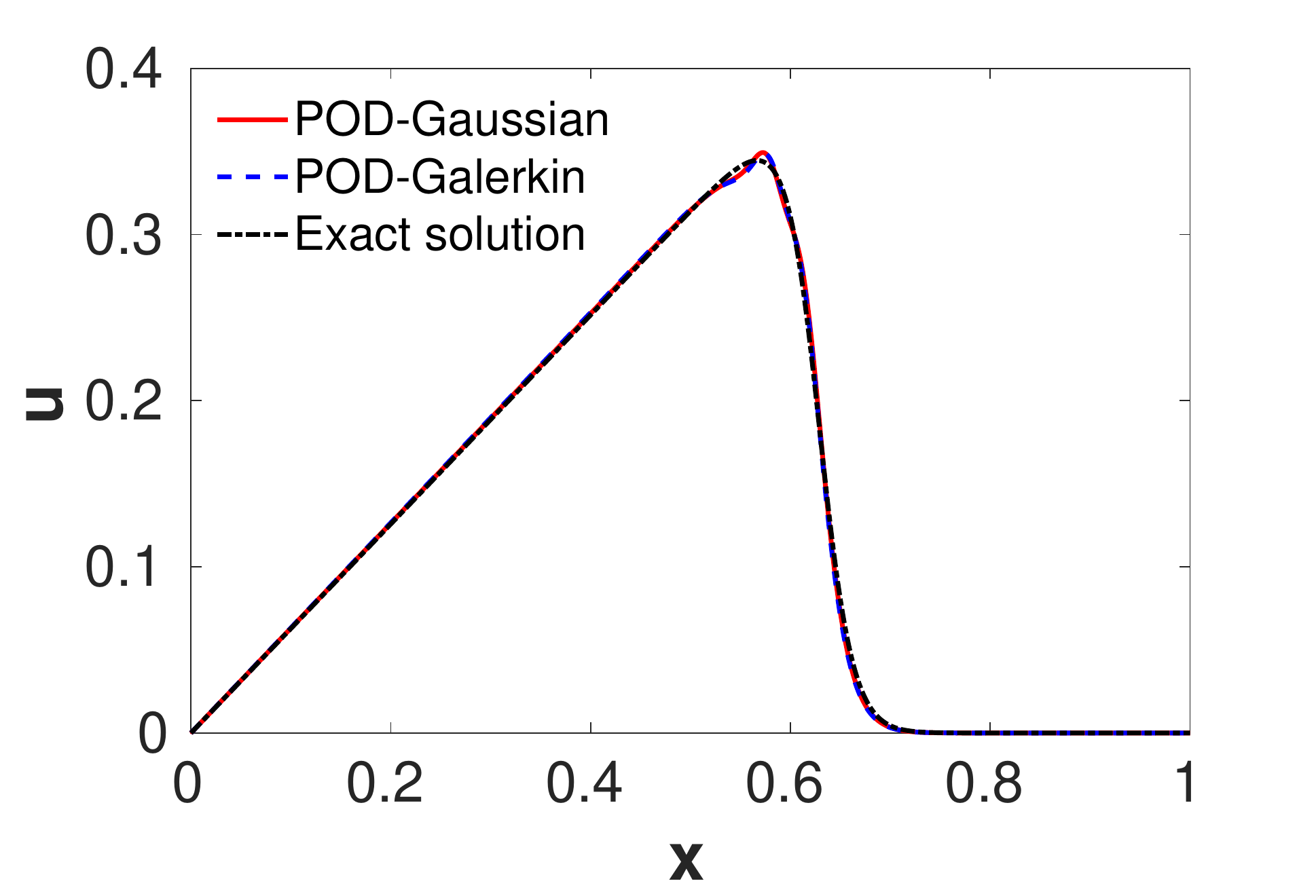}
}
\quad
\subfigure[$Re=500$, $R = 4$]{
\includegraphics[width=6cm]{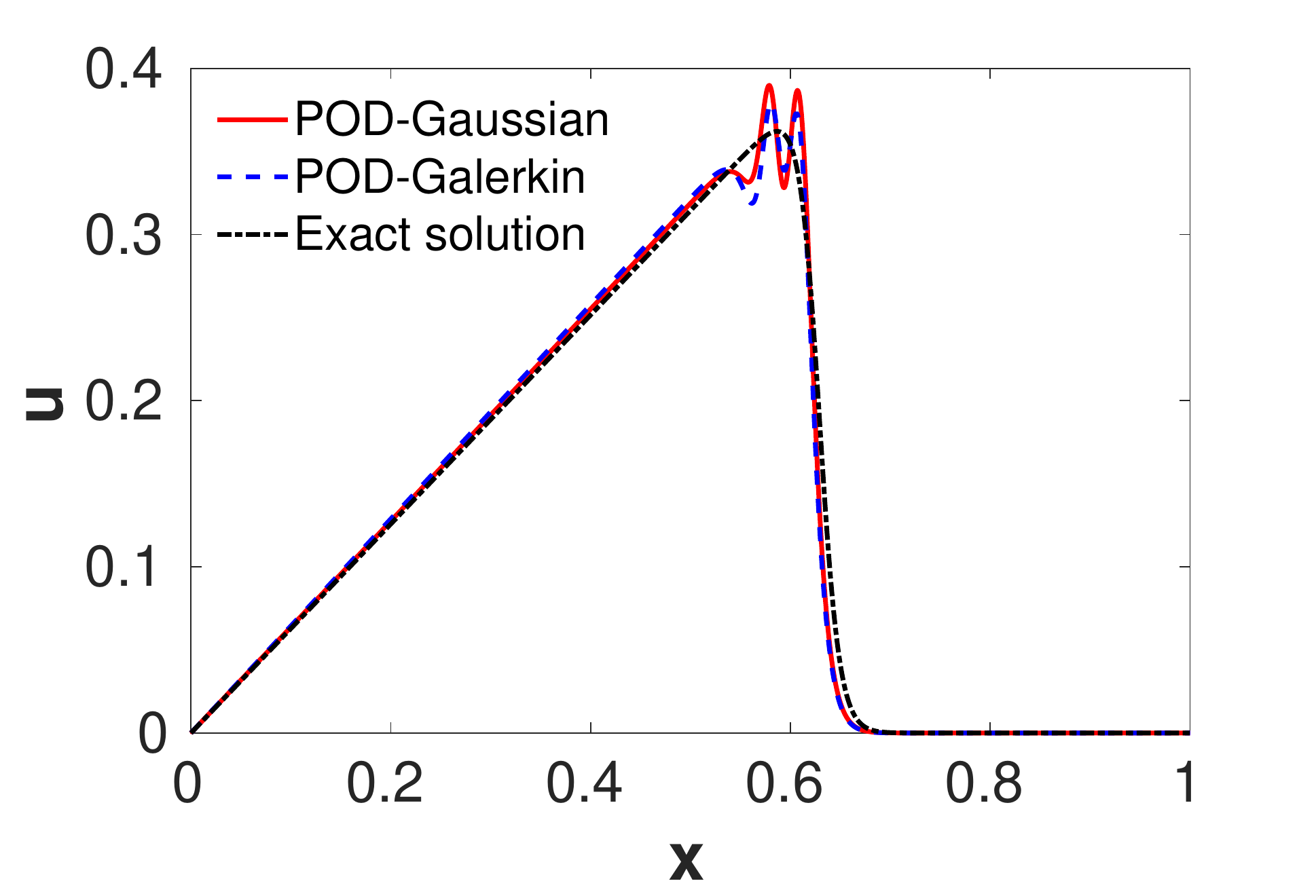}
}
\caption{Burgers equation: The solution of $u(x,t)$ at a future time $t^*=0.6$ predicted by the ROM built from POD and Gaussian process (POD-Gaussian) or Galerkin projection (POD-Galerkin), compared with the exact solution.}\label{fig:Burgers_u}
\end{figure}

The analytical solution (Eq. \eqref{equ:Burgers_analytical}) of the Burgers equation was used to generate the snapshot data to construct the ROM with the time step $\delta t = 10^{-2}$ and spatial grid length $\delta x = 10^{-3}$. In particular, 20 snapshots from $t=0.3$ to $t=0.5$ were used to extract the dominant POD modes. We set the truncation threshold $\alpha^\text{POD} = 0.01$, which means the POD modes retained dominate at least $99.99\%$ of the fluctuating kinetic energy. To satisfy this threshold, the number of POD modes retained ($R$) varies with the Reynolds number, which is indicated in Figure \ref{fig:Burgers_u}. 

To construct the temporal coefficients $a_k(t)$, we employed both the Galerkin projection and GPR and compared their performance. With the constructed $a_k(t)$, we predicted the solution $u(\mathbf{x},t^*)$ of the Burgers equation at a future time $t^*=0.6$ for different Reynolds numbers from $Re=1$ to $Re=500$. The results are shown in Figure \ref{fig:Burgers_u}. The formulation used in the Galerkin projection to determine $a_k(t)$ is given as below: 
 \begin{equation}\label{equ:Burgers_Garlerkin}
    \dot{a}_k(t) = (\frac{1}{{{\mathop{\rm Re}\nolimits} }}\frac{{{\partial ^2}u}}{{\partial {x^2}}} - u\frac{{\partial u}}{{\partial x}},{\phi _k}) = {B_k} + \sum\limits_{i = 1}^R {{L_{ik}}{a_i(t)}} {\rm{ + }}\sum\limits_{i = 1}^R {\sum\limits_{j = 1}^R {{N_{ijk}}{a_i(t)}{a_j(t)}} } , \ \  \ \ \text{for} \ \ k = 1,2,...,R \; ,
 \end{equation}
 where ${B_k} = (\frac{1}{{{\mathop{\rm Re}\nolimits} }}\frac{{{\partial ^2}\bar u}}{{\partial {x^2}}} - \bar u\frac{{\partial \bar u}}{{\partial x}},{\phi _k})$, ${L_{ik}} = (\frac{1}{{{\mathop{\rm Re}\nolimits} }}\frac{{{\partial ^2}{\phi _i}}}{{\partial {x^2}}} - {\phi _i}\frac{{\partial \bar u}}{{\partial x}} - \bar u\frac{{\partial {\phi _i}}}{{\partial x}},{\phi _k})$, and ${N_{ijk}} = ( - {\phi _i}\frac{{\partial {\phi _j}}}{{\partial x}},{\phi _k})$. 
%%%%%%%%%%%%%%%%%%%%%%
%%%%%%%%%%%%%%%%%%%%%%

From Figure \ref{fig:Burgers_u}, we find that the solution forecasted at the future time $t^*=0.6$ by the constructed ROM is accurate for lower Reynolds numbers, either based on the Gaussian process or Galerkin projection. For higher Reynolds numbers, the accuracy deteriorates, and oscillations occur near the shock wave singularity. This is because the dominant spatial bases $\phi_k(x)$ would change rapidly upon the development of shock wave, which violates the assumption of POD-based model order reduction: the dominant spatial bases are the same for the training data and forecasted solutions. The accuracy of the ROM constructed using the GPR is comparable with that using the Galerkin projection. However, while the Galerkin projection relies on both the data and knowledge of the Burgers equation to establish the ODEs (Eq. \eqref{equ:Burgers_Garlerkin}) for the temporal coefficients $a_k(t)$, the Gaussian process only requires data.

\begin{figure}[htbp]
\centering
\subfigure[The relative error $\epsilon_r^\text{ROM}$ of the ROM solution at $t^* =0.6 $ for varying number of retained POD modes ]{\label{subfig:Burgers_errorPOD_new}
\includegraphics[width=6cm]{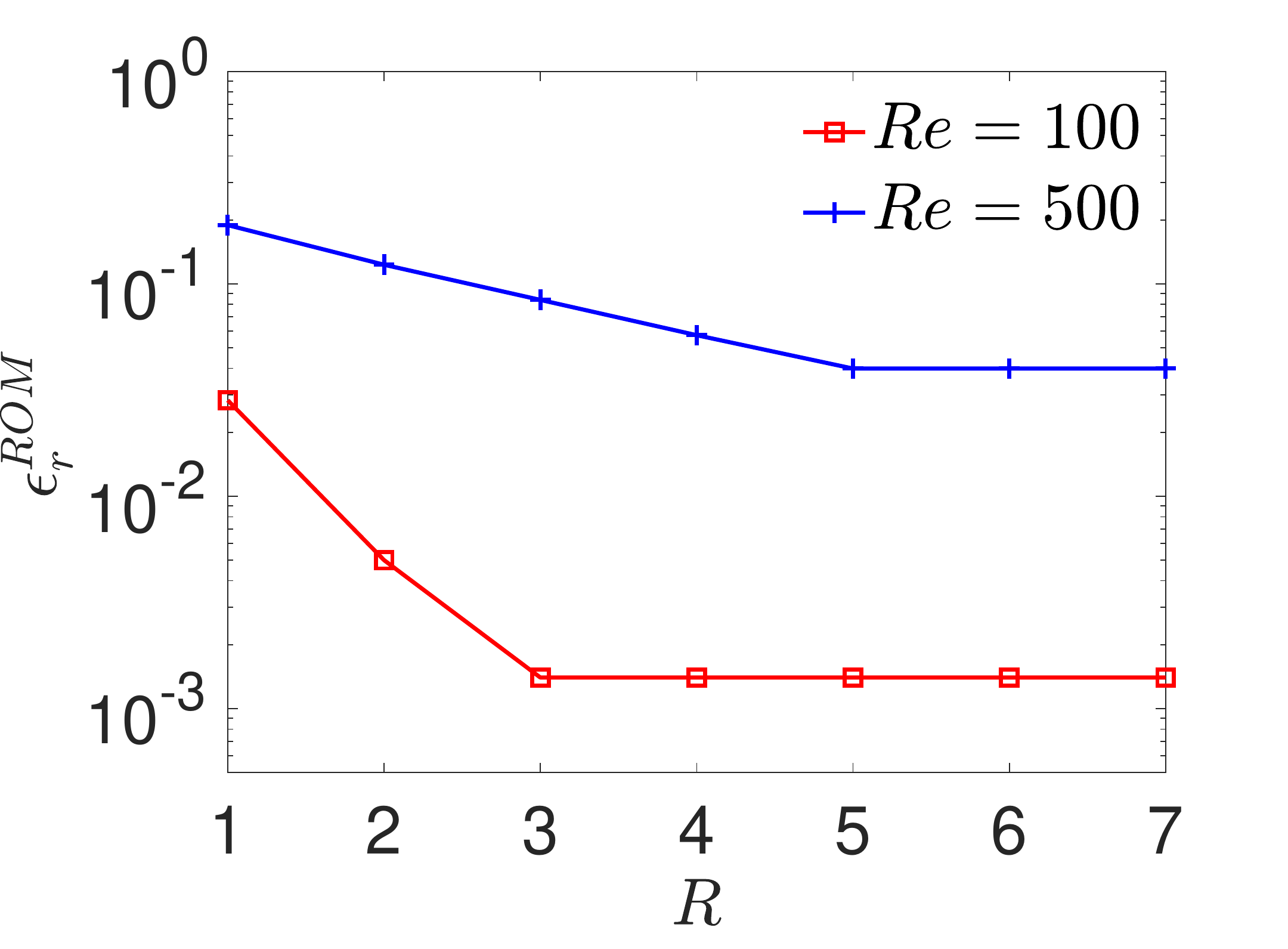}
}
\quad
\subfigure[Relative information content for ${\mathop{\rm Re}\nolimits} = 1, 100, 300, 500$]{\label{fig:Burgers_lambda}
\includegraphics[width=6cm]{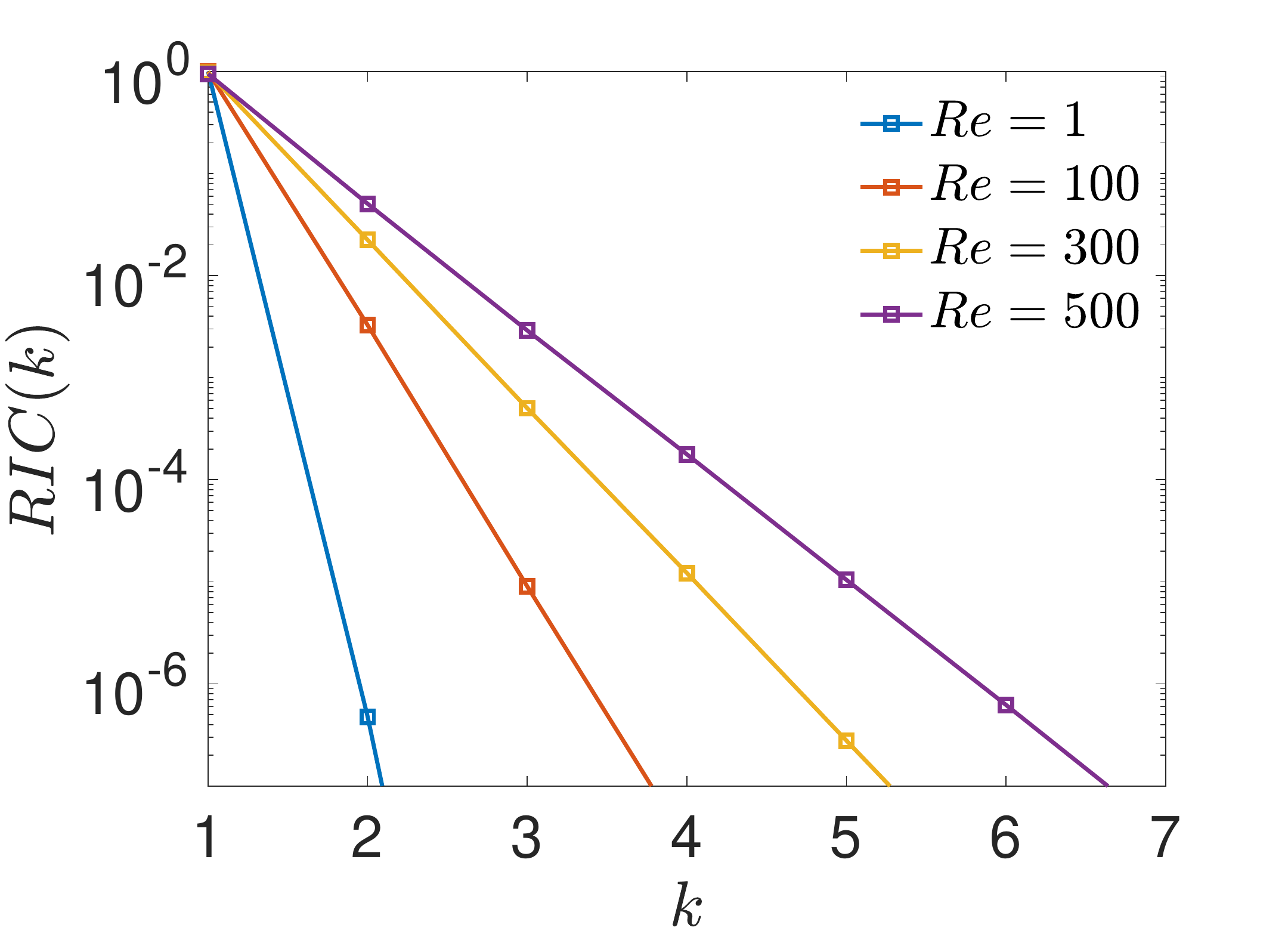}
}
\caption{Burgers equation: Error analysis for different number of POD modes retained.}
\label{fig:Burgers_error}
\end{figure}

We further examined the behavior of relative error $\epsilon_r^\text{ROM}$ (Eq. \eqref{equ:relative_error}) of the ROM constructed using the POD and GPR. The solution predicted at $t^*=0.6$ with the snapshots from $t_1=0.3$ to $t_M=0.5$ was considered. As expected, the constructed ROM is more accurate with more POD modes retained, as depicted in Figure \ref{subfig:Burgers_errorPOD_new} for ${\mathop{\rm Re}\nolimits} = 100$ and ${\mathop{\rm Re}\nolimits}=500$. However, the error decay stalls when $R$ (the number of POD modes retained) reaches a certain number, which is because $\epsilon^\text{GPR}$ begins to dominate when $\epsilon^\text{POD}$ becomes small.
%%%%%%%%%%%%%
%%%%%%%%%%%%%
In addition, the error decreases with respect to $R$ at a faster rate for lower Reynolds number (${\mathop{\rm Re}\nolimits} = 100$) and at a slower rate for higher Reynolds number (${\mathop{\rm Re}\nolimits} = 500$). This can be explained through the analysis of eigenvalues. Recall the truncation error of POD is determined from the eigenvalues of the correlation matrix of the snapshot data,  as specified in Eq. \eqref{equ:POD_truncation_error}. The energy contributed by each mode in the total energy can be evaluated by the relative information content (RIC), expressed as:
%%%%%%%%%%%%%%%%
\begin{equation}
    RIC(k) = ( \frac{\lambda_k}{\sum_{j=1}^{M} \lambda_j} ) \times 100\% \; .
\end{equation}
%%%%%%%%%%%%%%%%
In Figure \ref{fig:Burgers_lambda}, we compare the RIC of $k$-th POD mode for different Reynolds numbers. 
The RIC demonstrates how the contribution of each POD mode decays. The smaller the Reynolds number is, the decay slope is more steep, and thereby the POD truncation error decreases faster with respect to $R$. The decay rate of eigenvalues, as defined in Eq. \eqref{equ:Predict_POD_limit} is also used to determine the furthest forecast time. Figure \ref{fig:Burgers_lambda} indicates smaller Reynolds number permits longer time extrapolation (i.e., larger $\Delta t^*=t^*-t_M$) given the dataset range $[t_1, t_M]$ is fixed. 

\begin{figure}[htbp]
\centering
\includegraphics[width=6cm]{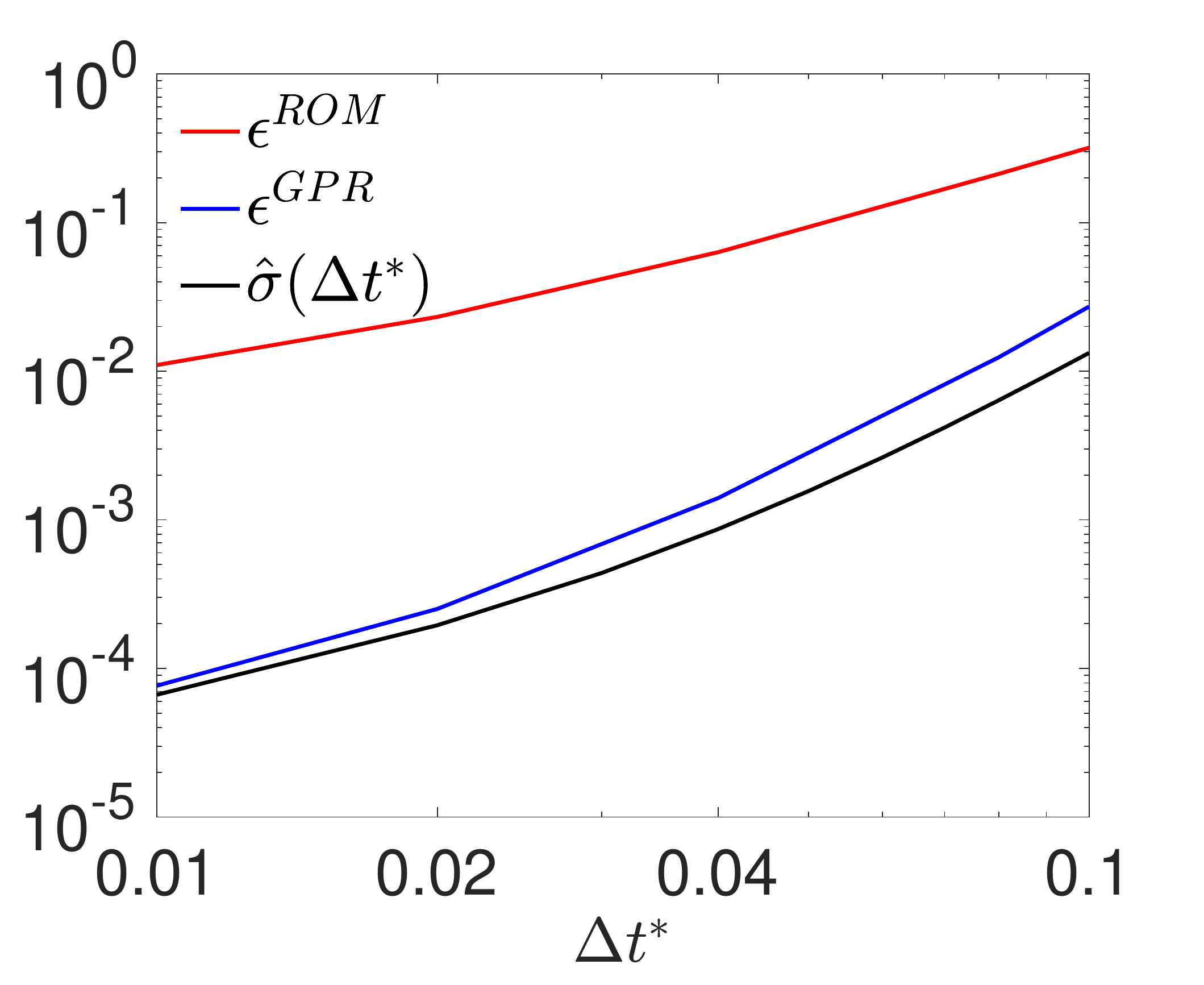}
\caption{Burgers equation: $\epsilon^\text{ROM}$ at different forecast times beyond the dataset range $\Delta t^*=t^*-t_M$. Here, $\epsilon^{\text{GPR}}$ and $\hat{\sigma}$ are also shown for comparison.}
\label{fig:Burgers_errorGaussian_new}
\end{figure}

\begin{figure}[htbp]
\centering
\subfigure[$a_1$]{
\includegraphics[width=6cm]{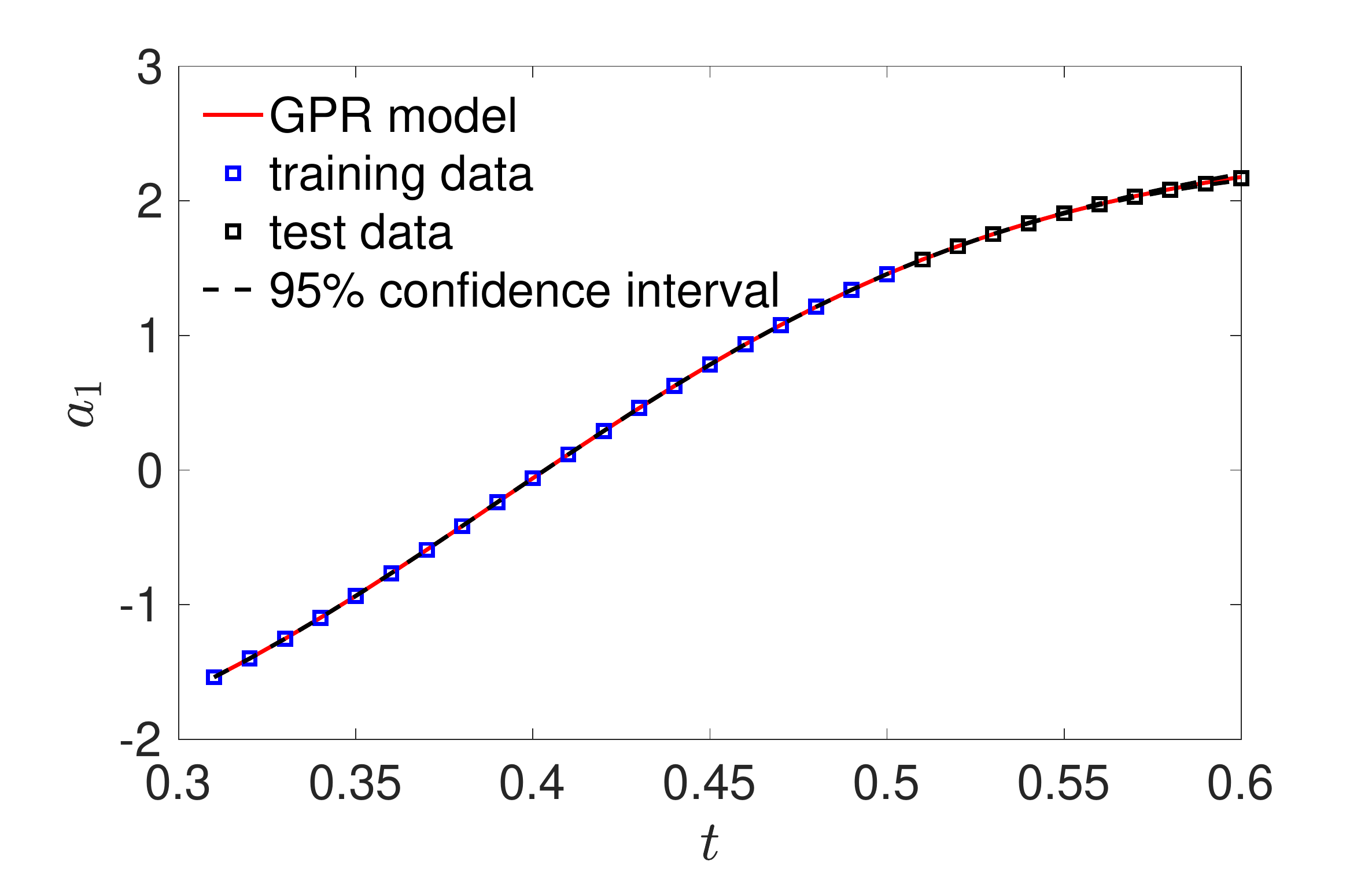}
}
\quad
\subfigure[$a_2$]{
\includegraphics[width=6cm]{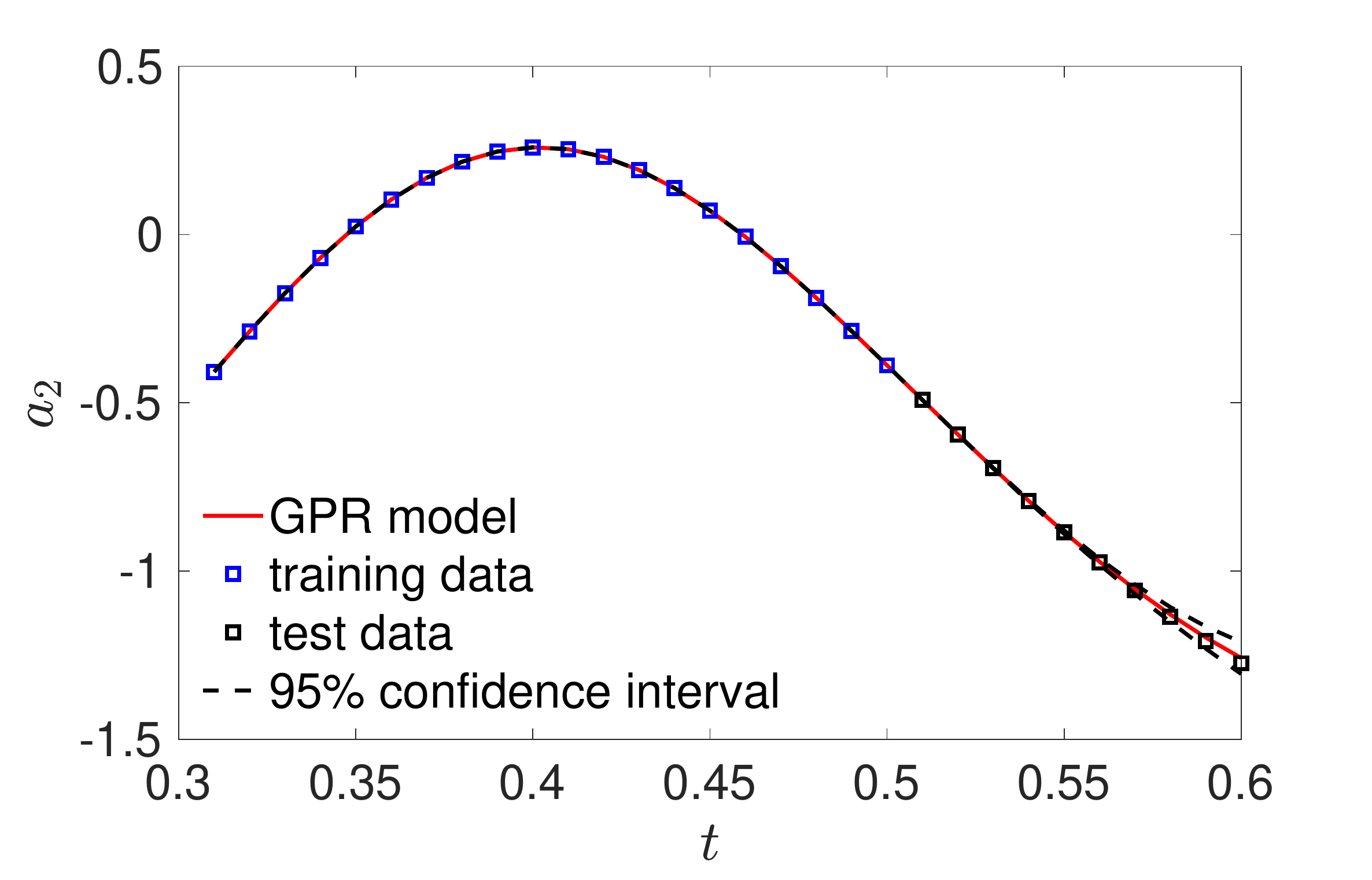}
}
\quad
\subfigure[$a_3$]{
\includegraphics[width=6cm]{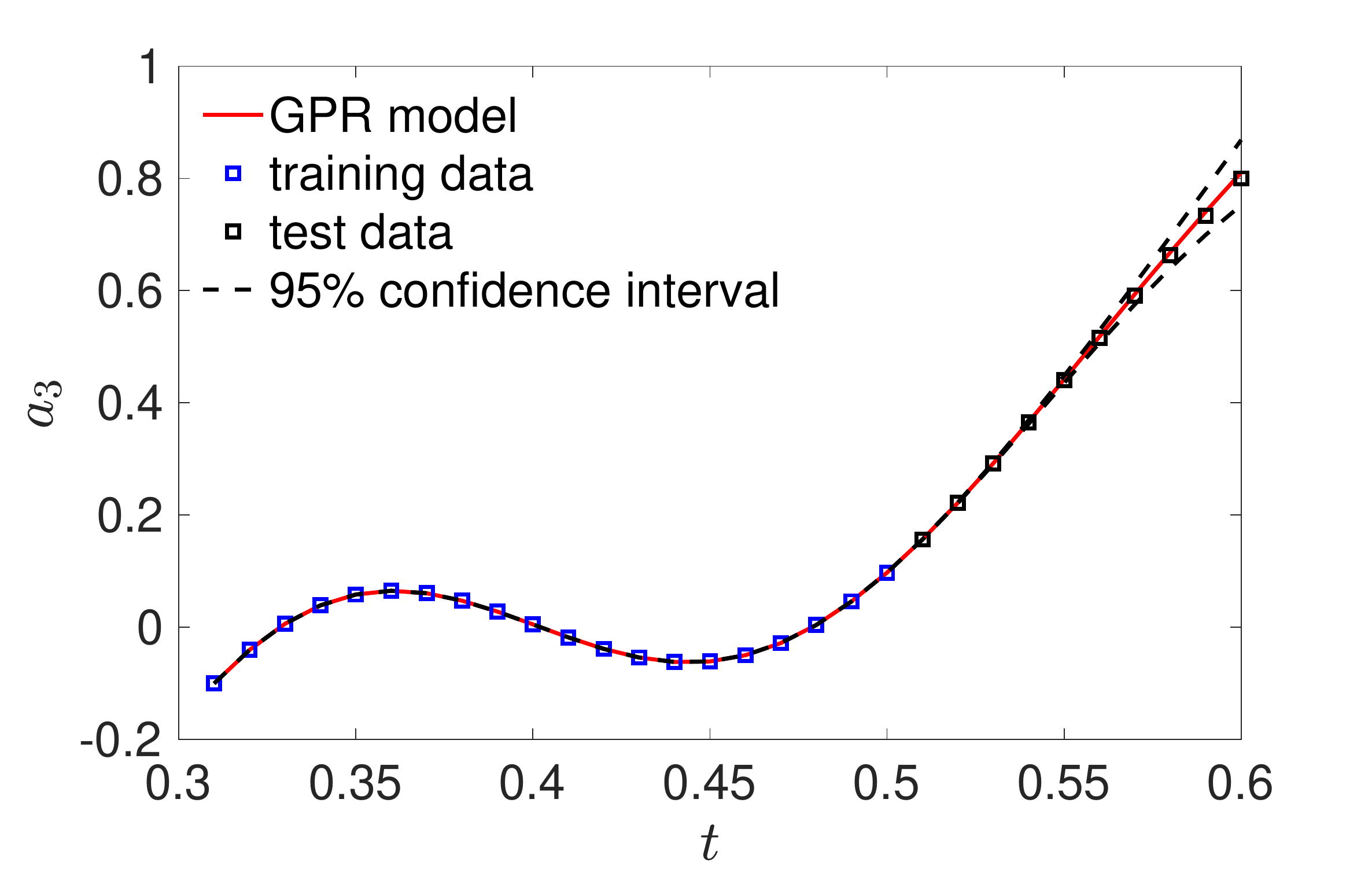}
}
\quad
\subfigure[$a_4$]{
\includegraphics[width=6cm]{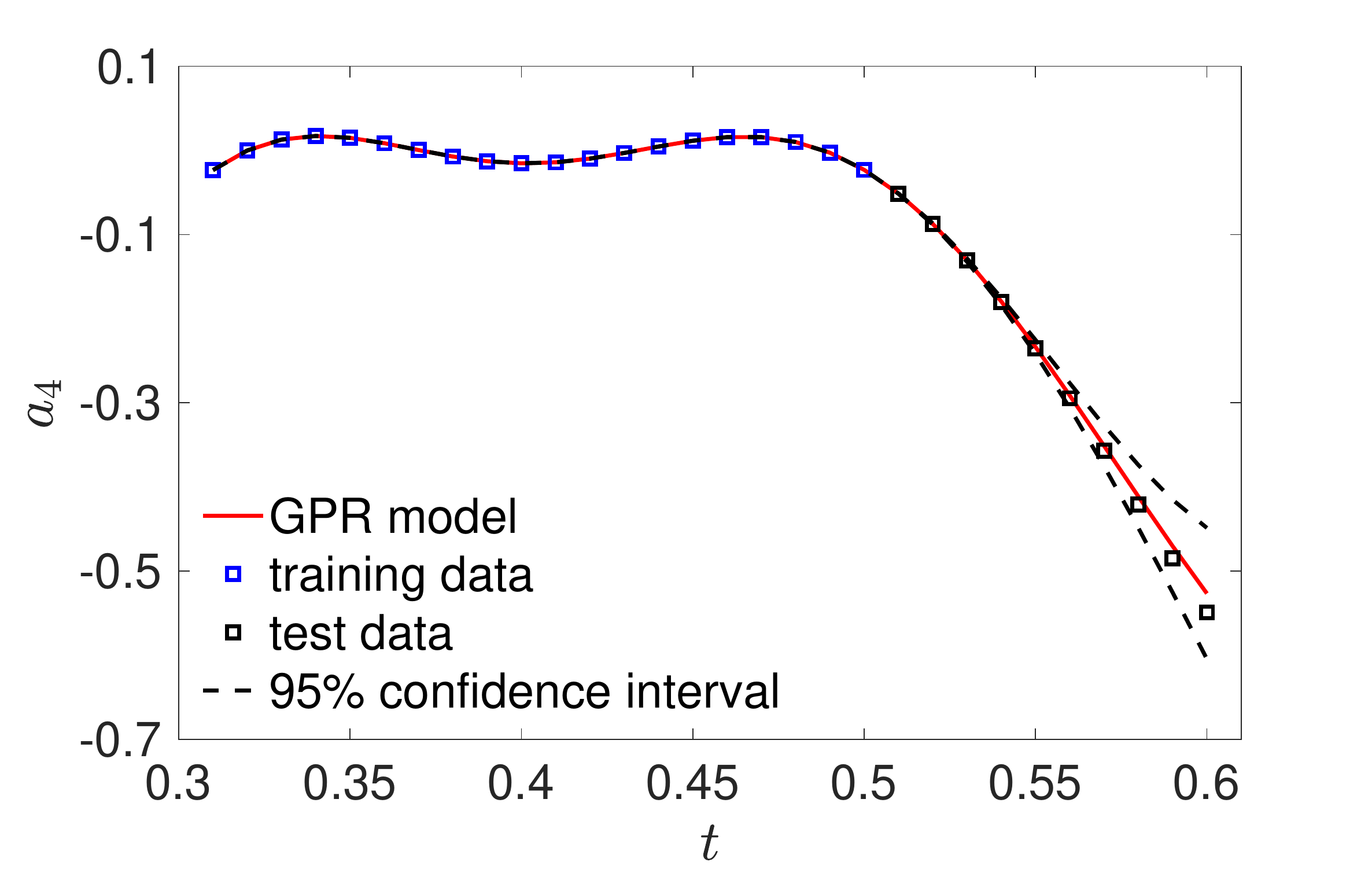}
}
\caption{Burgers equation: GPR for each dominant temporal coefficient $a_k(t)$.}\label{fig:Burgers_a(t)}
\end{figure}

In Figure \ref{fig:Burgers_errorGaussian_new}, we assessed the errors $\epsilon^\text{ROM}$ and $\epsilon^\text{GPR}$ and the uncertainty level of GPR $\hat{\sigma}$ at different forecast times $t^*$ with the same snapshots from $t_1 = 0.3$ to $t_M =0.5$. Here, $Re = 500$ was considered and 4 POD modes were retained. The growth of $\hat{\sigma}$ as $\Delta t^*$ increases can be explained by Figure \ref{fig:Burgers_a(t)}, where the GPR model trained for each dominant $a_k(t)$ is presented. It can be seen that the 95\% confidence interval enlarges, indicating the uncertainty level (or $\hat{\sigma}(t^*)$) increases, as time extrapolation goes further. This result also supports the use of $\hat{\sigma}(t^*)$ as a criterion to adaptively determine the furthest forecast time $t^*$, as given in Eq. \eqref{equ:Gaussian_presetToL}. $\epsilon^\text{GPR}$ varies following the trend of $\hat{\sigma}(\Delta t^*)$. Finally, $\epsilon^\text{ROM}$ grows with $\Delta t^*$, i.e., with the forecast time $t^*$ going further beyond the dataset range, resulting from the behavior of $\epsilon^\text{GPR}$.
% and the change of dominant POD bases.  
%%%%%%%%%%
%%%%%%%%%%

\subsection{Lid-driven cavity flow}

In this section, we studied the 2D lid-driven cavity flow, where a square cavity $\Omega= [0,1]\times [0,1]$ consists of three stationary walls and a lid moving with a constant tangential velocity. To obtain the velocity and pressure fields for the lid-driven cavity flow, numerical simulations are usually employed for solving the full Navier-Stokes (NS) equation. 
Here, we assessed the ability of the proposed Gaussian process-enabled nonintrusive model order reduction to predict the velocity and pressure fields for this flow. To construct the ROM, the snapshot data were generated via simulation which is based on the staggered finite-volume spatial discretization, projection method to enforce incompressibility, and treating the nonlinear advection term explicitly and the viscous term implicitly \cite{NS_Chorin1968,NSnumerical_Griebel1998}. 

\begin{figure}[htbp]
\centering
\subfigure[Velocity field (with streamlines) by numerical simulations]{
\includegraphics[width=4.8cm]{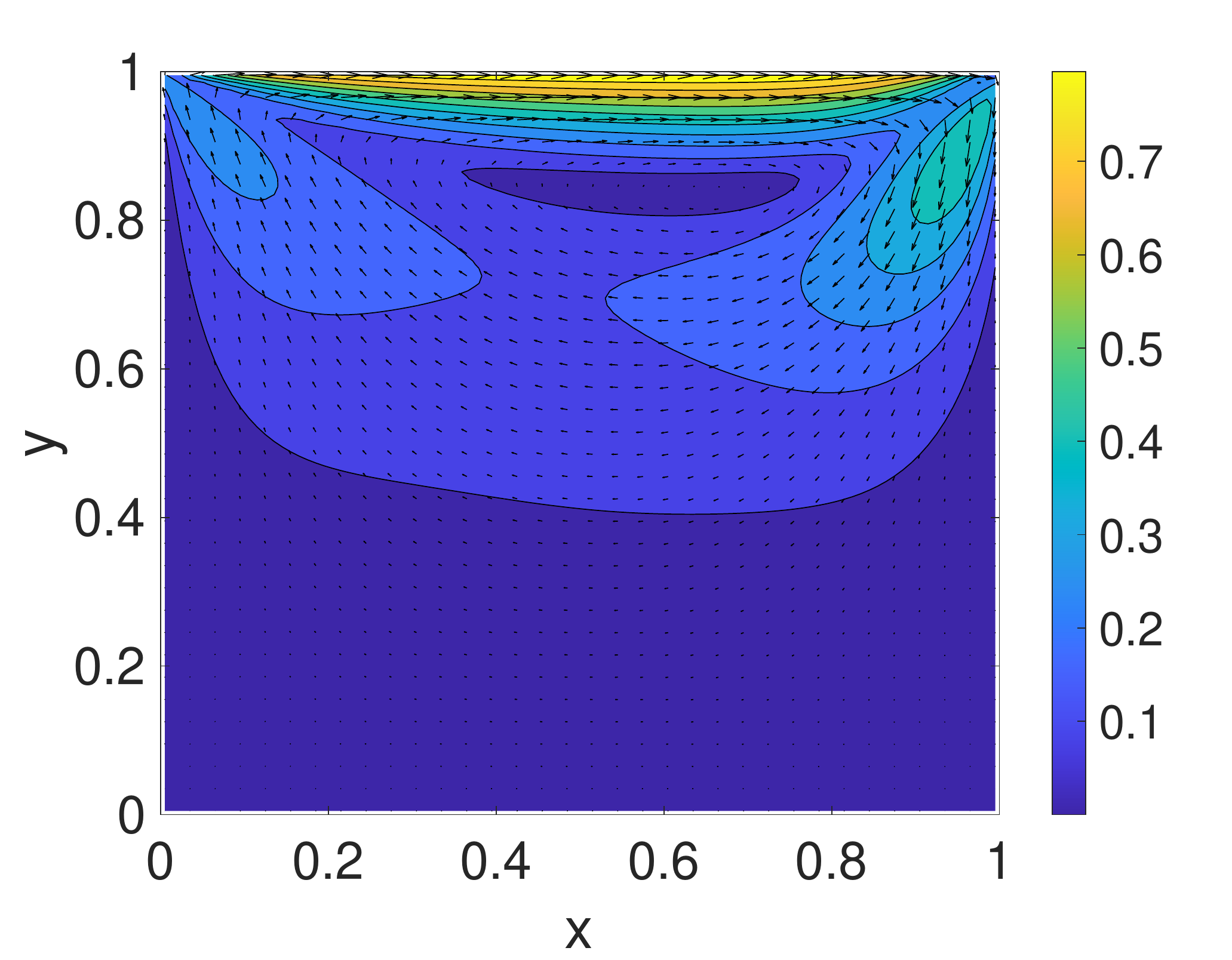}
}
\quad
\subfigure[Velocity field (with streamlines) by ROM]{
\includegraphics[width=4.8cm]{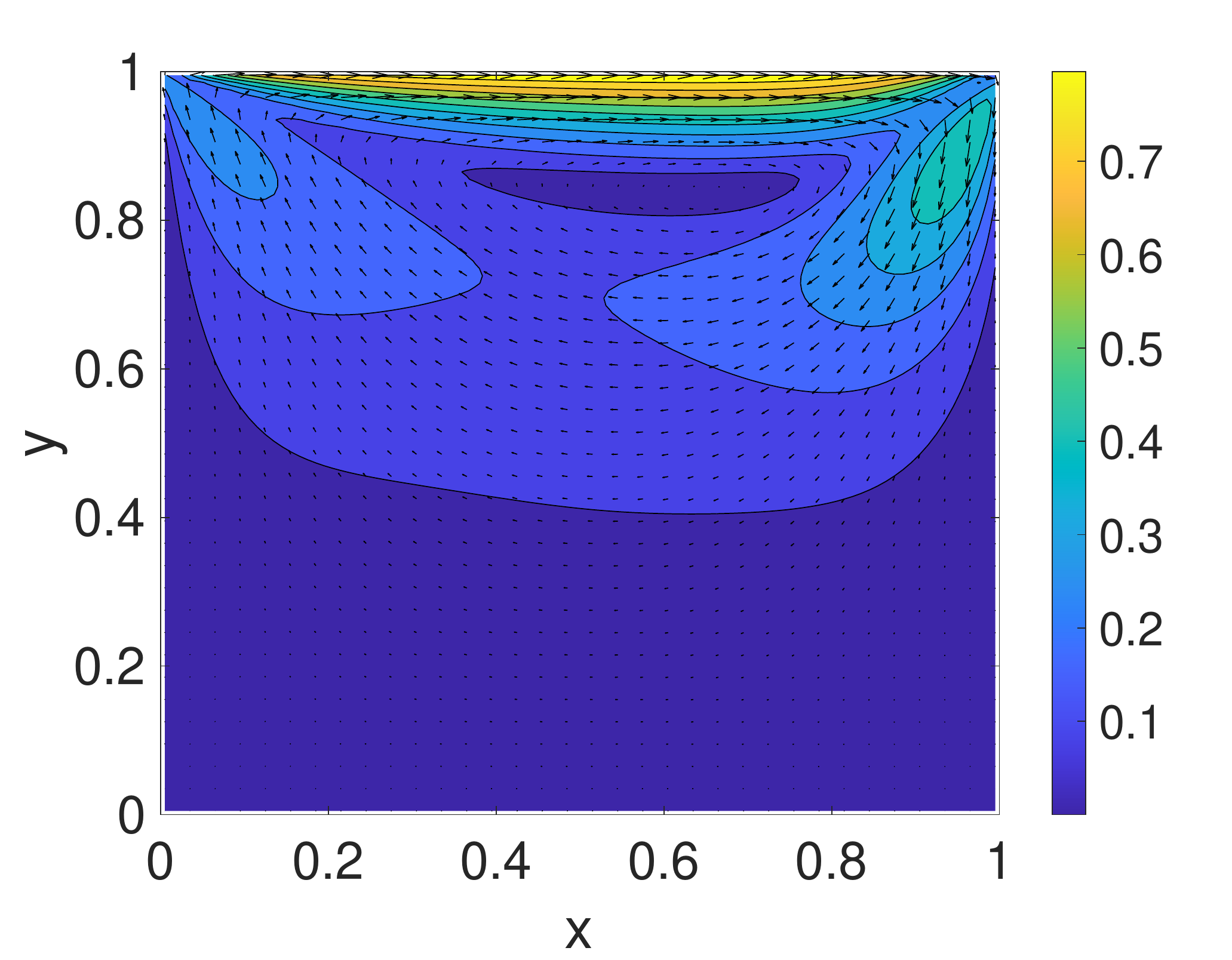}
}
\quad
\subfigure[Pointwise error for velocity]{
\includegraphics[width=4.8cm]{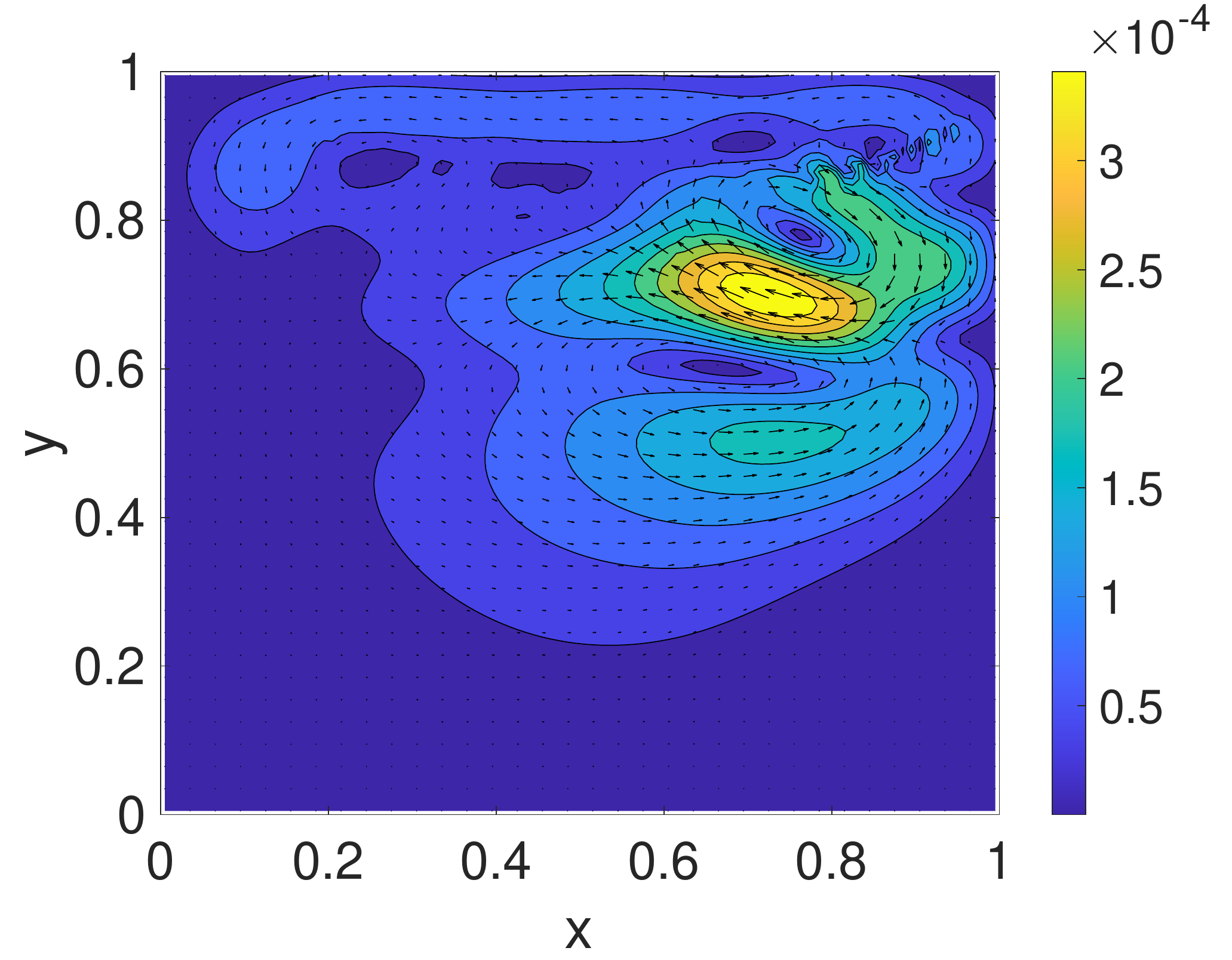}
}
\quad
\subfigure[Pressure field by numerical simulations]{
\includegraphics[width=4.8cm]{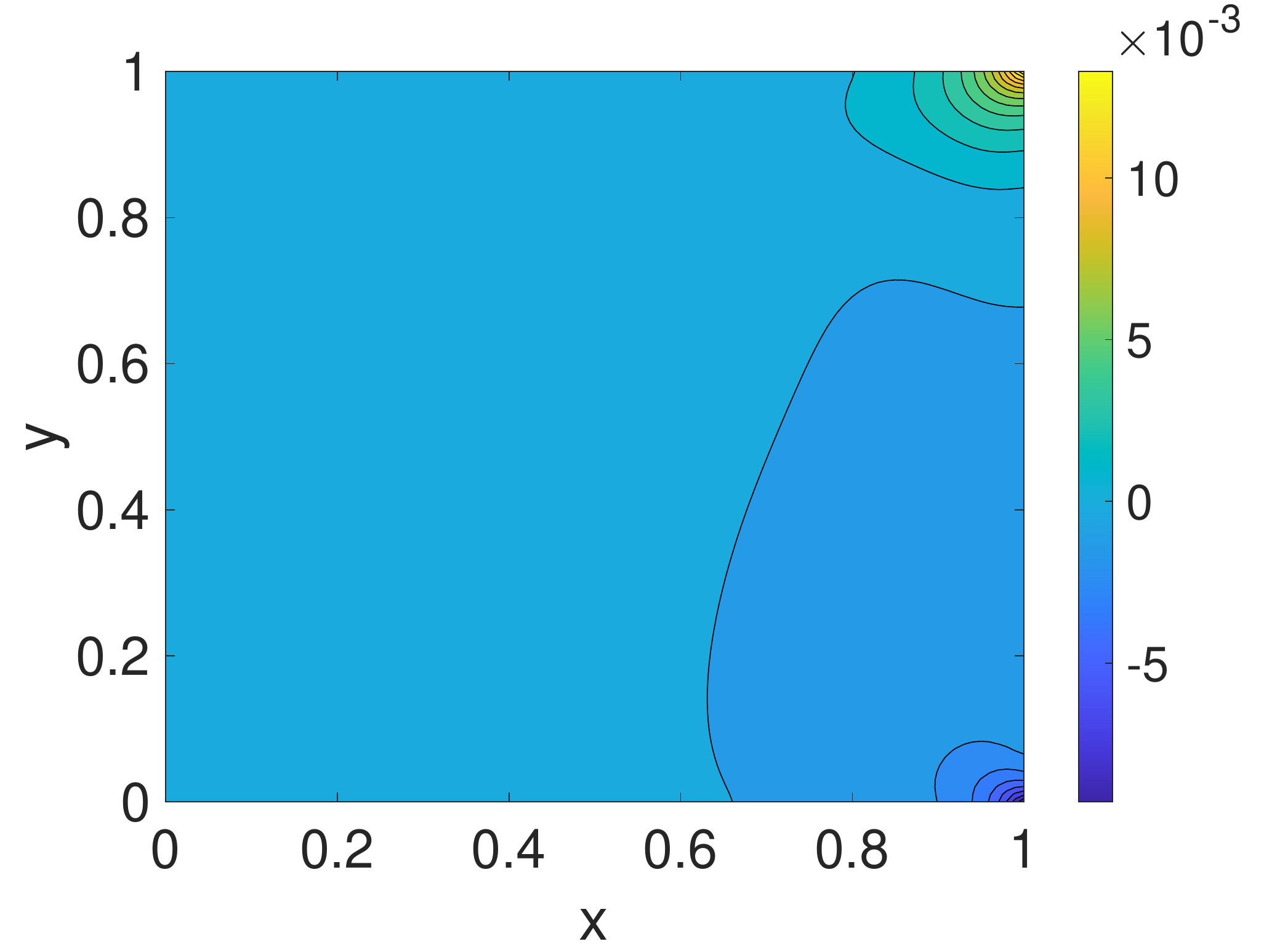}
}
\quad
\subfigure[Pressure field by ROM]{
\includegraphics[width=4.8cm]{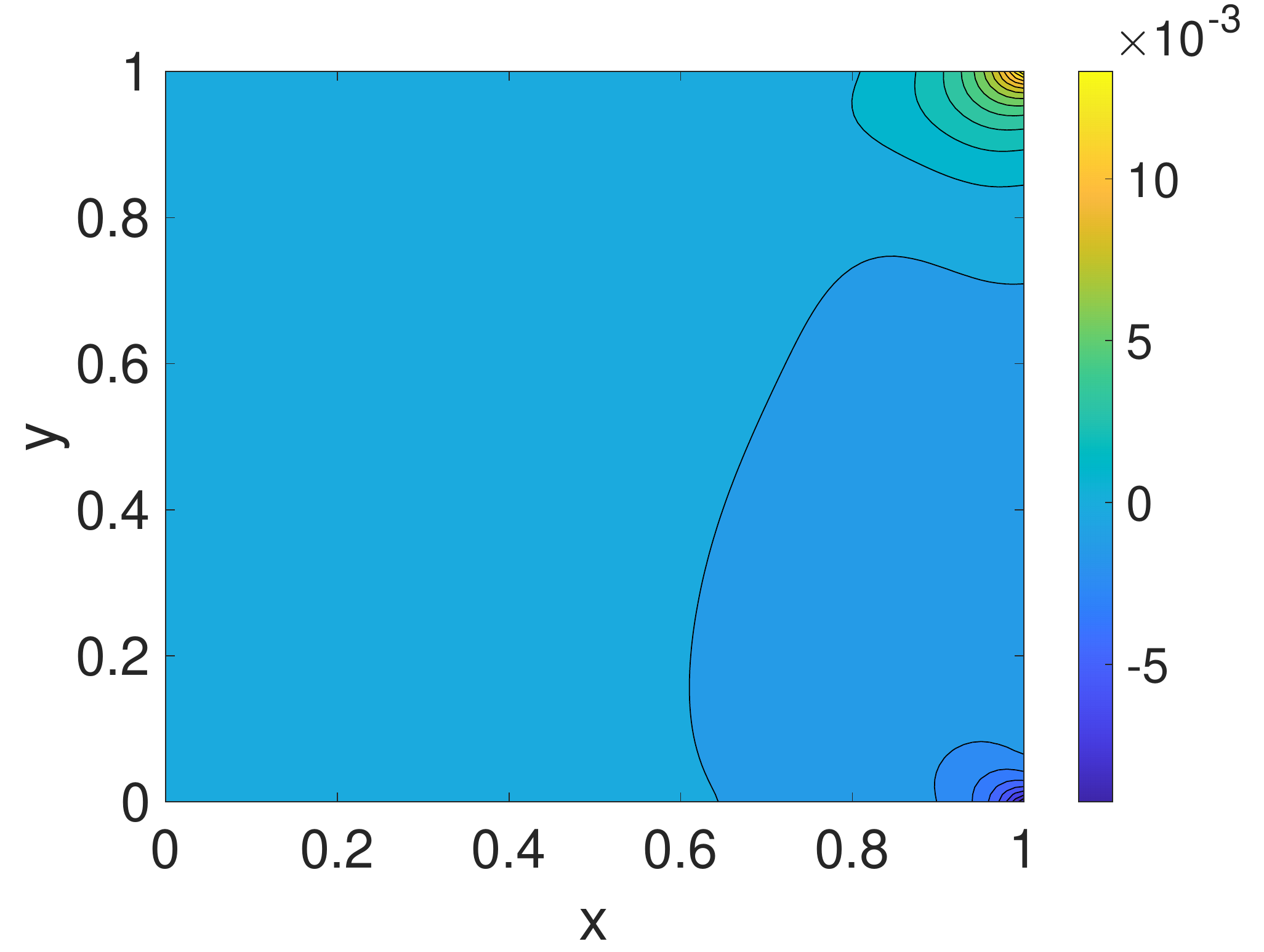}
}
\quad
\subfigure[Pointwise error for pressure]{
\includegraphics[width=4.8cm]{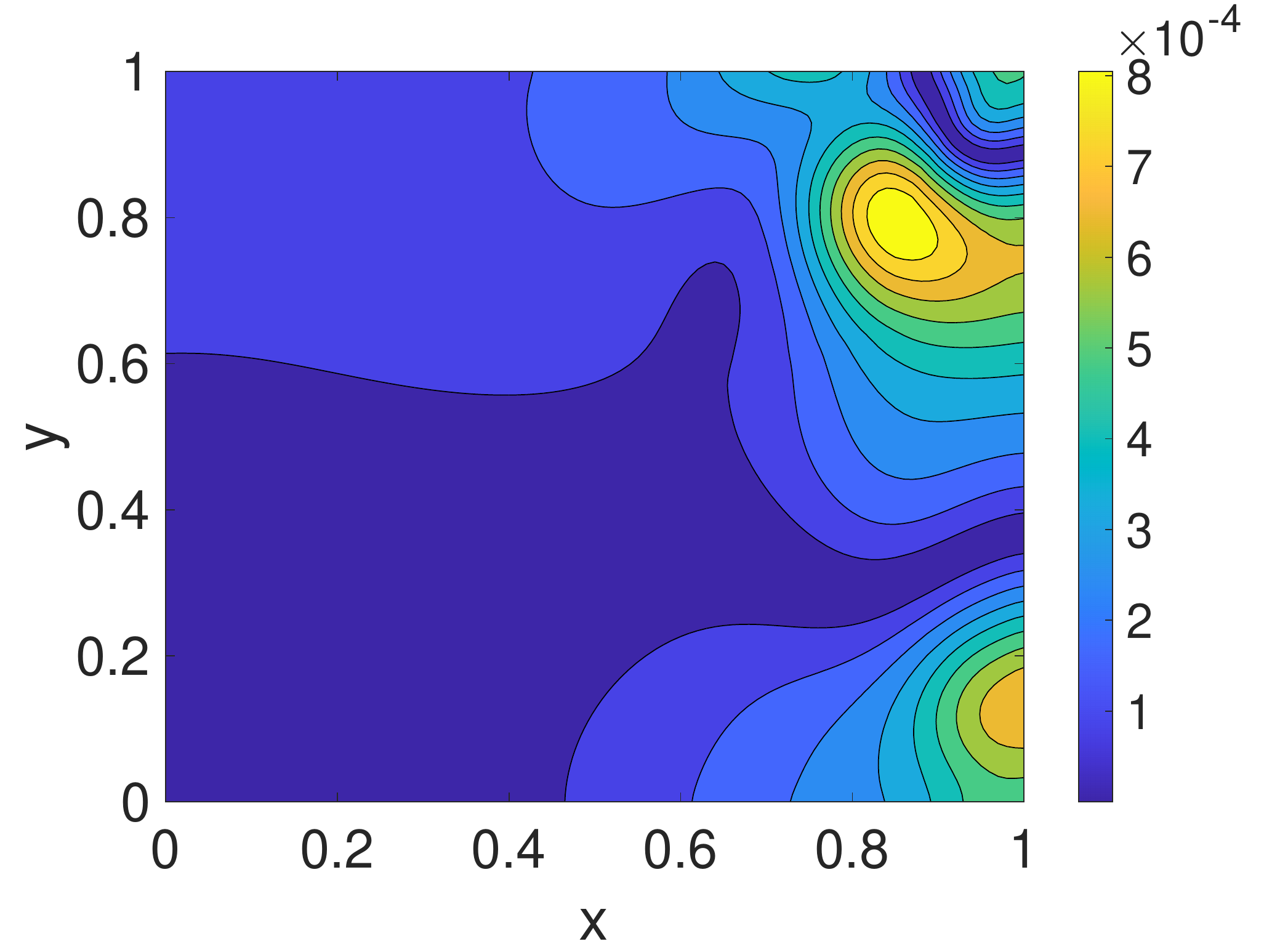}
}
\caption{Lid-driven cavity flow: Comparison of the solutions predicted by the ROM with the solutions obtained from numerical simulations at $t=1.0$ for $Re=100$. Here, $R=2$, i.e., the first 2 POD modes are retained.}
\label{fig:lidcavity_t01_Re100}
\end{figure}

\begin{figure}[htbp]
\centering
\subfigure[Velocity field (with streamlines) by numerical simulations]{
\includegraphics[width=4.8cm]{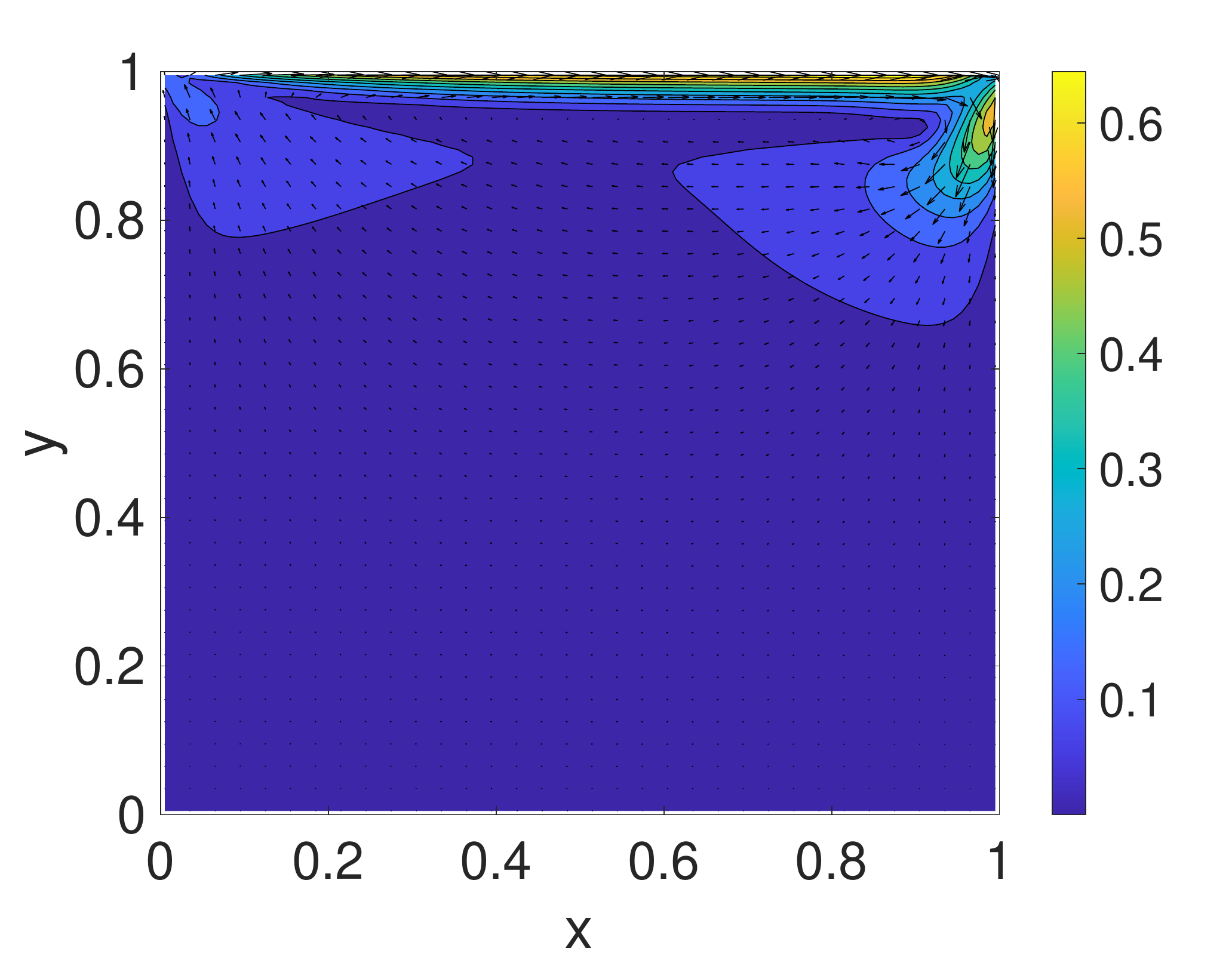}
}
\quad
\subfigure[Velocity field (with streamlines) by ROM]{
\includegraphics[width=4.8cm]{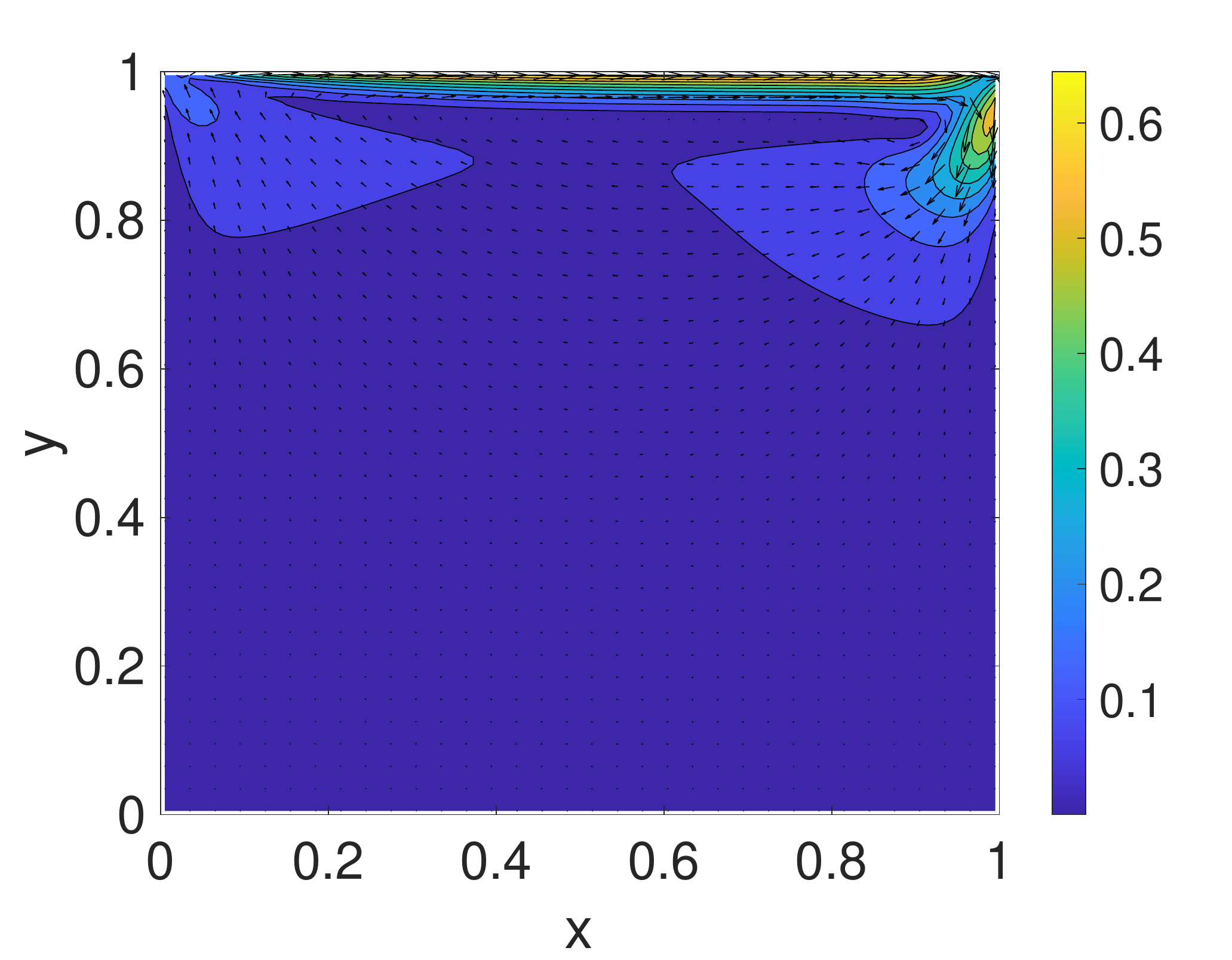}
}
\quad
\subfigure[Pointwise error for velocity]{
\includegraphics[width=4.8cm]{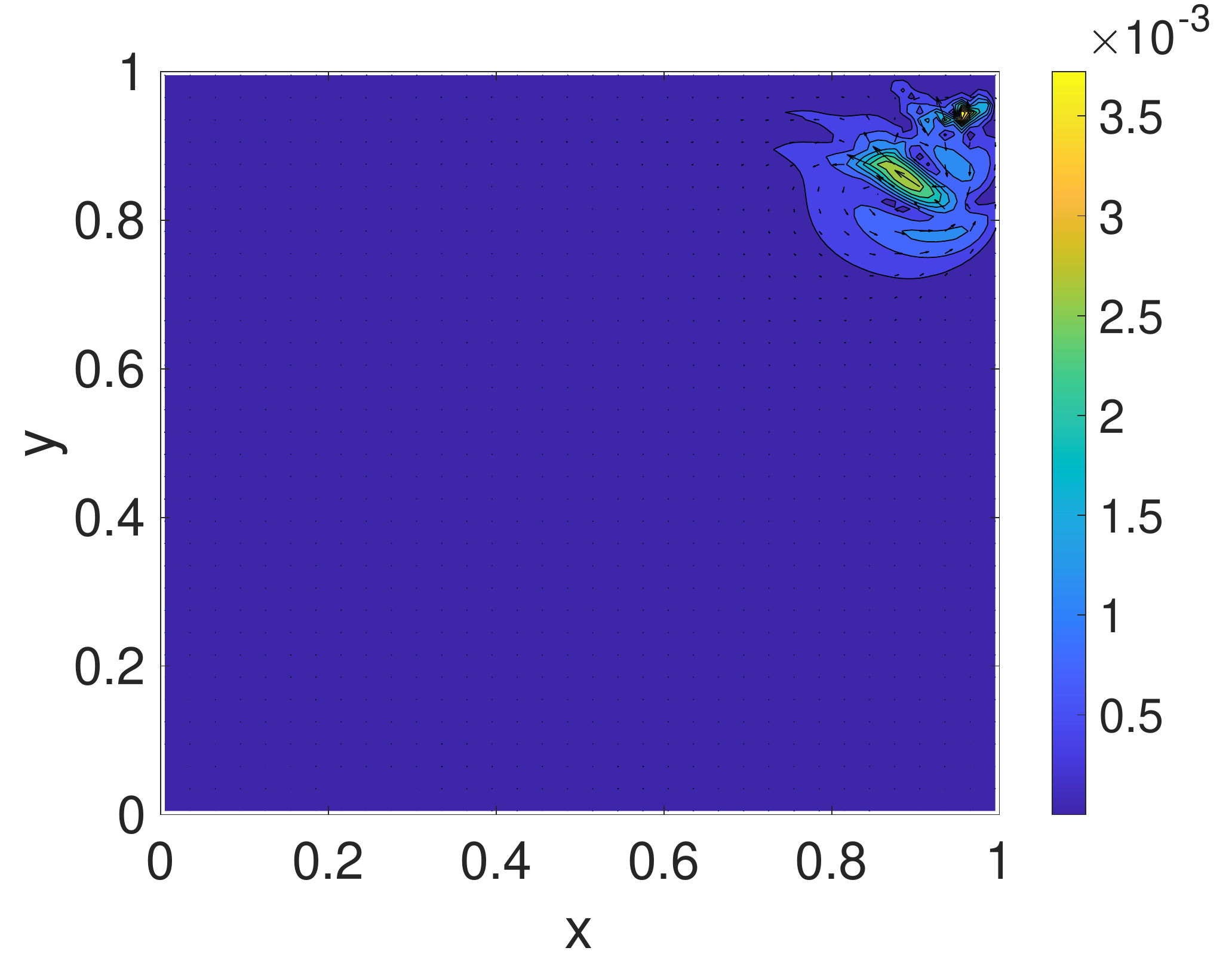}
}
\quad
\subfigure[Pressure field by numerical simulations]{
\includegraphics[width=4.8cm]{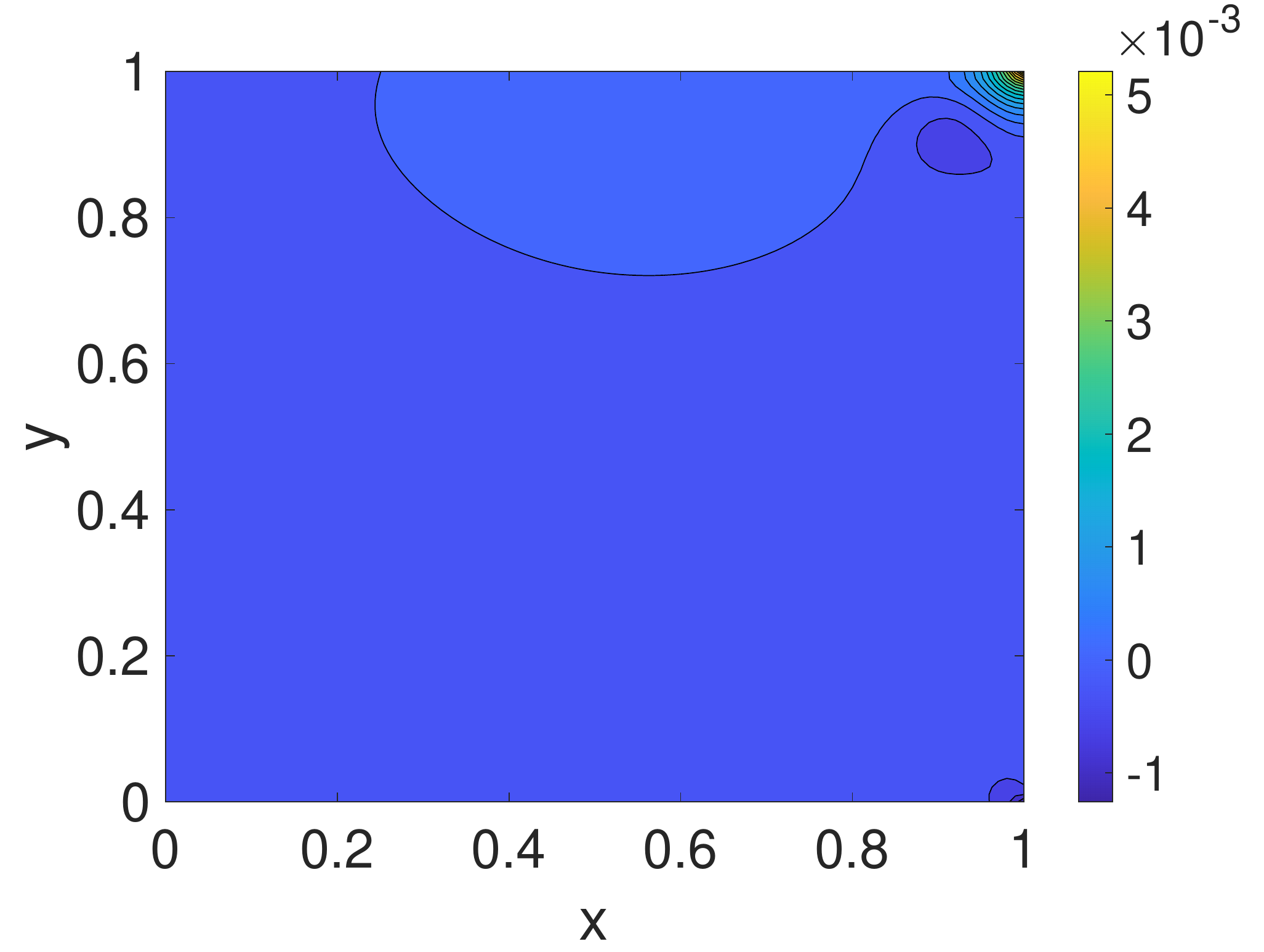}
}
\quad
\subfigure[Pressure field by ROM]{
\includegraphics[width=4.8cm]{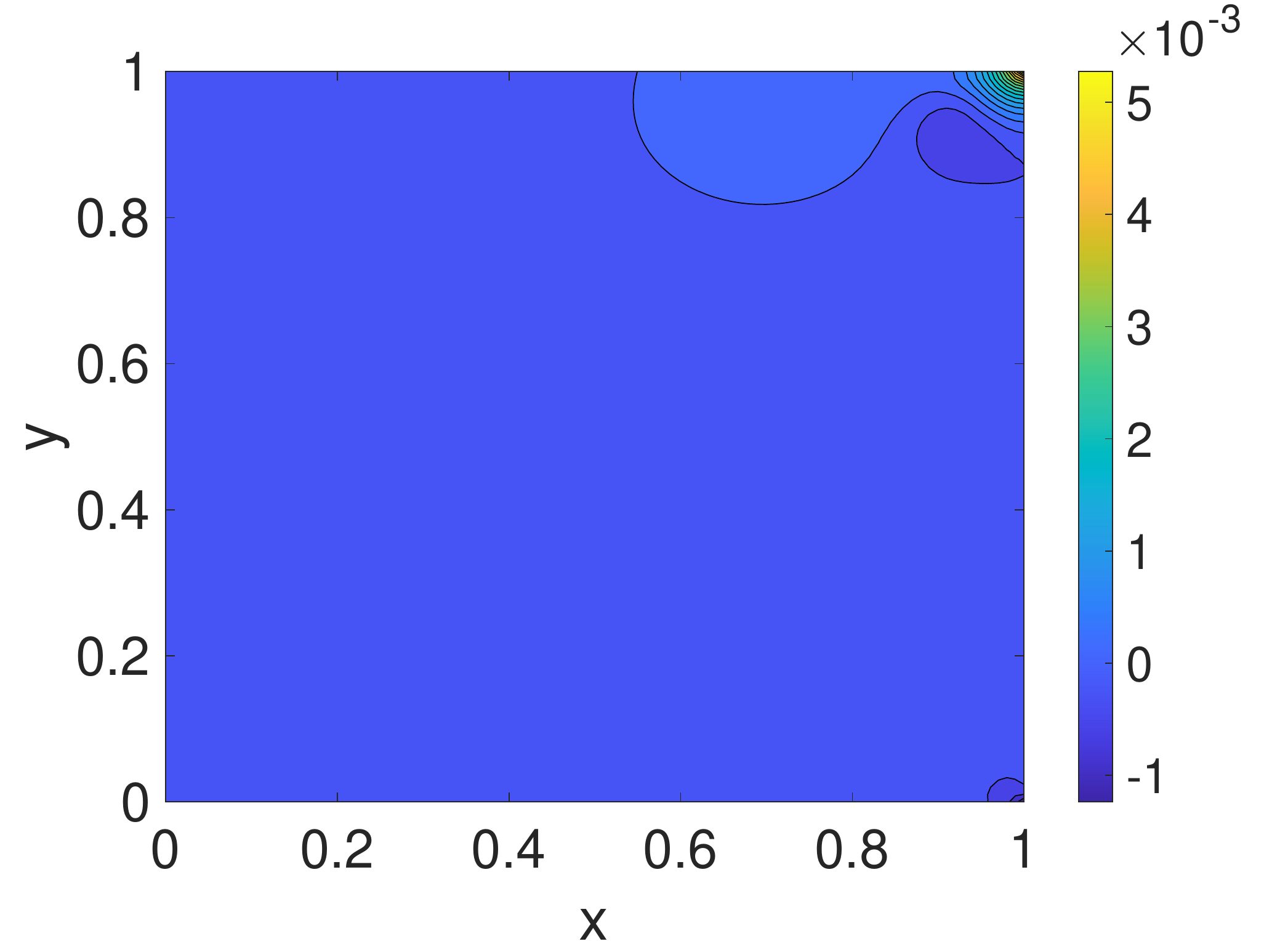}
}
\quad
\subfigure[Pointwise error for pressure]{
\includegraphics[width=4.8cm]{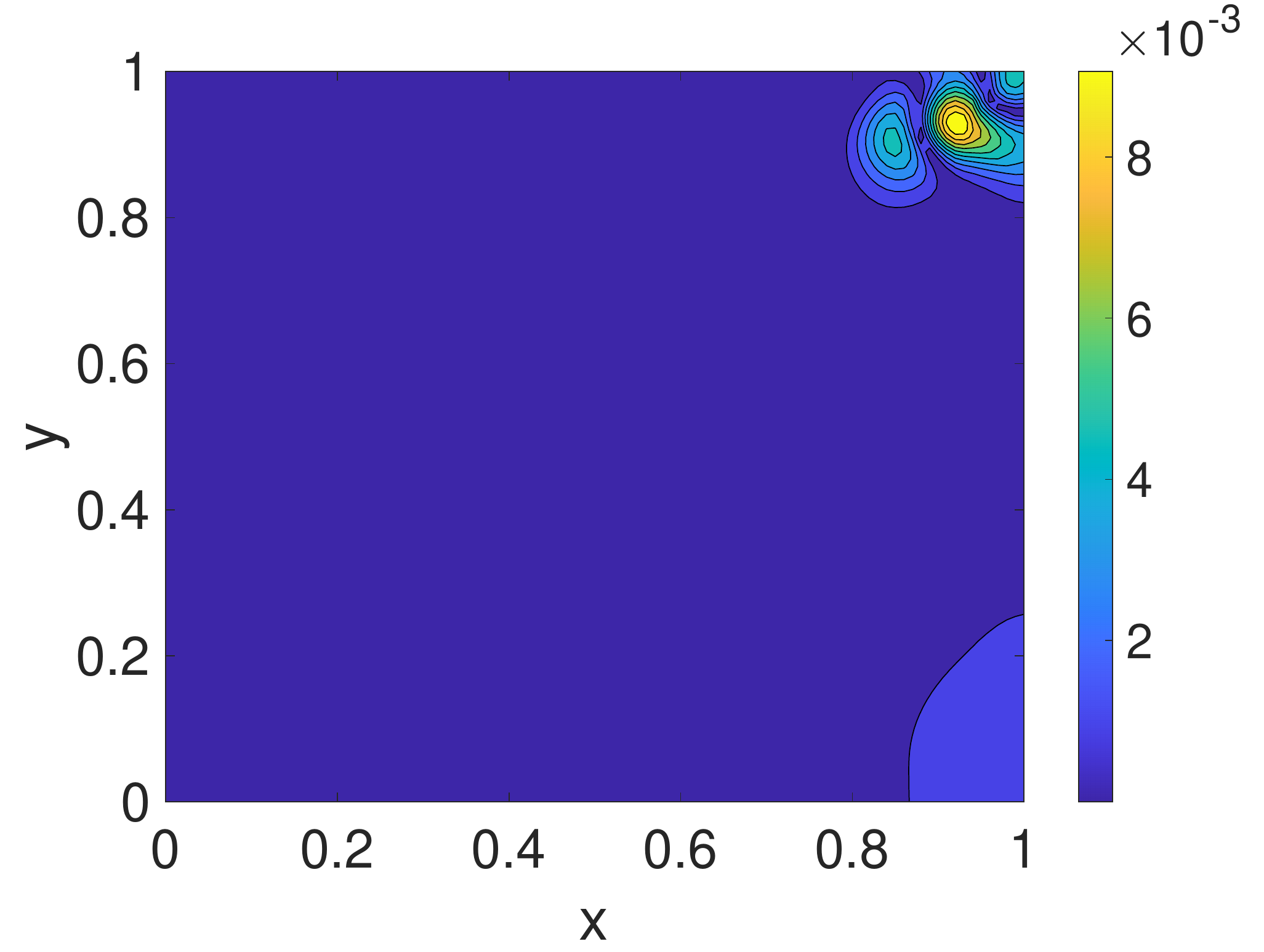}
}
\caption{Lid-driven cavity flow: Comparison of the solutions predicted by the ROM with the solutions obtained from numerical simulations at $t=1.0$ for $Re=1000$. Here, $R=3$, i.e., the first 3 POD modes are retained.}
\label{fig:lidcavity_t01_Re1000}
\end{figure}

Two different Reynolds numbers were considered: $Re=100$ and $Re=1000$. In numerical simulations, the time step was set as $\delta t=0.01$, and $100\times 100$ grids were used in the spatial discretization. 
%The DNS was conducted until $t=0.8$, and 
We took the numerical solutions between $t=0.6 \sim 0.8$ as the snapshot data to construct the ROM using the POD and GPR. The POD truncation threshold was set as $\alpha^\text{POD} = 0.01$.
The constructed ROM was then employed to predict the velocity and pressure fields of the flow for $t>0.8$. Figures \ref{fig:lidcavity_t01_Re100} and \ref{fig:lidcavity_t01_Re1000} present the predicted solutions at $t^*=1.0$ for $Re=100$ and $Re=1000$, respectively. This forecast time $t^*$ was determined from Eqs. \eqref{equ:Predict_POD_limit} and \eqref{equ:Gaussian_presetToL} with $\beta ^{\text{POD}} = 0.3$ and $\beta ^{\text{GPR},a} = 0.01$. By comparison with the full-order solutions by simulations, we find the prediction of the constructed ROM is very accurate. The relative errors $\epsilon_r^\text{ROM}$ for both velocity and pressure are less than $2\%$ for either $Re=100$ or $Re=1000$.
%%%%%%%%%%%%%%
%%%%%%%%%%%%%%
%%%%%%%%%%%%%%
%%%%%%%%%%%%%%%%%%%%%%%

\begin{figure}[htbp]
\centering
\subfigure[Velocity field (with streamlines) solely by numerical simulations]{
\includegraphics[width=4.8cm]{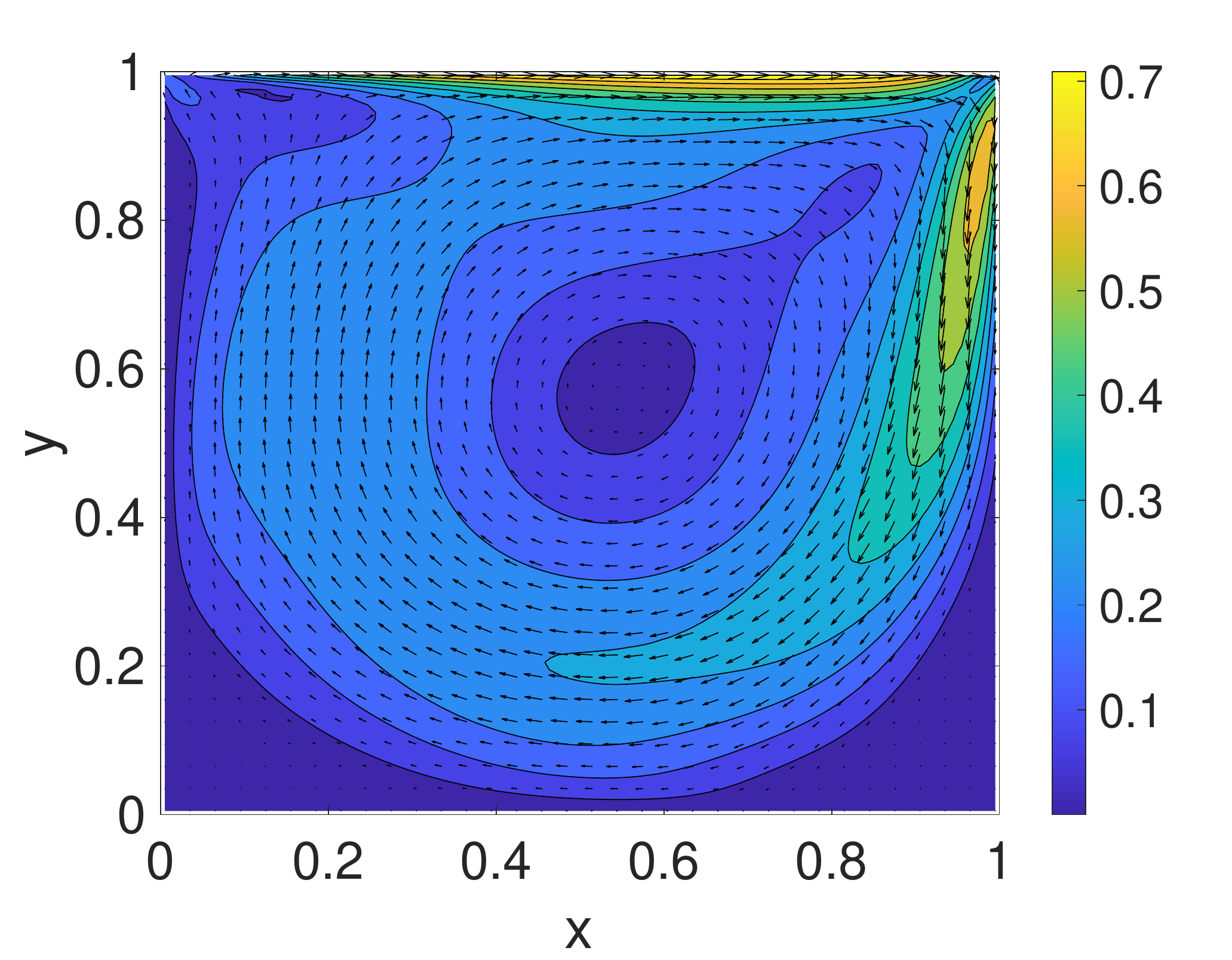}
}
\quad
\subfigure[Velocity field (with streamlines) by combining numerical simulations and ROM]{
\includegraphics[width=4.8cm]{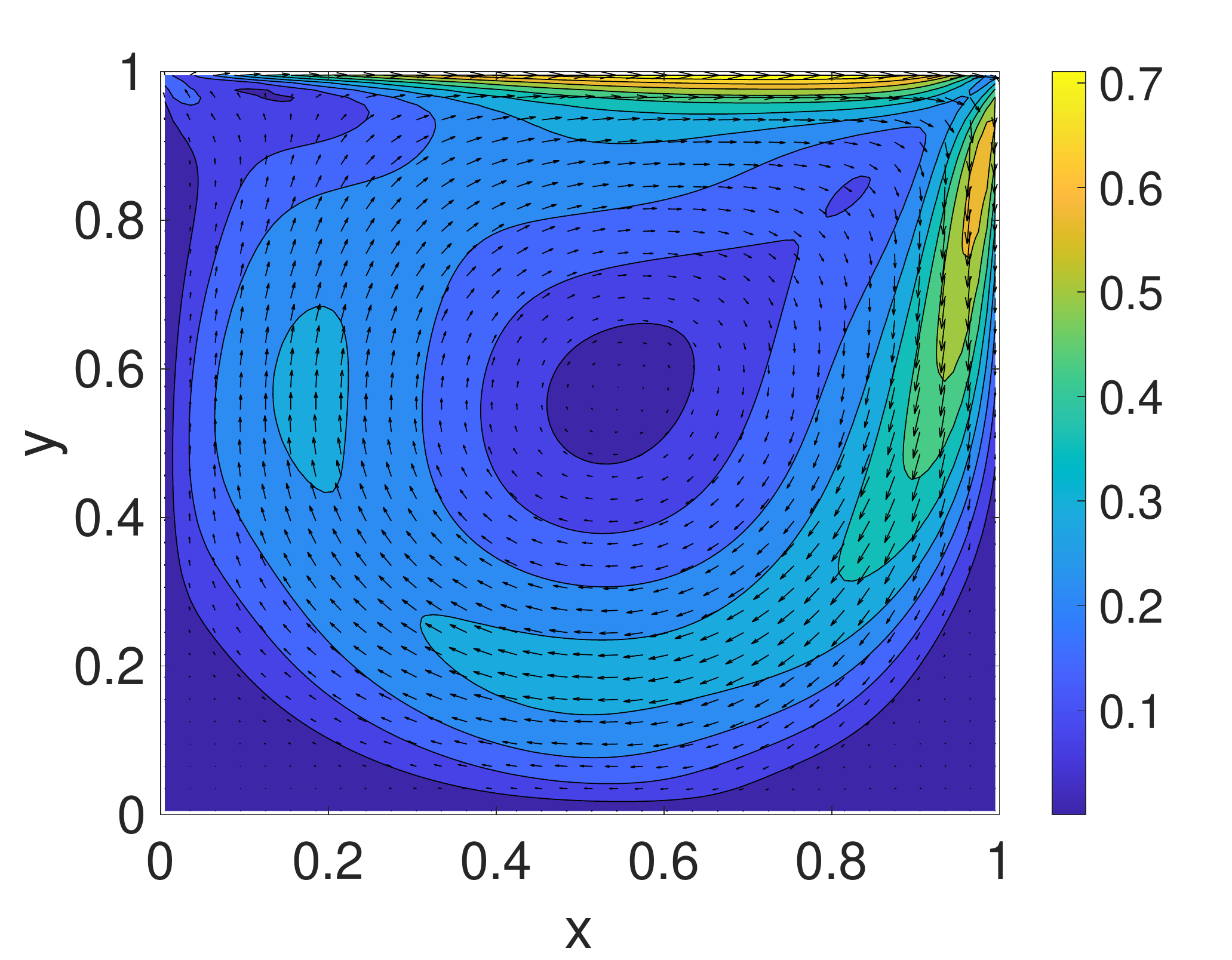}
}
\quad
\subfigure[Pointwise error for velocity]{
\includegraphics[width=4.8cm]{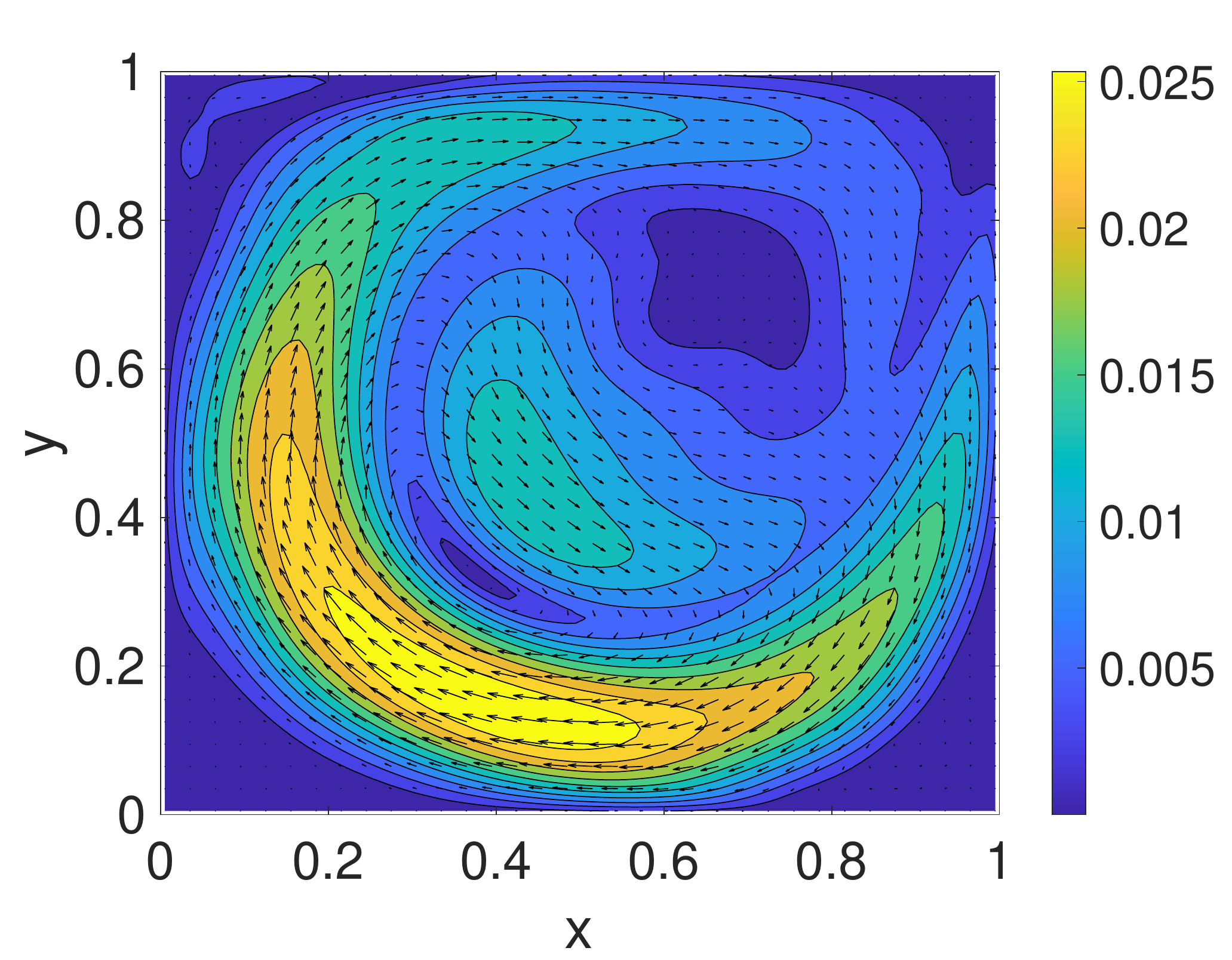}
}
\quad
\subfigure[Pressure field solely by numerical simulations]{
\includegraphics[width=4.8cm]{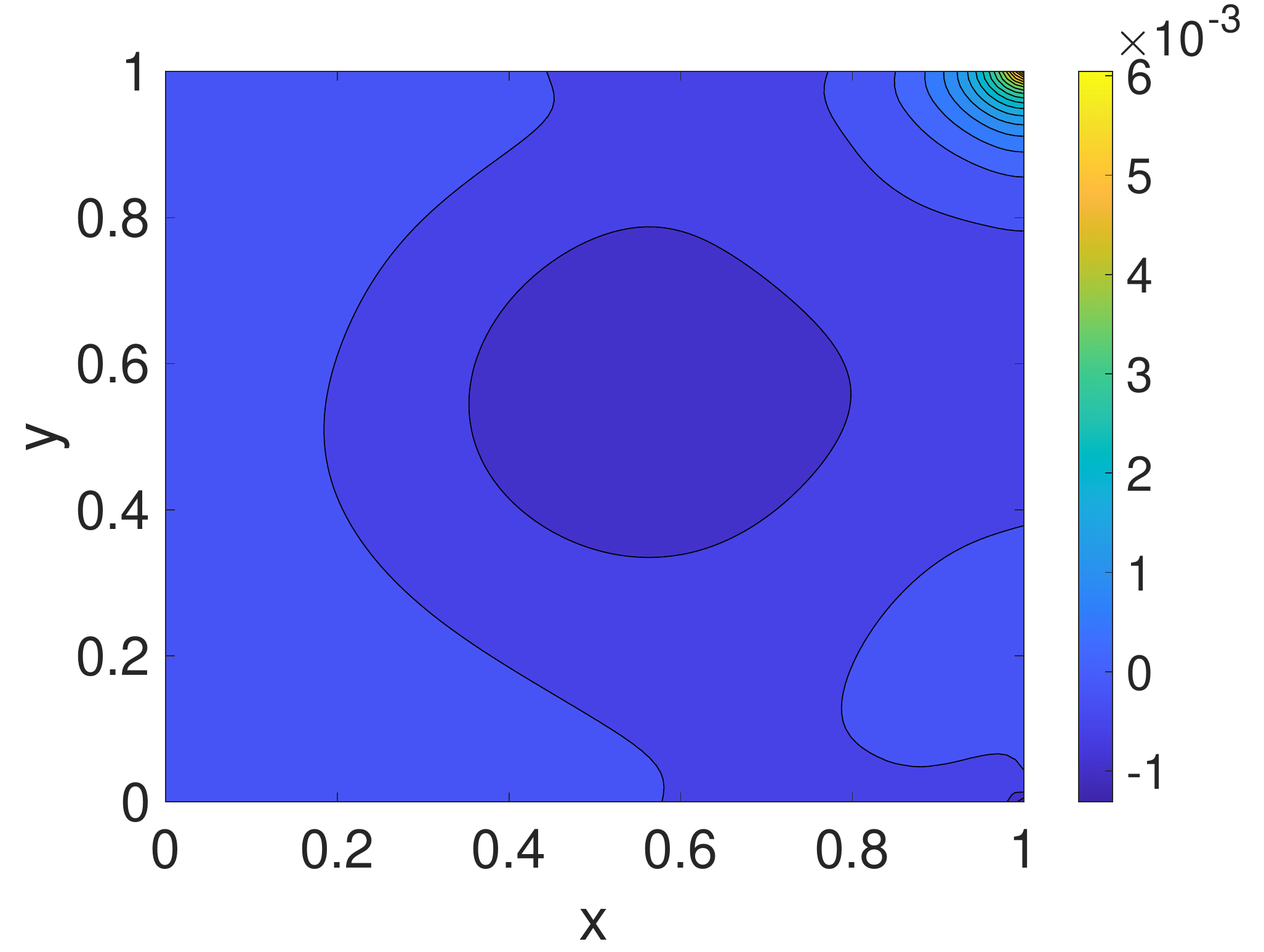}
}
\quad
\subfigure[Pressure field by combining numerical simulations and ROM]{
\includegraphics[width=4.8cm]{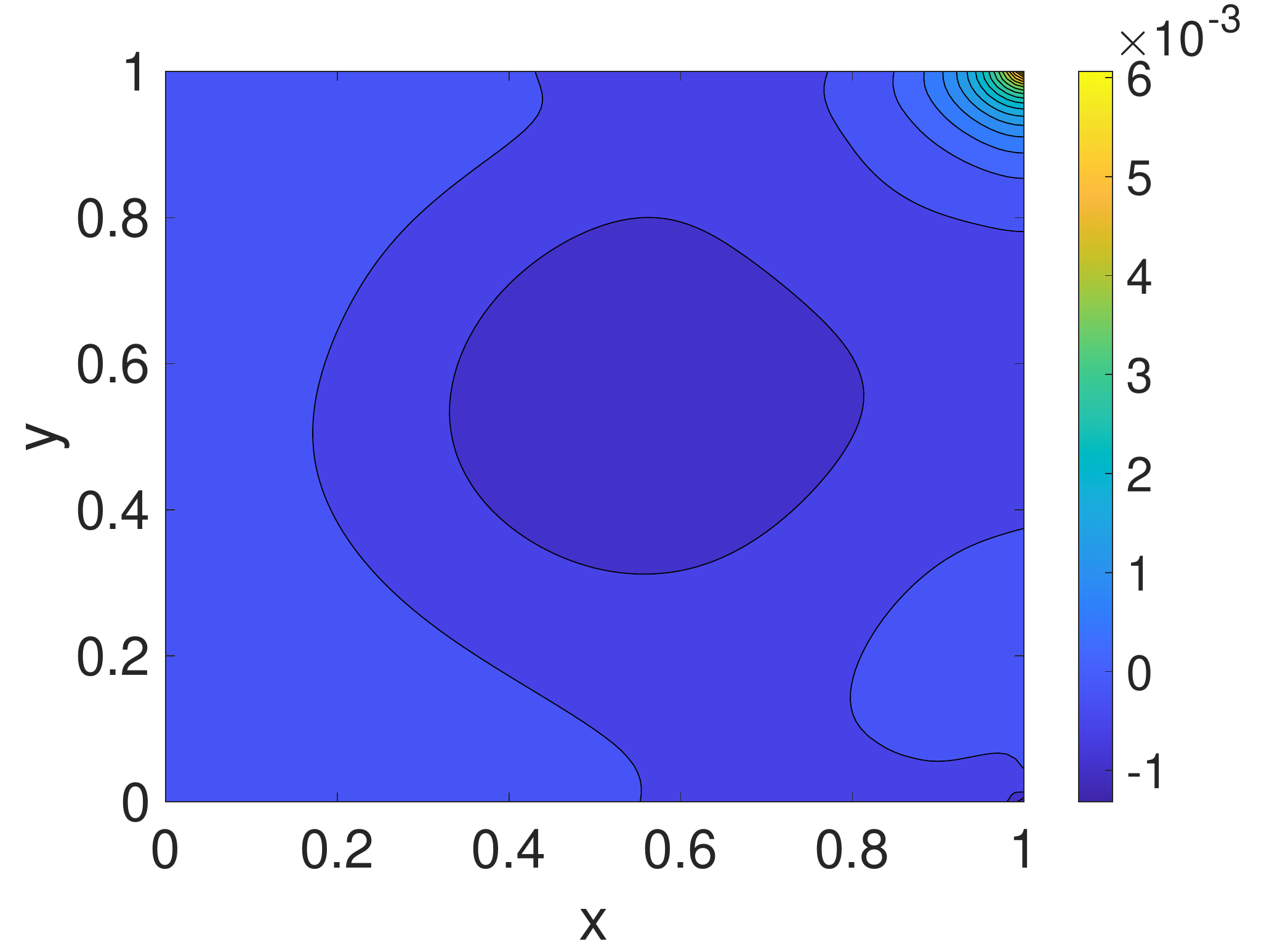}
}
\quad
\subfigure[Pointwise error for pressure]{
\includegraphics[width=4.8cm]{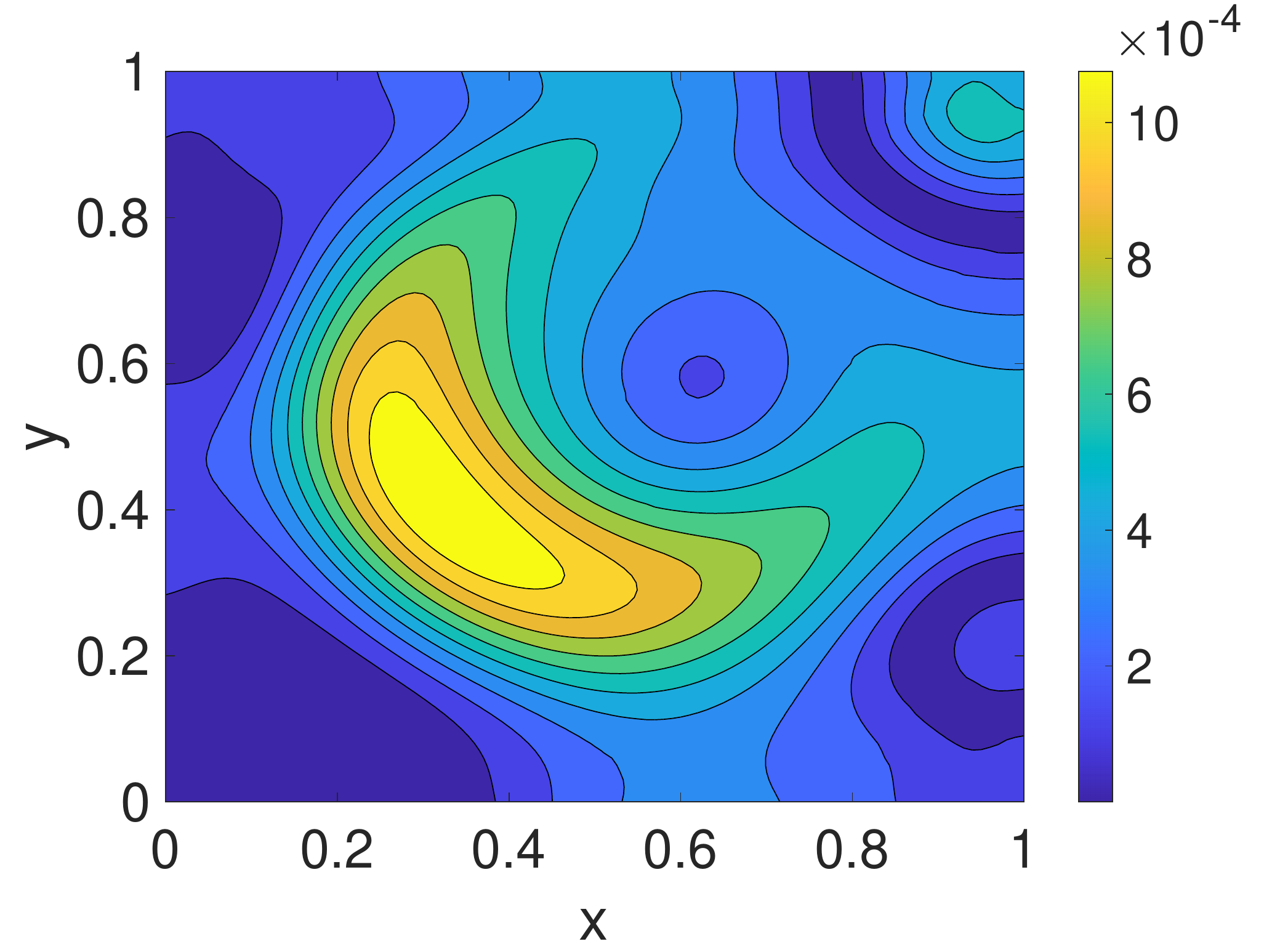}
}
\caption{Lid-driven cavity flow: Comparison of the solutions predicted by adaptively combining the numerical simulations and ROM with the solutions obtained solely from numerical simulations at $t=20.0$ for $Re=1000$.}
\label{fig:lidcavity_t20_Re1000}
\end{figure}

Next, we examined our implementation of the automated process that adaptively combines the numerical simulations and ROM (as described in \S \ref{subsubsec:algorithm_ROMGaussian}) for long-time prediction of the solutions for the lid-driven cavity flow. Figure \ref{fig:lidcavity_t20_Re1000} shows the velocity and pressure fields predicted at $t=20.0$ for $Re=1000$ by combining the numerical simulations and ROM, compared with the solutions obtained solely from numerical simulations. Good accuracy is achieved with the relative errors $\epsilon_r^\text{ROM}$ for both velocity and pressure less than $5\%$.
%For example, the solution at $t=1.0$ which is predicted by the solution from $t=0.6$ to $t=0.8$ could be used as initial condition to restart the numerical methods to get the solution from $t=1.0$ to $t=1.2$. Then these steps' solution could be used t o predict the solution at $t=1.4$ or some other time, where the prediction length is determined by the covarience in Gaussian model. Through repeating this procedure to the steady state, many steps to solve Navier-Stokes equations by numerical method could be saved. And we find that the prediction length could be large when it is close to the steady state. Fig. 4 shows the predicted fluid field at $t=20.0$ predicted by the data from $t=15.0$ to $15.2$. Because the spatial basis changed slowly near the steady state, we can save numerical computational costs of more time steps by using the POD-Gaussian prediction. 
%%%%%%%%%%%%%%%%%%%
%%%%%%%%%%%%%%%%%%%

\subsection{Heterogeneous fracture deformation}\label{subsec:2DExp}
In this section, we applied the proposed nonintrusive model order reduction method to experimental data, where the full-order model is unknown. The experimental data are the 2D displacement fields for heterogeneous fracture deformation. From the snapshots of the displacement fields at different times, we constructed the ROM using the POD and GPR. The constructed ROM was then employed to forecast the displacement fields of future times beyond the snapshot data. 

In experiments, the data of displacement fields were obtained by employing the augmented-Lagrangian digital image correlation (ALDIC) method \cite{2Dexp_ALDIC_Jin2019} on the images from the experiment of Avellar \cite{2Dexp_image_AvellarPhDthesis}. The images recorded the gray scale of a speckle pattern painted on the surface of a material specimen. Upon deformation, the gray scale of the speckle pattern changed. Hence, by comparison of the gray scale in the images before and after deformation, the displacement field can be determined by solving an optimization problem. Although effective, the ALDIC method is quite expensive and requires   storage of massive image data. Instead, the ROM, once constructed, can efficiently reproduce the displacement field at any given time until the furthest forecast time $t^*$ allowed by the ROM is reached. To demonstrate the proposed reduced order modeling, we chose to consider the data for a material specimen with heterogeneous stiffness and thereby complex displacement fields \cite{2Dexp_ALDIC_Jin2019}, for which the saving of computational cost by using the ROM compared with the ALDIC method is even more remarkable.

\begin{figure}[htbp]
\centering
\subfigure[ALDIC]{
\includegraphics[width=4.8cm]{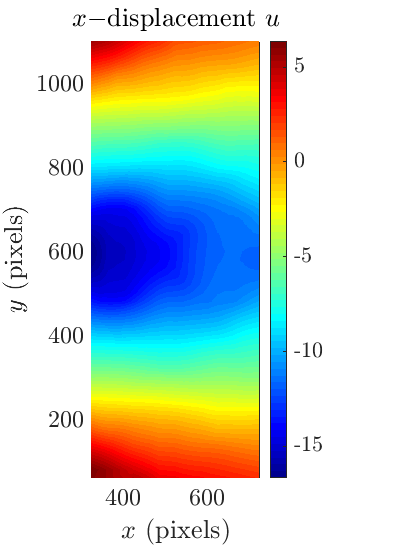}
}
\quad
\subfigure[ROM]{
\includegraphics[width=4.8cm]{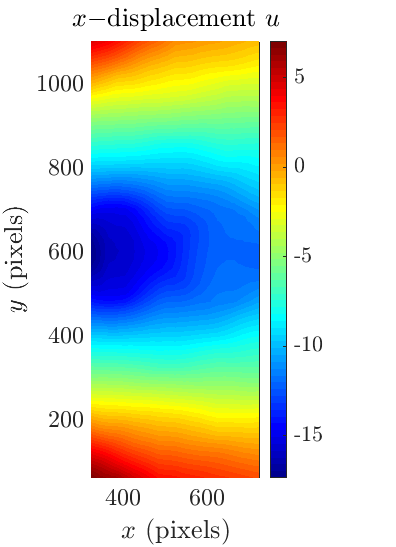}
}
\quad
\subfigure[Pointwise error]{
\includegraphics[width=4.8cm]{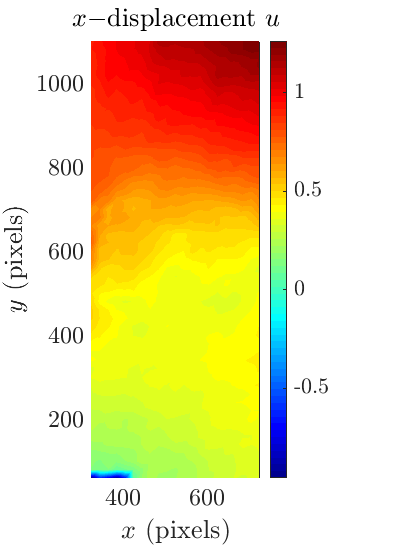}
}
\quad
\subfigure[ALDIC]{
\includegraphics[width=4.8cm]{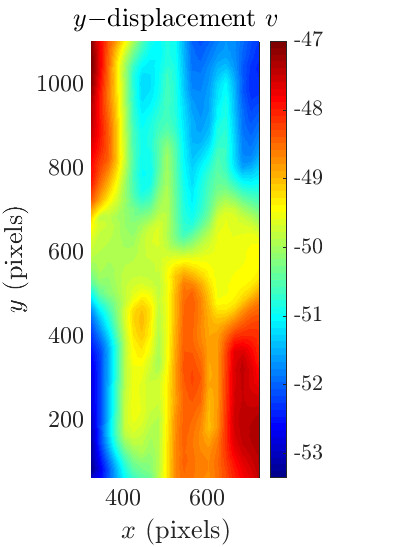}
}
\quad
\subfigure[ROM]{
\includegraphics[width=4.8cm]{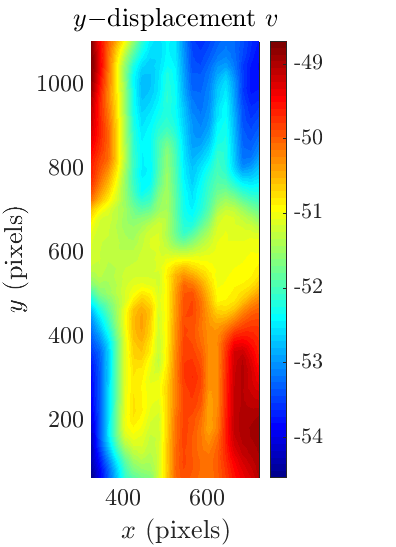}
}
\quad
\subfigure[Pointwise error]{
\includegraphics[width=4.8cm]{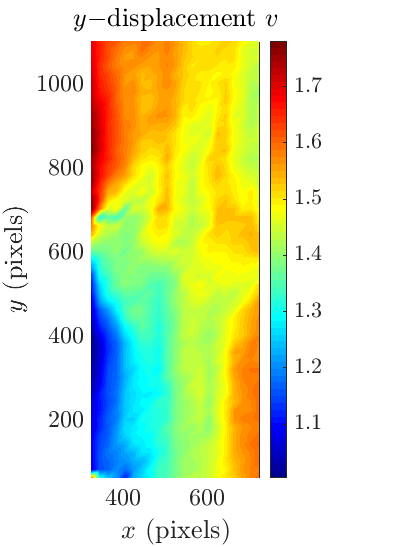}
}
\caption{Heterogeneous fracture deformation: The ((a)-(c)) $x$- and ((d)-(f)) $y$-displacement fields predicted by the ROM at $t^*=25$, compared with the results by the ALDIC method.}
\label{fig:fracture_displacement}
\end{figure}

More specifically, we used 15 snapshots of the displacement fields from $t_1 = 0$ to $t_M = 15$ to construct the ROM, which was then employed to predict the displacement field at $t^*=25$. Here, the first 3 POD modes were retained ($R=3$) with the POD truncation threshold $\alpha^\text{POD} = 0.01$. The furthest forecast time $t^*$ was determined from Eqs. \eqref{equ:Predict_POD_limit} and \eqref{equ:Gaussian_presetToL} with $\beta ^{\text{POD}} = 0.3$ and $\beta ^{\text{GPR},a} = 0.1$. The results are presented in Figure \ref{fig:fracture_displacement}. For comparison, we also show the displacement fields determined using the ALDIC method. From the comparison, we demonstrate the good accuracy of the predictions by the ROM. Taking the results by the ALDIC method as the ``exact" solutions, the relative error $\epsilon_r^\text{ROM}$ is less than 8\% for the prediction of $x$-displacement field and less than 2\% for the prediction of $y$-displacement field.
%%%%%%%%%%%%%%%%%%%%%%%%%
%%%%%%%%%%%%%%%%%%%%%%%%%

In regard to the computational cost, it typically takes the ALDIC method about 53s to generate the displacement field from two images \cite{2Dexp_ALDIC_Jin2019}, which was tested in {\sc Matlab} using a workstation with Intel (R) Xeon(R) CPU E5-2650 v3 2.30 GHz (2 Processors). It only takes the ROM about 0.02s to predict the displacement field using {\sc Matlab} and comparable hardware. And the cost for constructing the ROM is about 0.70s. Thus, integrating the ALDIC method with the proposed reduced order modeling can significantly reduce the computational cost in experiments to determine the time-varying displacement fields.
%And for the proposed POD-Gaussian method, it only need 1.03s, which is performed by {\sc Matlab} on Intel(R) Core(TM) i5-6500 CPU @ 3.20GHz.

\subsection{Bubble cavitation in hydrogel}\label{subsec:buble_cavity}
\begin{figure}[htbp]
\centering
\includegraphics[width=6cm]{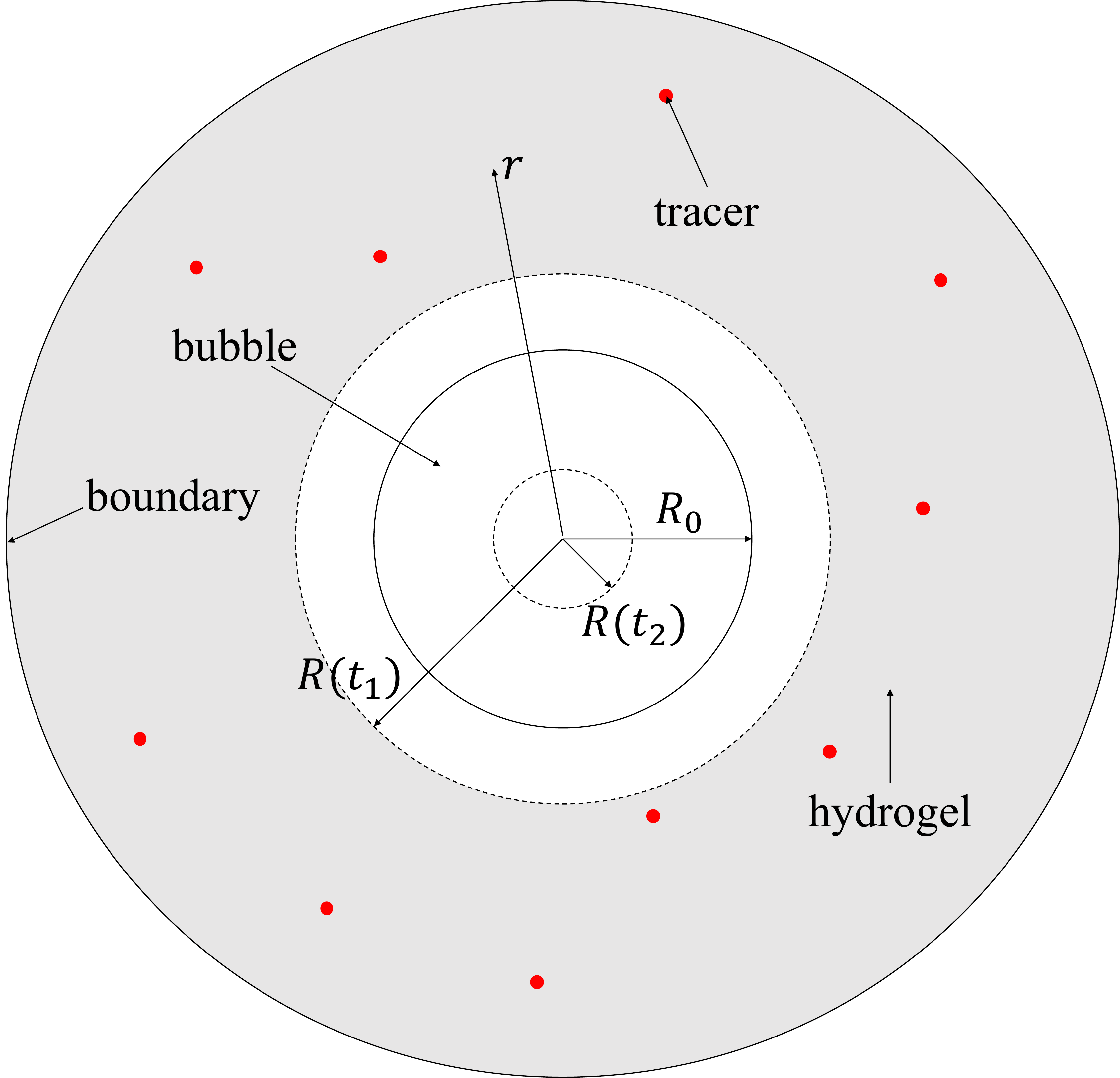}
\caption{Schematic of a bubble cavitation in hydrogel. The bubble can expand (e.g. at $t_1$) or shrink (e.g., at $t_2$).}
\label{fig:Bubble}
\end{figure}

After the three benchmarks used to validate the nonintrusive model order reduction method based on the POD and GPR, we next considered a problem involving moving boundary. In particular, it concerned bubble cavitation in hydrogel. The data used to construct and validate the ROM were from the experiments \cite{1Dexp_cavity_Jin2020}, where a spherical bubble cavitation was formed in the center of a hydrogel material. Assume the initial radius of the bubble is $R_0$. Due to the plasma recombination and the volume change of the vapor and  non-condensable gas within the bubble, the radius of the bubble is a function of time $R(t)$; as a result, the interface between the bubble and hydrogel is a moving boundary. The expansion or shrinkage of the bubble led to compression or tension on the surrounding hydrogel material. A schematic of this problem is illustrated in Figure \ref{fig:Bubble}. 
%%%%%%%%%%%%%%%%%%%%%%%%
%%%%%%%%%%%%%%%%%%%%%%%%
The distribution of the elastic strain $S$ in the surrounding hydrogel as well as the bubble's radius $R(t)$ were measured in the experiments. The elastic strain $S$ was measured at each tracer position. Due to the spherical symmetry of this problem, we simplified it to a one-dimensional problem. Hence, the strain field only depends on $r$, the radial distance to the center of the bubble. 

\begin{figure}[htbp]
\centering
\includegraphics[width=12cm]{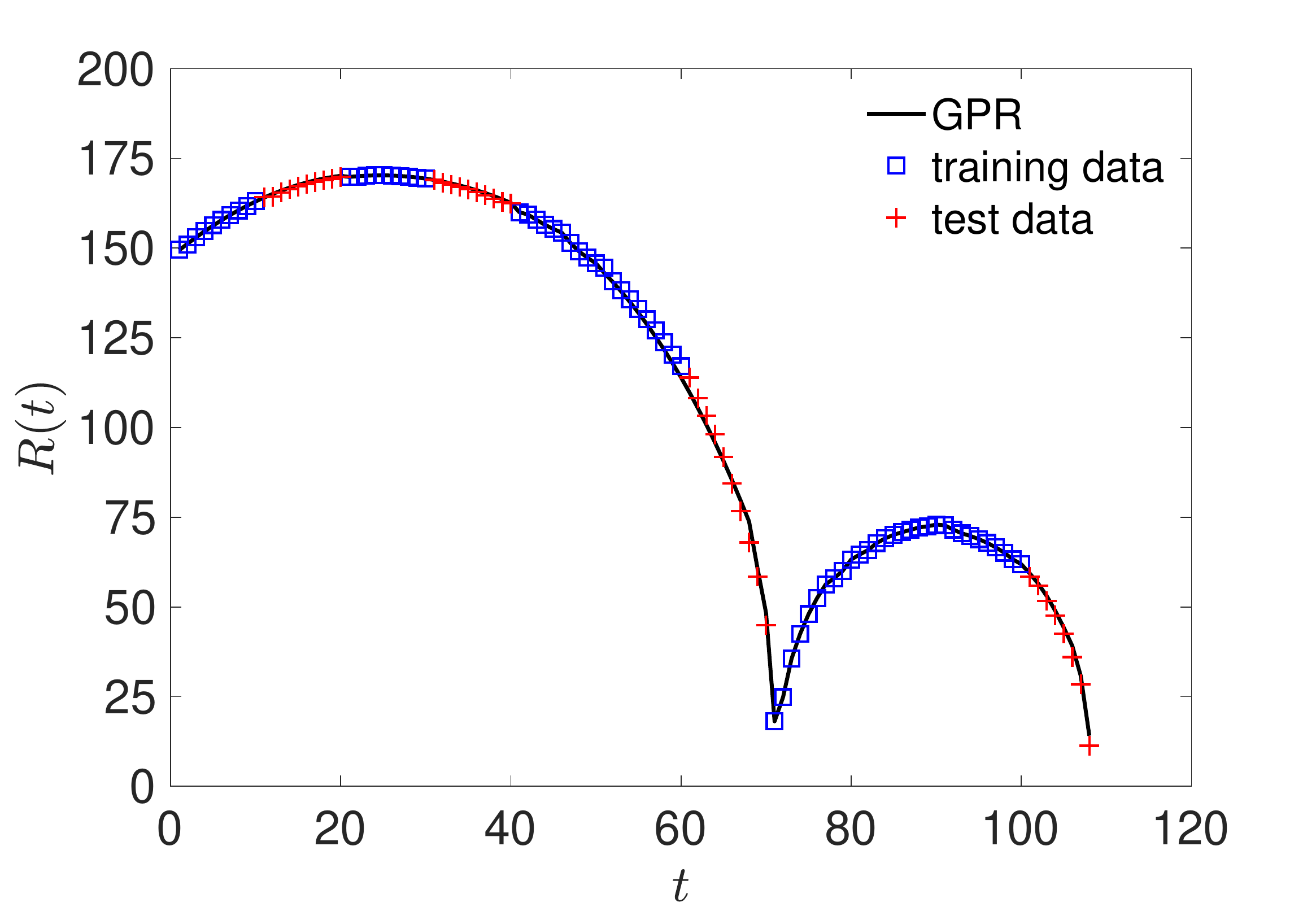}
\caption{Bubble cavitation in hydrogel: The prediction of the bubble cavitation's radius $R(t)$.}
\label{fig:bubbleradius}
\end{figure}

From $M$ snapshots of the experimental data, we first constructed a GPR model for $R(t)$, the bubble's radius as a function of time. The constructed GPR model can forecast the bubble's radius at a given time beyond the range of data. The furthest forecast time $t^*$ was determined from Eqs. \eqref{equ:Predict_POD_limit}, \eqref{equ:Gaussian_presetToL} and \eqref{equ:predict_boudary_limit} with $\beta^{\text{POD}}=0.3$, $\beta^{\text{GPR},a}=0.1$ and $\beta^{\text{GPR},\Gamma_m}=0.1$. After the furthest forecast time $t^*$ was reached, new $M$ snapshots of experimental data were collected and used to infer another GPR model to forecast $R(t)$ for further times. Repeating this process and the resulting adaptive combination of experimental measurements and GPR modeling allowed us to efficiently predict the long-time evolution of the radius of the bubble cavitation with a complex dynamics of expansion and shrinkage, as shown in Figure  \ref{fig:bubbleradius}. Good agreement was achieved between the GPR model's predictions and the test data from experiments. 
%%%%%%%%%%%%%%%%%%%%%%%%%
%%%%%%%%%%%%%%%%%%%%%%%%%
%The peak bubble size builds large elastic stresses in the surrounding material, which drive primary bubble collapse. The radius of bubble decreases as Fig.\ref{subfig:Bubblecollapse} shown. Later, lower-amplitude bubble oscillation is dominated by material viscous effects, finally reaching mechanical equilibrium. Thus with the oscillation of bubble, the strain field in the material will be influenced. In experiment, there are some random impurities, which are shown as the small circle in Fig.\ref{fig:Bubble}, embedded in the surrounding hydrogels and moving along with the bubble dynamics. The impurities are used as tracers to visualize how the surrounding hydrogel deforms. The position of impurities could be tracked to get the strain field in the hydrogel.

\begin{figure}[htbp]
\centering
\subfigure[$t^*=20$]{\label{subfig:Bubblestraincompression}
\includegraphics[width=7.5cm]{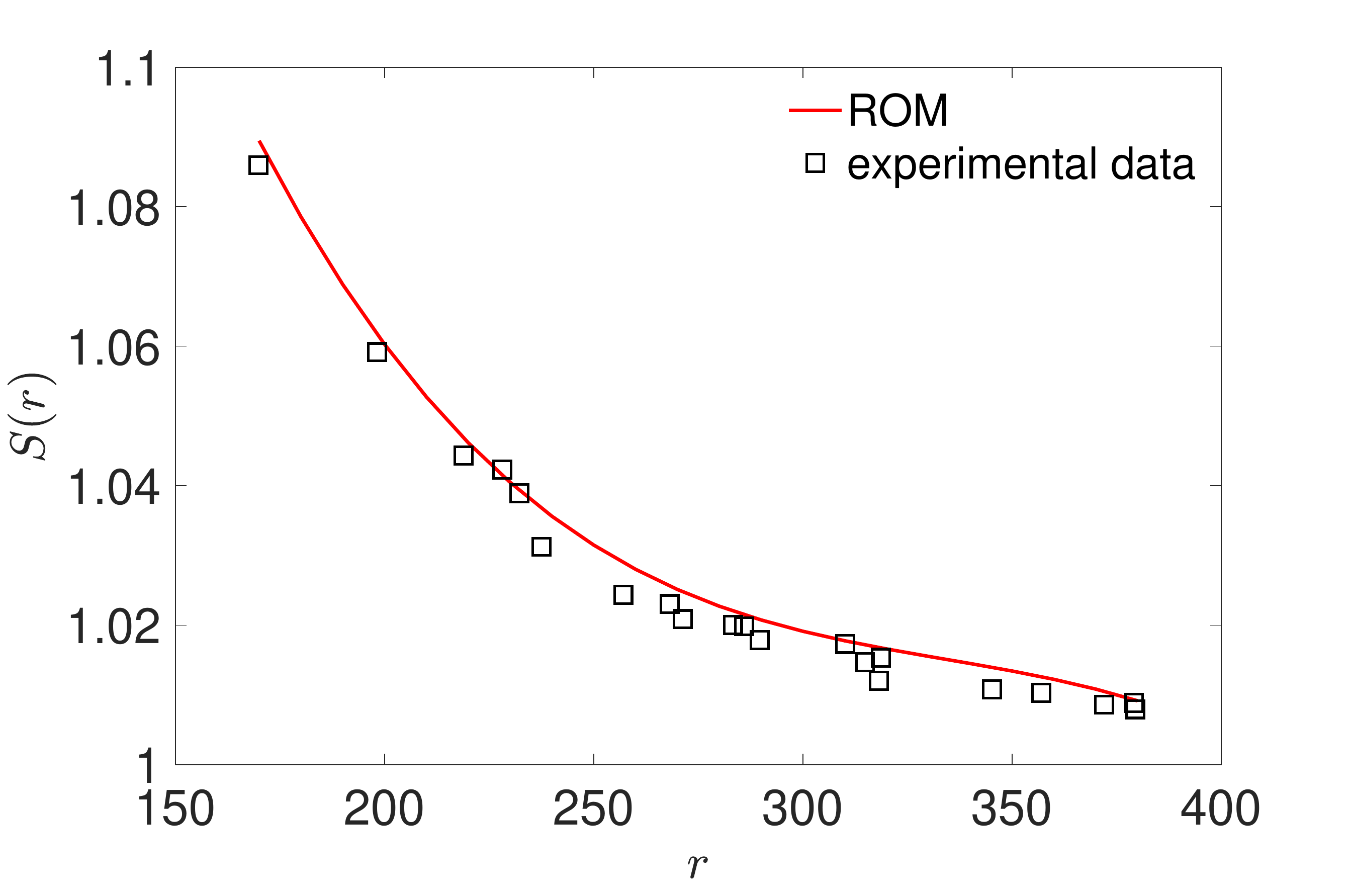}
}
\quad
\subfigure[$t^*=70$]{\label{subfig:Bubblestraintension}
\includegraphics[width=7.5cm]{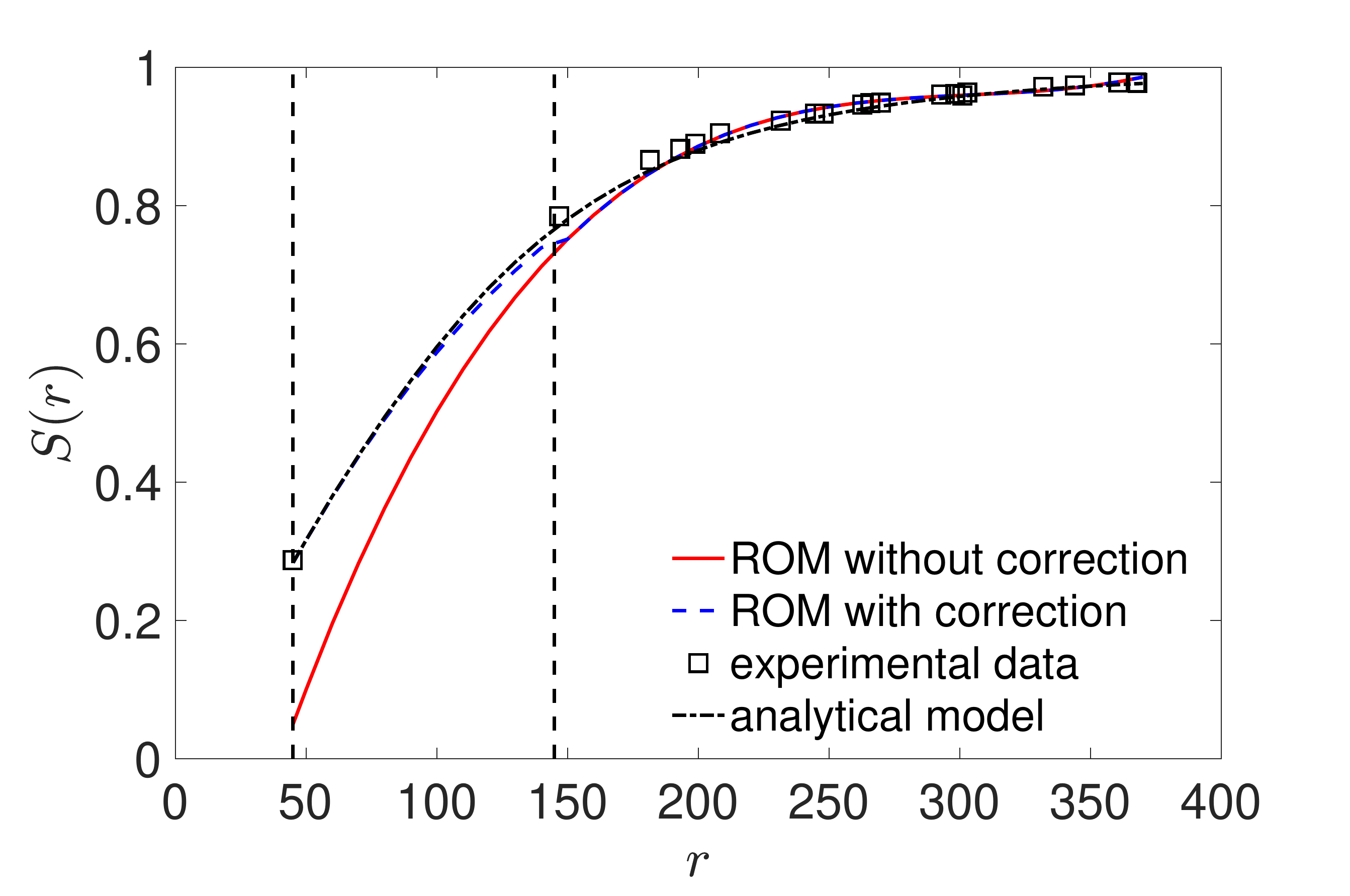}
}
\caption{Bubble cavitation in hydrogel: The prediction of the elastic strain $S(r,t)$ in hydrogel.}
\label{fig:Bubblestrain}
\end{figure}

Besides the bubble cavitation's radius, we also predicted the elastic strain field $S(r,t)$ in the surrounding hydrogel. To this end, the ROM for $S(r,t)$ was constructed using the POD and GPR from the experimental snapshot data. For demonstration, we predicted for two different times: one during the expansion of the bubble and the other during the bubble's shrinkage. More specifically, we used 10 snapshot data from $t_1 =1$ to $t_M=10$ to construct a ROM, which was then employed to forecast $S(r,t)$ at $t^*=20$; and we used 10 snapshot data from $t_1=51$ to $t_M=60$ to construct another ROM to predict $S(r,t)$ at $t^*=70$. 
%The experimental data was interpolated onto uniform grids by least squares. 
Each prediction of $S(r,t)$ by the constructed ROM, along with the comparison with the experimental data, is presented in Figure \ref{fig:Bubblestrain}. After $t>50$, the bubble cavitation experienced shrinkage until $t=70$. During this period of time, the surrounding hydrogel was stretched toward the bubble's center. Hence, the strain field $S(r,t>t_M)$ included the range of $r$ that was not covered in the snapshot data (from $t_1=51$ to $t_M=60$). To construct the ROM covering that range of $r$, we extrapolated by least squares the snapshot data of $S(r)$ until the minimum possible value of $R$. When we forecasted $S(r,t)$ at $t^*=70$ by the constructed ROM, for the region of $r$ not covered in the snapshot data (i.e., the region between two vertical dash lines in Figure \ref{subfig:Bubblestraintension}), the correction was made using the method proposed in \S\ref{subsubsec:correction_MB}. To validate the  prediction of $S(r)$ in that region, since no experimental data were available, we relied on the following analytical model \cite{1Dexp_cavity_Jin2020,1Dexp_modelYang2005} for validation: 
%%%%%%%%%%%%%%%%%%%
\begin{equation}
S(r,t) = (\frac{r}{r^3 + {R(\bar{t})}^3 - R(t)^3})\; , 
\label{equ:bubble_strainmodel}
\end{equation}
%%%%%%%%%%%%%%%%%%%
where $\bar{t}=5$.
%%%%%%%%%%%%%%%%%%%%%%%%%
%%%%%%%%%%%%%%%%%%%%%%%%%
As can be seen in Figure \ref{fig:Bubblestrain}, the predictions of the strain field $S(r,t)$ by the ROM reasonably agree with the experimental data and the analytical model. Thus, we anticipate the proposed reduced order modeling can be applied to predict the strain field for the regions and times that are not accessible in experiments or when the analytical model is not applicable (e.g., after the first collapse of the bubble). 

\subsection{Fluid-solid interactions}\label{subsec:FSI}
Finally, we studied the problems of fluid-solid interactions, where multiple solid bodies move in a fluid flow. Assume the domain $\Omega_m$ consists of $N_s$ solid bodies, and each of them has a boundary $\Gamma_m^n,\ \ n=1,...,N_s$, and a center-of-mass position $\bm{X}_n$ and orientation $\bm{\Theta}_n$. In this study, we predicted the dynamics of each solid body as well as the velocity and pressure fields in the fluid by the ROM constructed. The snapshot data used to construct the ROM were generated via numerical simulations, where the incompressible Stokes flow subject to moving solid boundaries was numerically solved using the generalized moving least squares discretization method \cite{GMLS_Hu2019}. For simplicity, the solid bodies are subject to rigid-body kinematics. The evolution of a moving solid boundary $\Gamma_m^n$ can be characterized by $[\bm{X}_n, \bm{\Theta}_n]$. 

% The flow is coupled to the colloid motion by the following Stokes equation:
% \begin{equation}
%   \begin{matrix}
%   \left\{
%     \begin{aligned}
%         & \frac{\nabla p}{\rho} - \mu \Delta^2 \bm{u}=0 ,\\
%         & \nabla \cdot \bm{u}=0,\\
%         & \bm{u}=\bm{w},\\
%         & \bm{u}=\dot{\bm{X}}_n+\dot{\bm{\Theta}}_n \times (\bm{x}-\bm{X}_n)
%     \end{aligned}
%   \right.
%   &
%   \begin{aligned}
%      & for \ \ x \in \Omega_f\\
%      & for \ \ x \in \Omega_f\\
%      & for \ \ x \in \Gamma_w\\
%      & for \ \ x \in \Gamma_n, \ \ n=1,...,N_c, 
%     \end{aligned}
%   \end{matrix}
% \end{equation}
% where $\mu$ is the kinematic viscosity of fluid; $\Gamma$ denotes the boundary of the domain which could be partitioned into disjoint union $\Gamma= \Gamma_w \cup \Gamma_1 \cdots \cup \Gamma_{N_c}$, here $\Gamma_w$ denotes the wall boundaries and $\Gamma_n$ denotes the boundaries of the colloids; $\bm{w}$ is the velocity of the wall boundaries, and $\bm{w}=\bm{0}$ for stationary walls. 

\begin{figure}[htbp]
\centering
\subfigure[Velocity in $y$ computed from numerical simulations]{
\includegraphics[width=4.8cm]{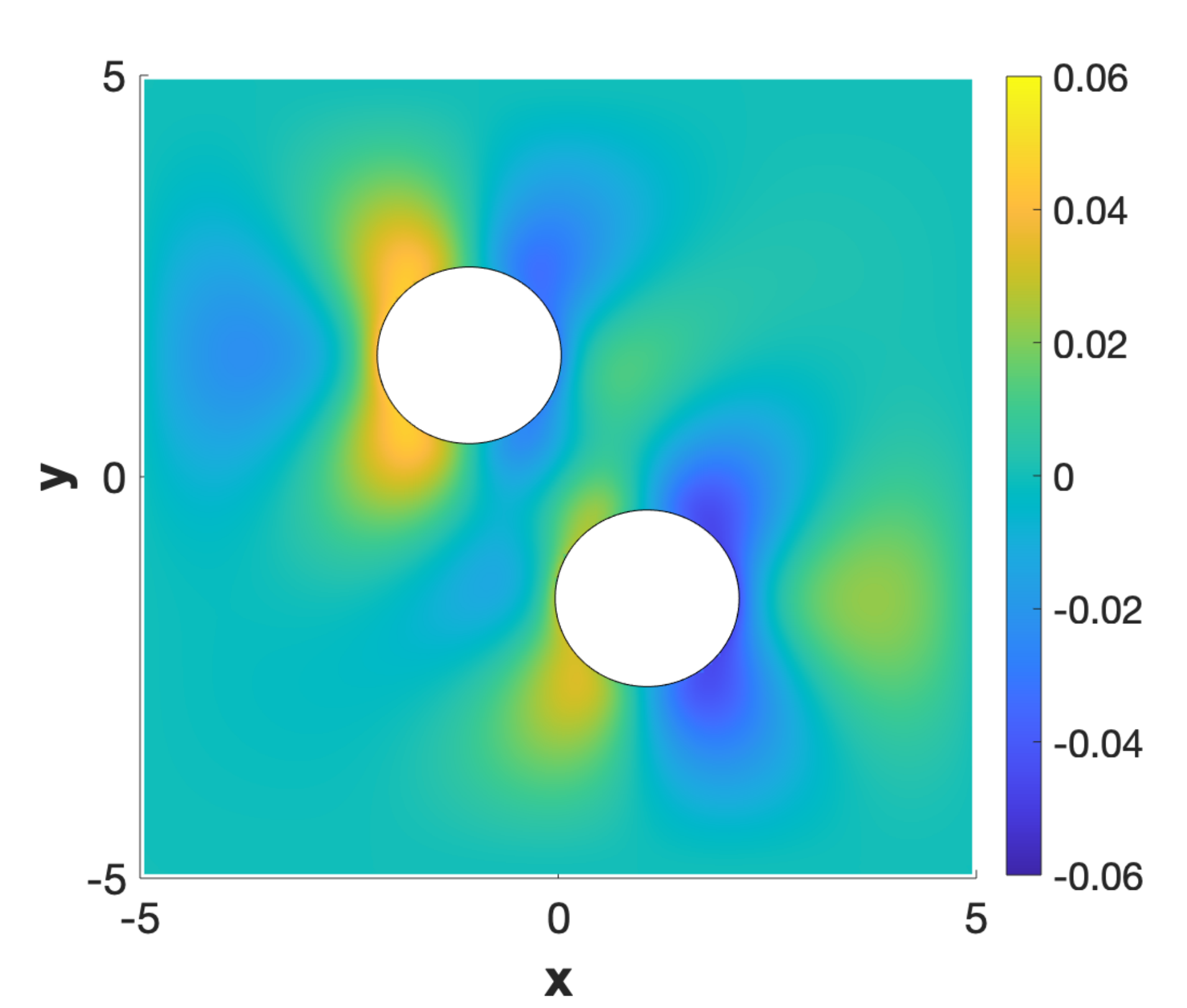}
}
\quad
\subfigure[Velocity in $y$ predicted by the ROM without correction]{
\includegraphics[width=4.8cm]{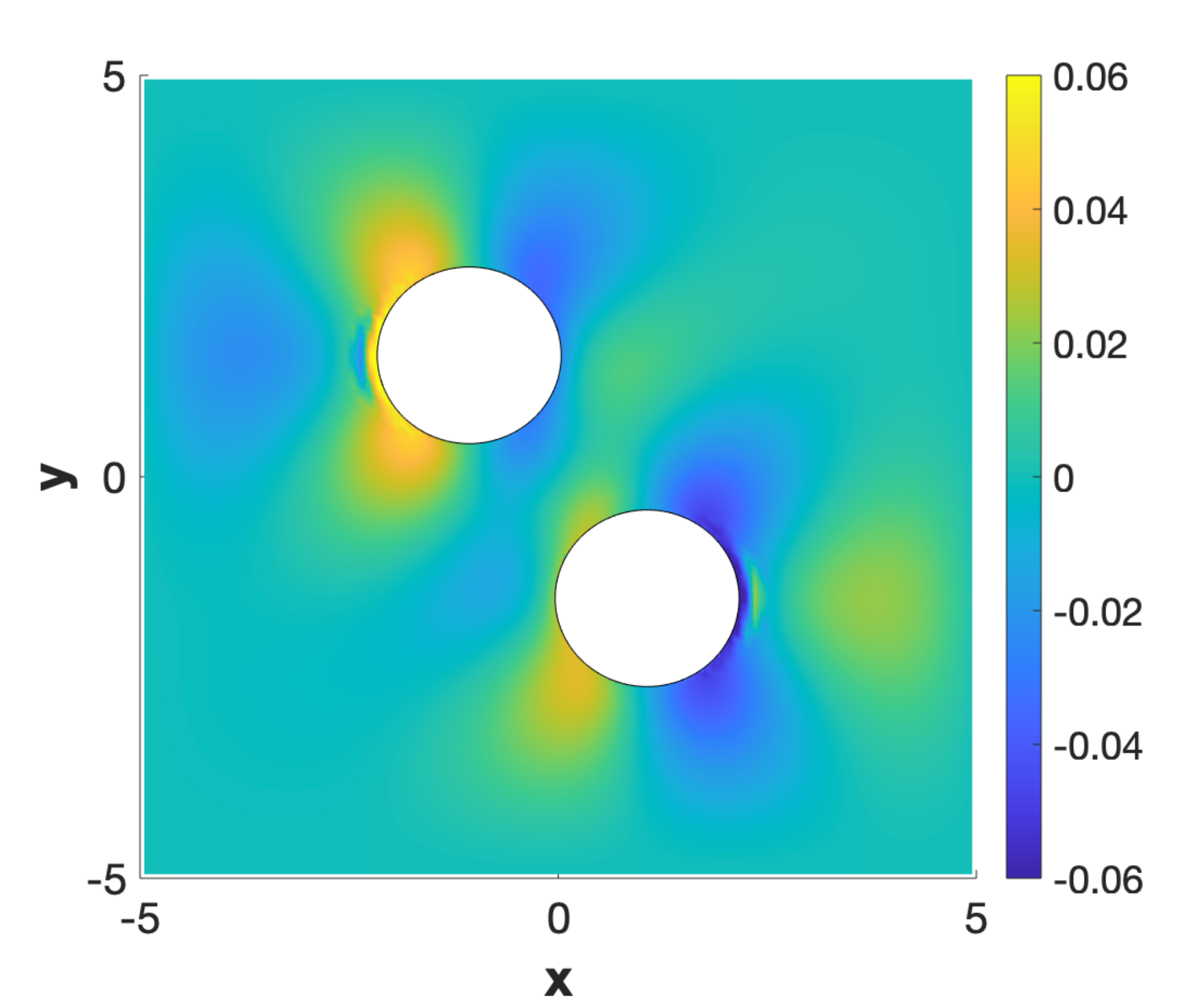}
}
\quad
\subfigure[Velocity in $y$  predicted by the ROM with correction]{
\includegraphics[width=4.8cm]{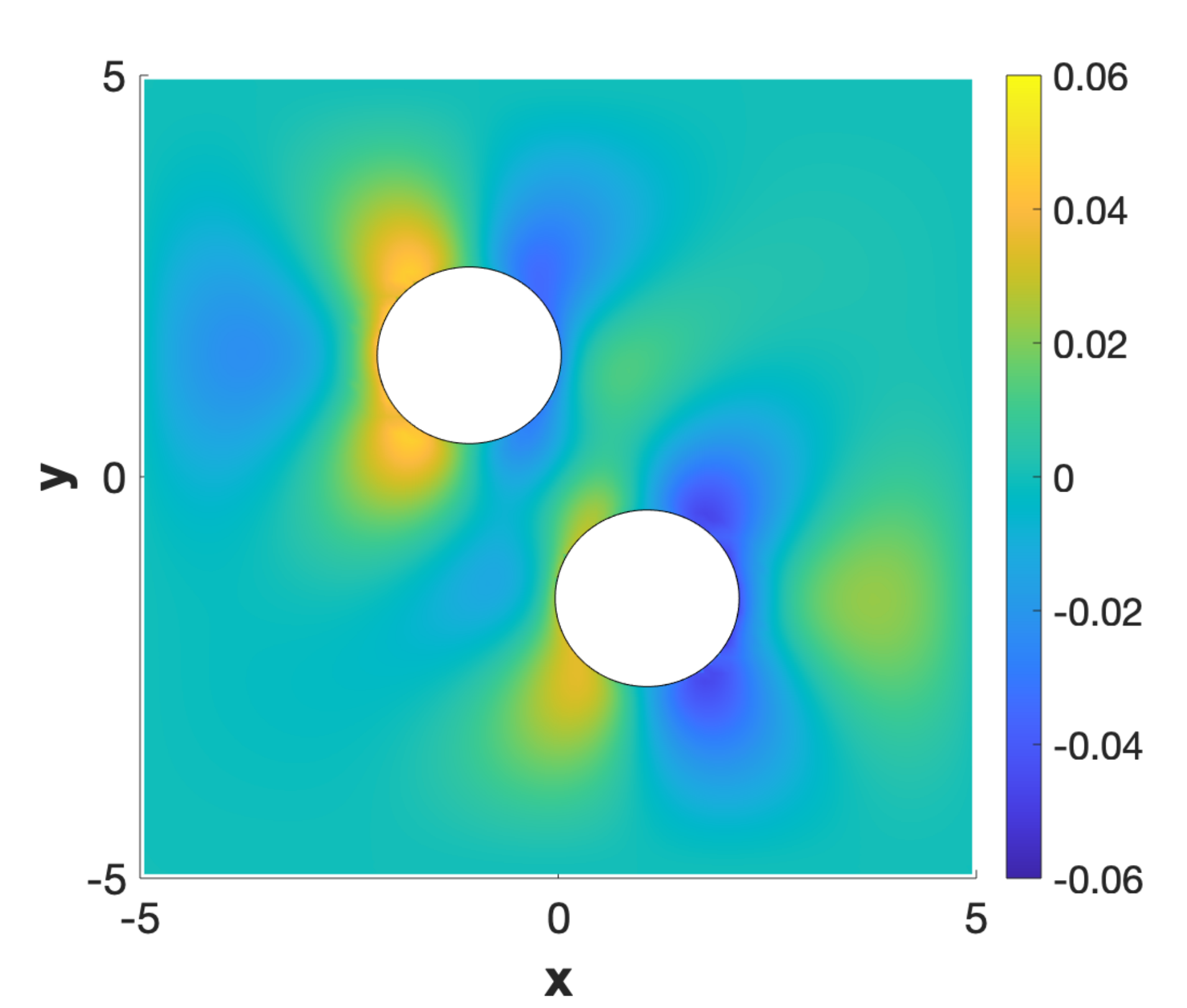}
}
\quad
\subfigure[Pressure field computed from numerical simulations]{
\includegraphics[width=4.8cm]{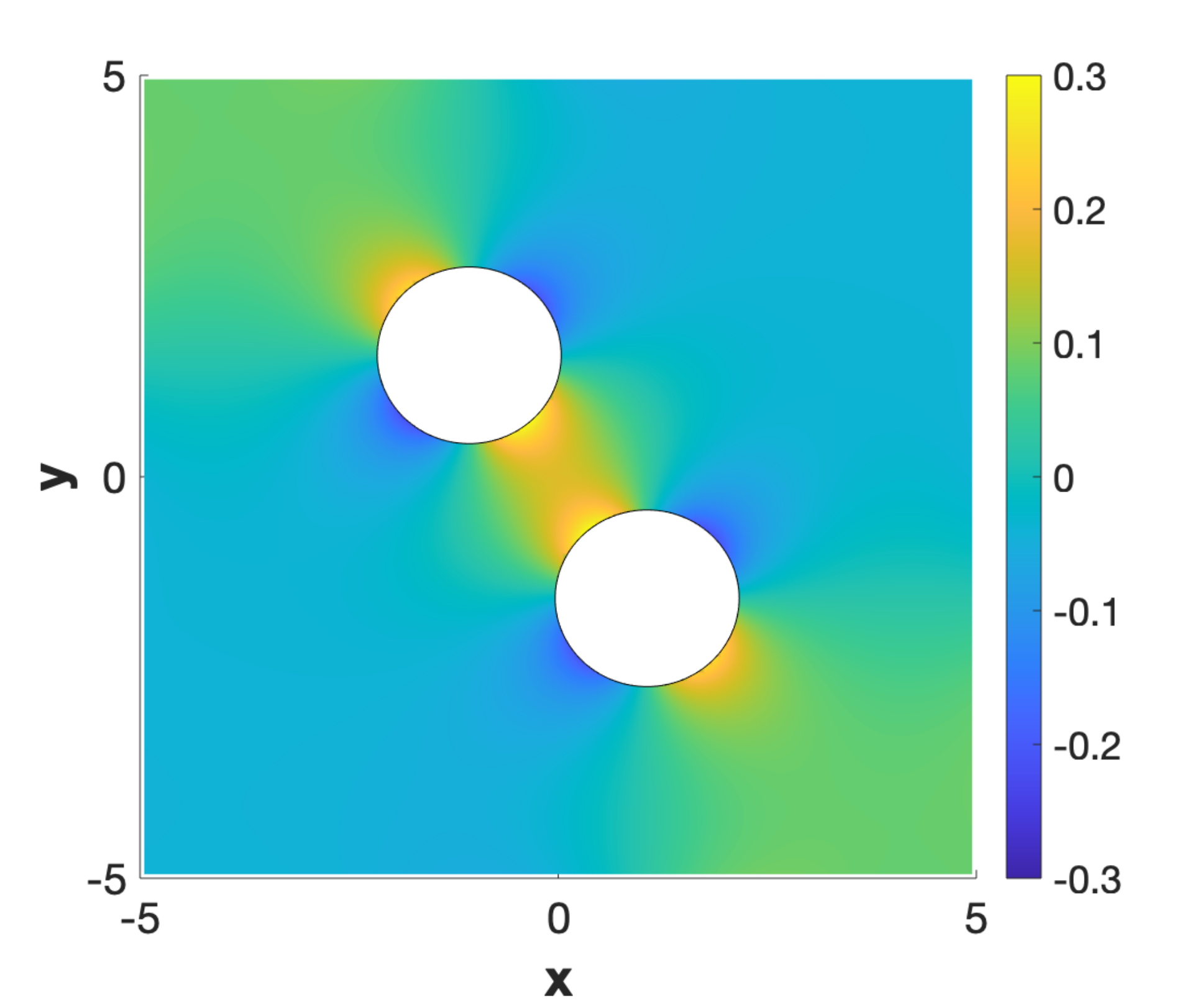}
}
\quad
\subfigure[Pressure field predicted by the ROM without correction]{
\includegraphics[width=4.8cm]{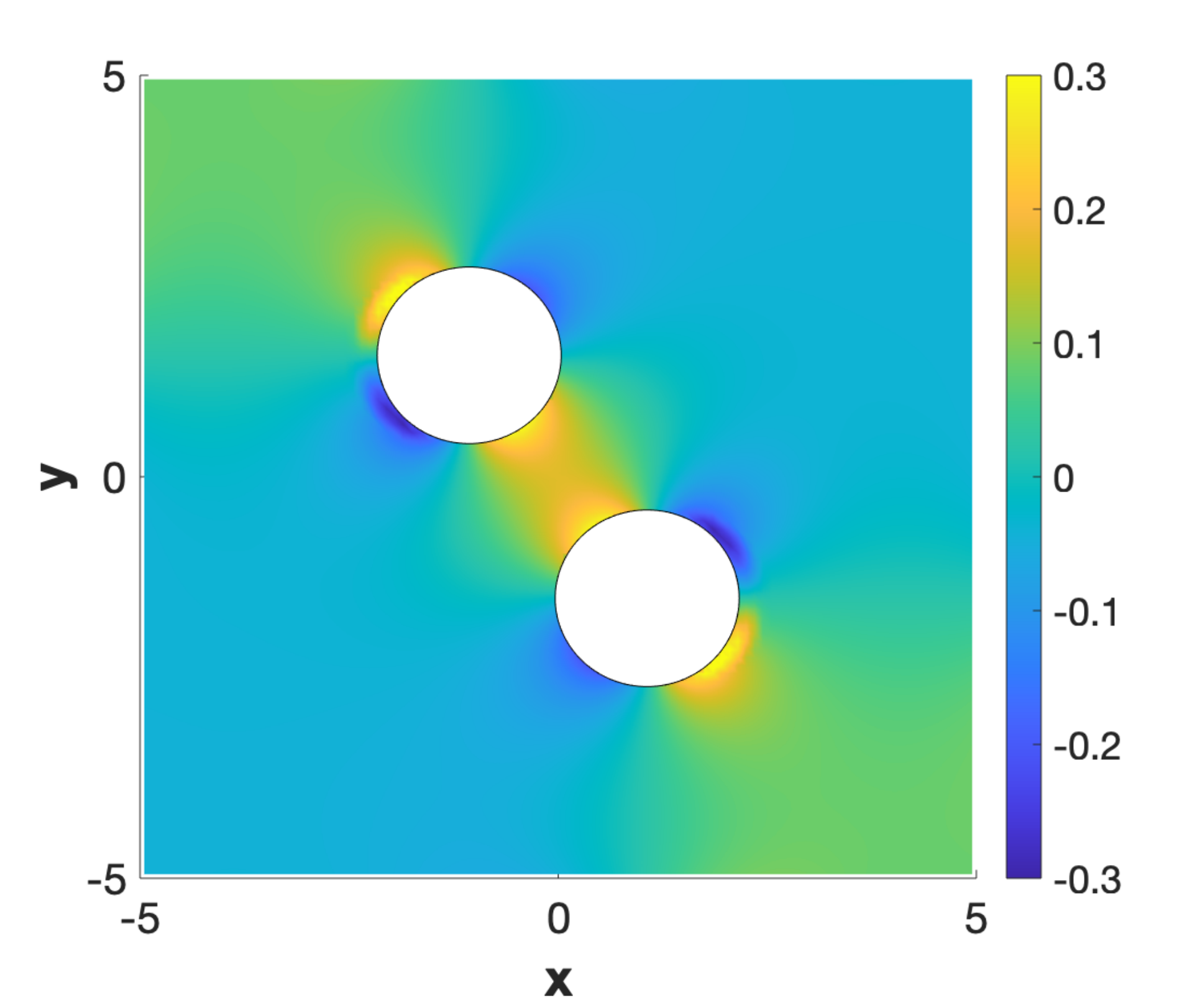}
}
\quad
\subfigure[Pressure field predicted by the ROM with correction]{
\includegraphics[width=4.8cm]{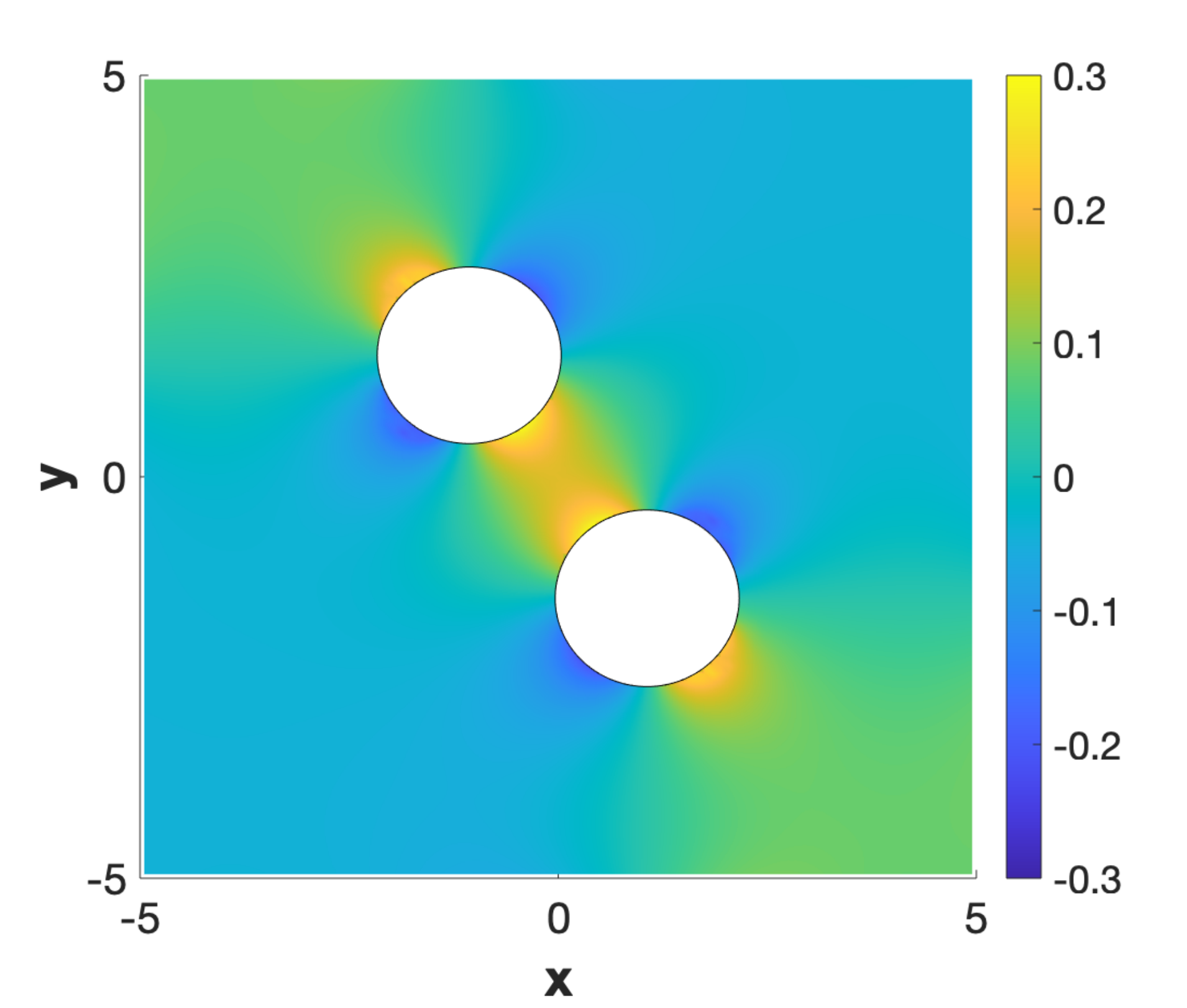}
}
\caption{Two cylinders under a shear flow: The velocity and pressure fields in the fluid predicted by the ROM at $t^*=3.0$, compared with the full-order solutions by numerical simulations.}
\label{fig:two_cylinders}
\end{figure}

\begin{figure}[htbp]
\centering
\subfigure[Velocity in $y$ at $x=-2$]{
\includegraphics[width=6.5cm]{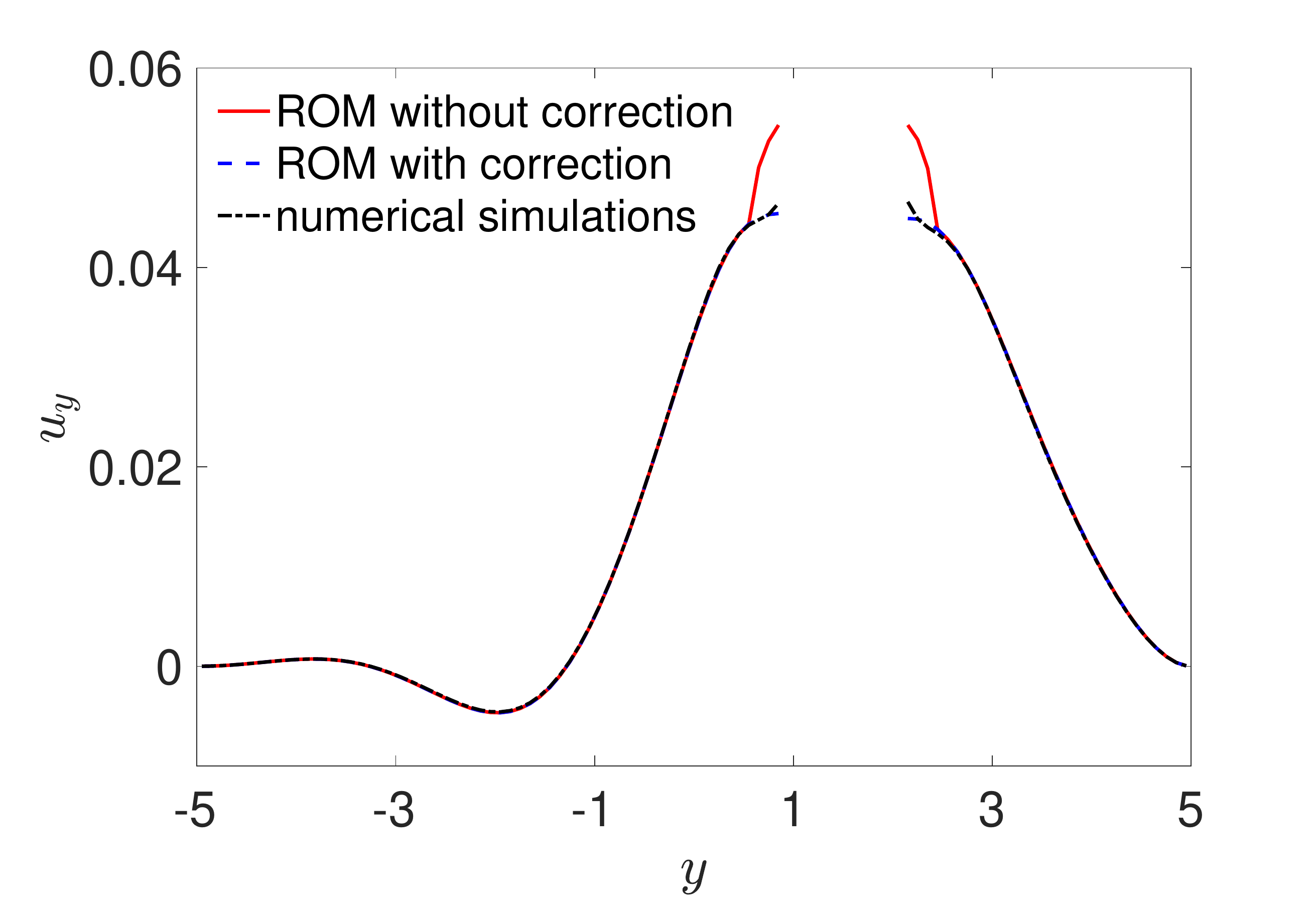}
}
\quad
\subfigure[Velocity in $y$ at $y=2$]{
\includegraphics[width=6.5cm]{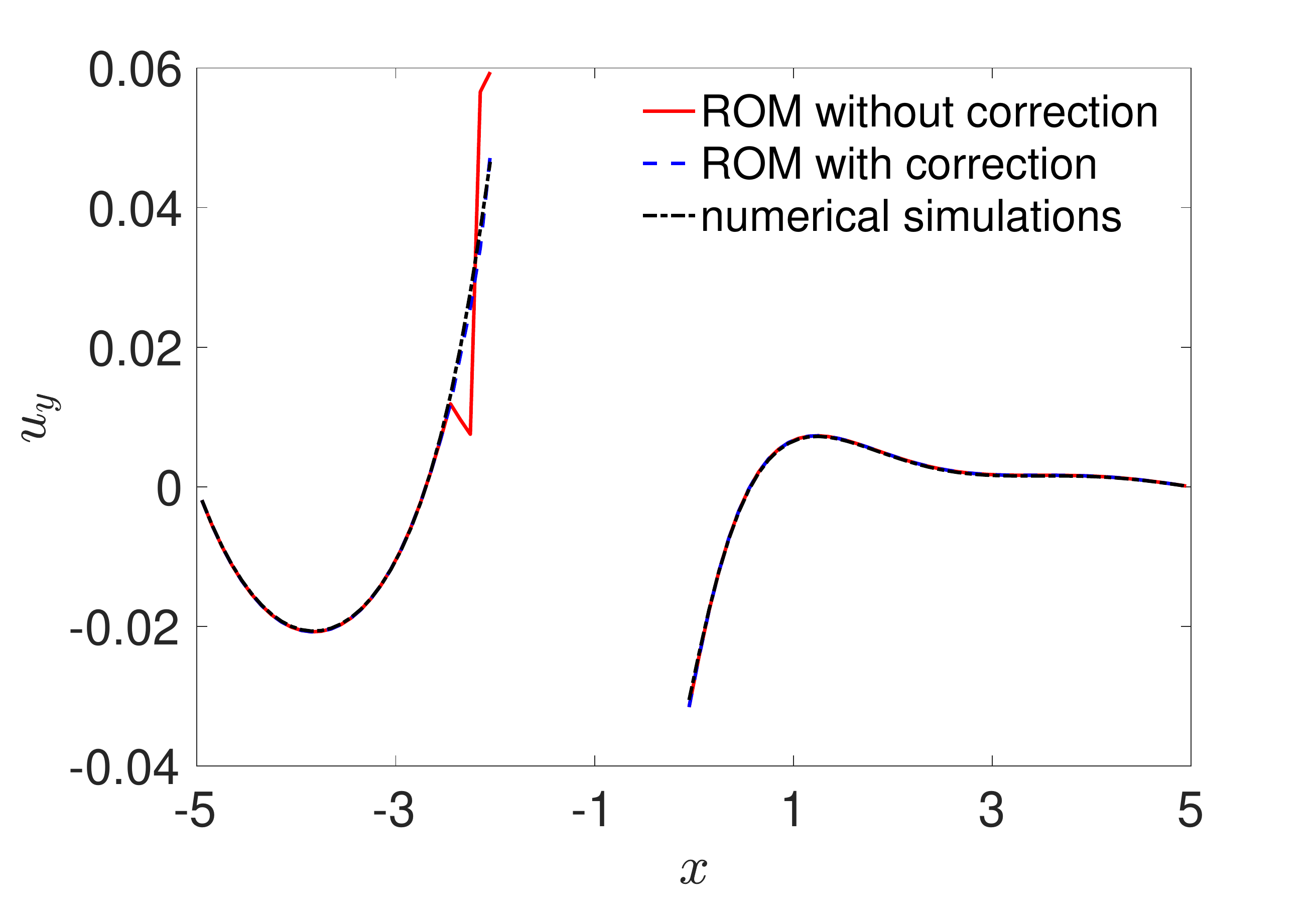}
}
\quad
\subfigure[Pressure at $x=-2$]{
\includegraphics[width=6.5cm]{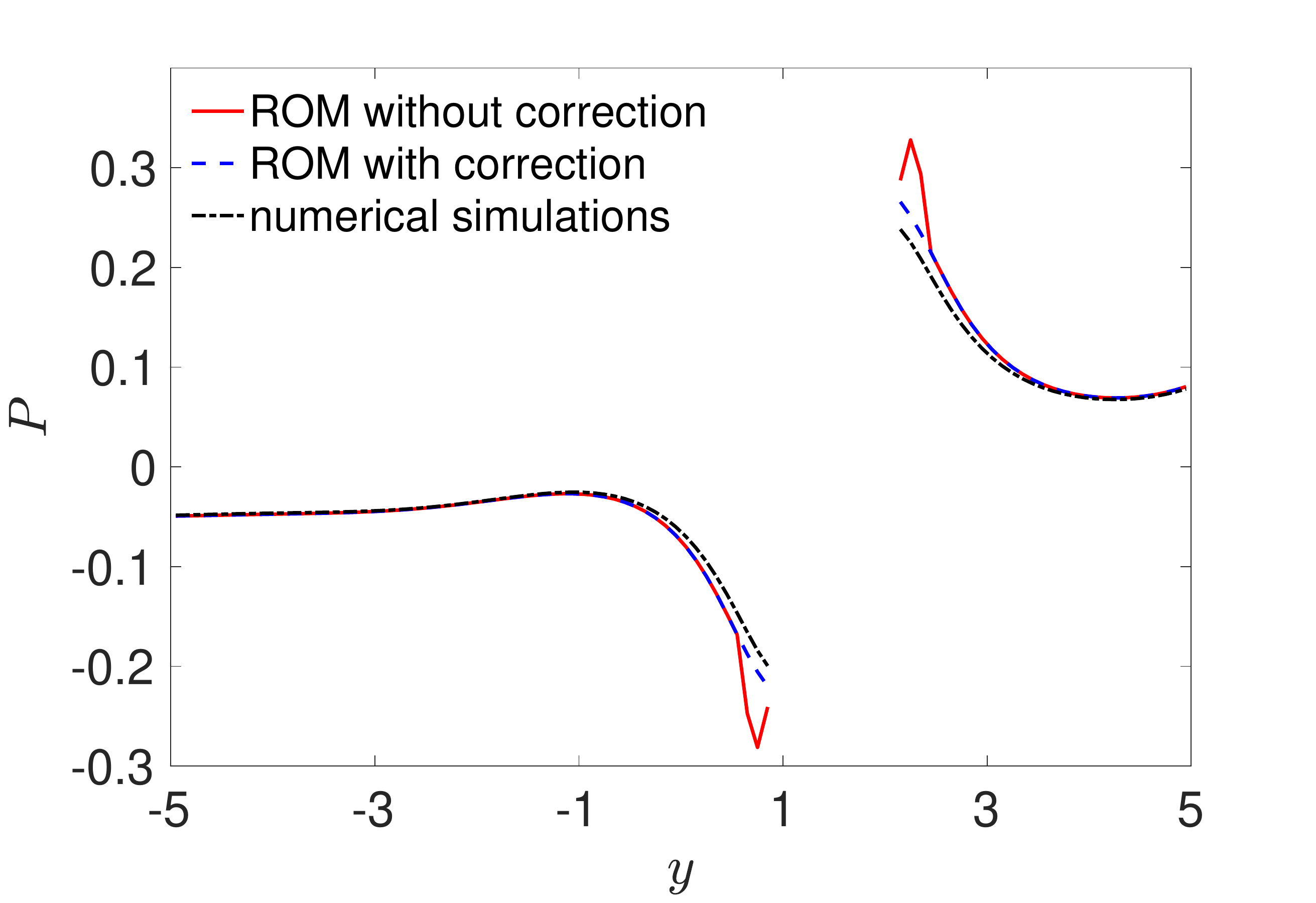}
}
\quad
\subfigure[Pressure at $y=2$]{
\includegraphics[width=6.5cm]{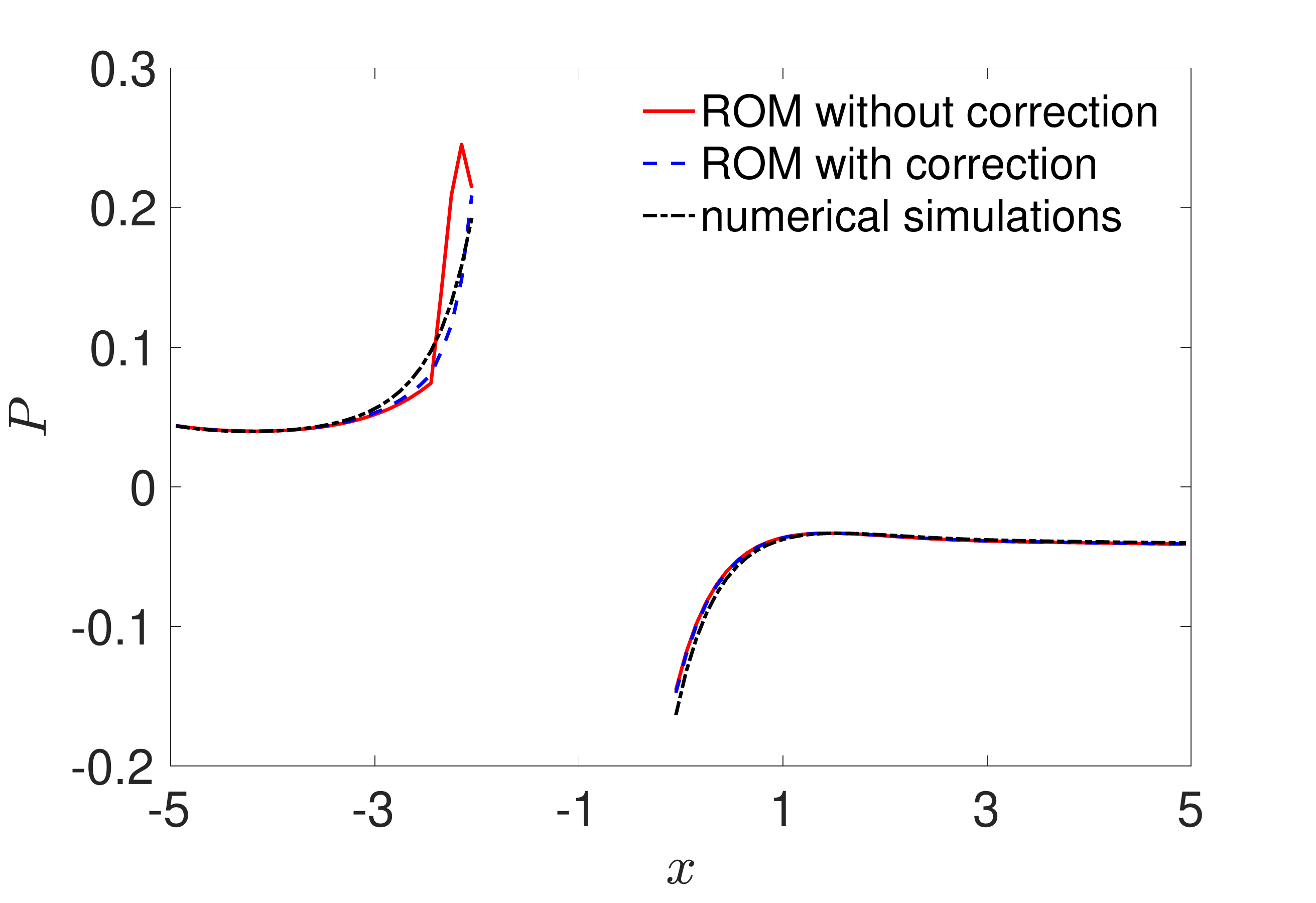}
}
\caption{Two cylinders under a shear flow: The velocity and pressure in the fluid along two lines $x=-2$ and $y=2$ at $t^*=3.0$ predicted by the ROM with and without correction, compared with the full-order solutions by numerical simulations.}
\label{fig:two_cylinders_lines}
\end{figure}

In all the cases studied herein, the fluid and solid bodies are confined in a 2D square box of $[-5,5] \times [-5,5]$. The top and bottom boundaries of the box are subject to identical velocities $u_0=0.5$ but along opposite $x$ directions so as to generate a shear flow. In numerical simulations, the time step was set $\delta t=0.1$, and the spatial discretization was uniform with the spacing $\delta x =0.1$. In the first case, there are two cylinders of equal size immersed in the fluid flow. We constructed the ROM from the first 20 snapshots from $t_1=0$ to $t_M=2.0$. The constructed ROM was then used to predict the velocity and pressure fields in the fluid and the positions of the two cylinders at $t^*=3.0$, which is the furthest forecast time determined from Eqs. \eqref{equ:Predict_POD_limit}, \eqref{equ:Gaussian_presetToL} and \eqref{equ:predict_boudary_limit} with $\beta^{\text{POD}} = 0.8$, $\beta^{\text{GPR},a} = 0.1$ and $\beta^{\text{GPR},\Gamma_m} = 0.1$. The first $4$ POD modes were retained in the ROM by setting $\alpha^{\text{POD}} = 0.05$. %Here a larger $\beta^{\text{POD}}$ was chosen because the correction method based on the MLS would reduce the error in some degree. 
Figure \ref{fig:two_cylinders} presents the velocity and pressure fields predicted by the ROM, compared with the full-order solutions by numerical simulations. Since the velocity in $x$ direction is dominated by linear shear flow, the comparison is made for the velocity along $y$ direction. We also compared the predictions with and without the correction near the moving solid boundaries, as discussed in \S\ref{subsubsec:correction_MB}. 
%%%%%%%%%%%%%%%%%%%%%%%
%%%%%%%%%%%%%%%%%%%%%%%
For a closer view, Figure \ref{fig:two_cylinders_lines} depicts the velocity and pressure along two lines at $x=-2 $ and $y=2$, respectively. It is clearly seen that the large errors appear in the regions near the moving solid boundaries on the downstream sides, which were previously occupied by the cylinders ($\Omega_m$) but not part of the fluid domain ($\Omega_f$). 
%The reason that causes the large errors is due to that in the snapshot data used to construct the ROM, the velocity for those regions was assumed the rigid-body motion of the cylinder, and the pressure was linearly interpolated from the surrounding fluid pressure, which are discontinuous in gradient with the solution in the fluid field. 
The correction method proposed in \S\ref{subsubsec:correction_MB} can effectively improve the accuracy of the ROM's predictions for those regions. 
%%%%%%%%%%%%%%%%%%%%%%%
%%%%%%%%%%%%%%%%%%%%%%%
  
\begin{figure}[htbp]
\centering
\includegraphics[width=12cm]{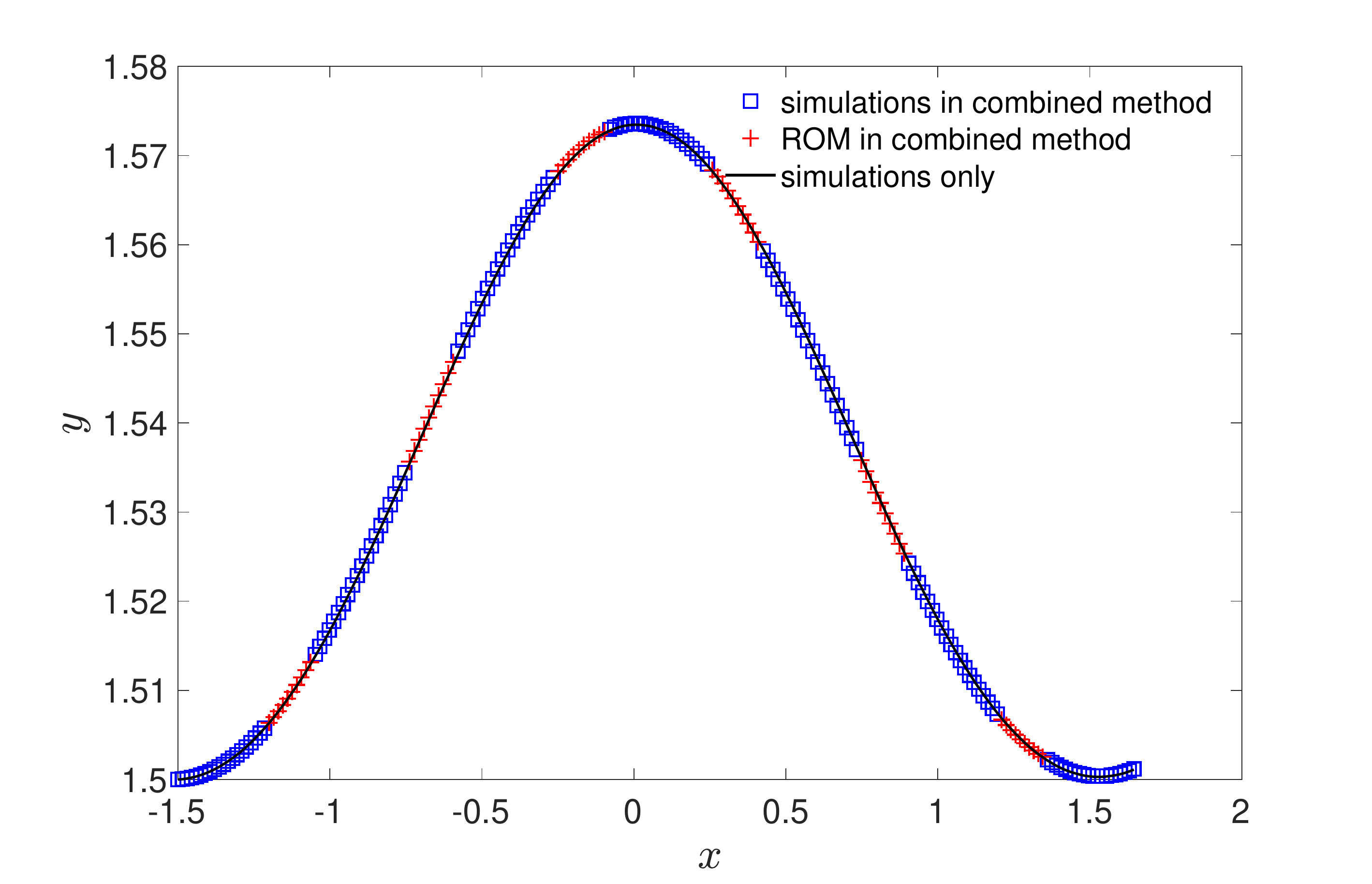}
\caption{Two cylinders under a shear flow: The trajectory of one cylinder predicted by the combination of the numerical simulations and ROM.}
\label{fig:two_cylinders_traj}
\end{figure}

For the problems of fluid-solid interactions, the numerical simulations can be rather demanding and expensive. Thus, we can adaptively combine the numerical simulations and ROM to save computational cost. In this case, we took the ROM's predictions (the velocity and pressure fields in the fluid and the positions of cylinders) at $t^*=3.0$ as the initial condition and restarted the numerical simulations until $t=5.0$. Using the numerical solutions for $t_1=3.1$ until $t_M=5.0$ as the snapshot data, we constructed a new ROM and predicted the solutions for $t^*=6.0$. By repeating this procedure until the target time $t=20.0$, we predicted the entire trajectory of one cylinder, as plotted in Figure \ref{fig:two_cylinders_traj}. By comparison with the result obtained solely from the numerical simulations, we find the prediction by adaptively combining the numerical simulations and ROM achieves good accuracy. In the meanwhile, the ROM's predictions replaced about $\frac 1 3$ of the simulation. Compared with the cost of the numerical simulations, the cost of constructing the ROM and employing it for prediction is trivial. 
%For example, it took $46.2s$ for the DNS to march from $t=2.0$ to $t=3.0$ with the time step $\delta t = 0.1$. 
%The DNS was run using C++ with OpenMPI-based parallel implementation on Intel(R) Xeon(R) ES-2698 V4 CPU @ 2.20GHz. 
For example, it took $21.2s$ for the numerical simulations to march from $t=2.0$ to $t=3.0$ with the time step $\delta t = 0.1$. 
The simulation was run using C++ with OpenMPI-based parallel implementation on one Intel(R) Xeon(R) E3-1231 v3 CPU @ 3.4GHz with 4 cores.
On the same hardware, it only needed 4.65s in total to construct the ROM and employ it for prediction by a serial {\sc Matlab} code.
%%%%%%%%%%%%%%%%%%%%%%%
%%%%%%%%%%%%%%%%%%%%%%%

\begin{figure}[htbp]
\centering
\subfigure[Velocity in $y$ computed from numerical simulations]{
\includegraphics[width=4.8cm]{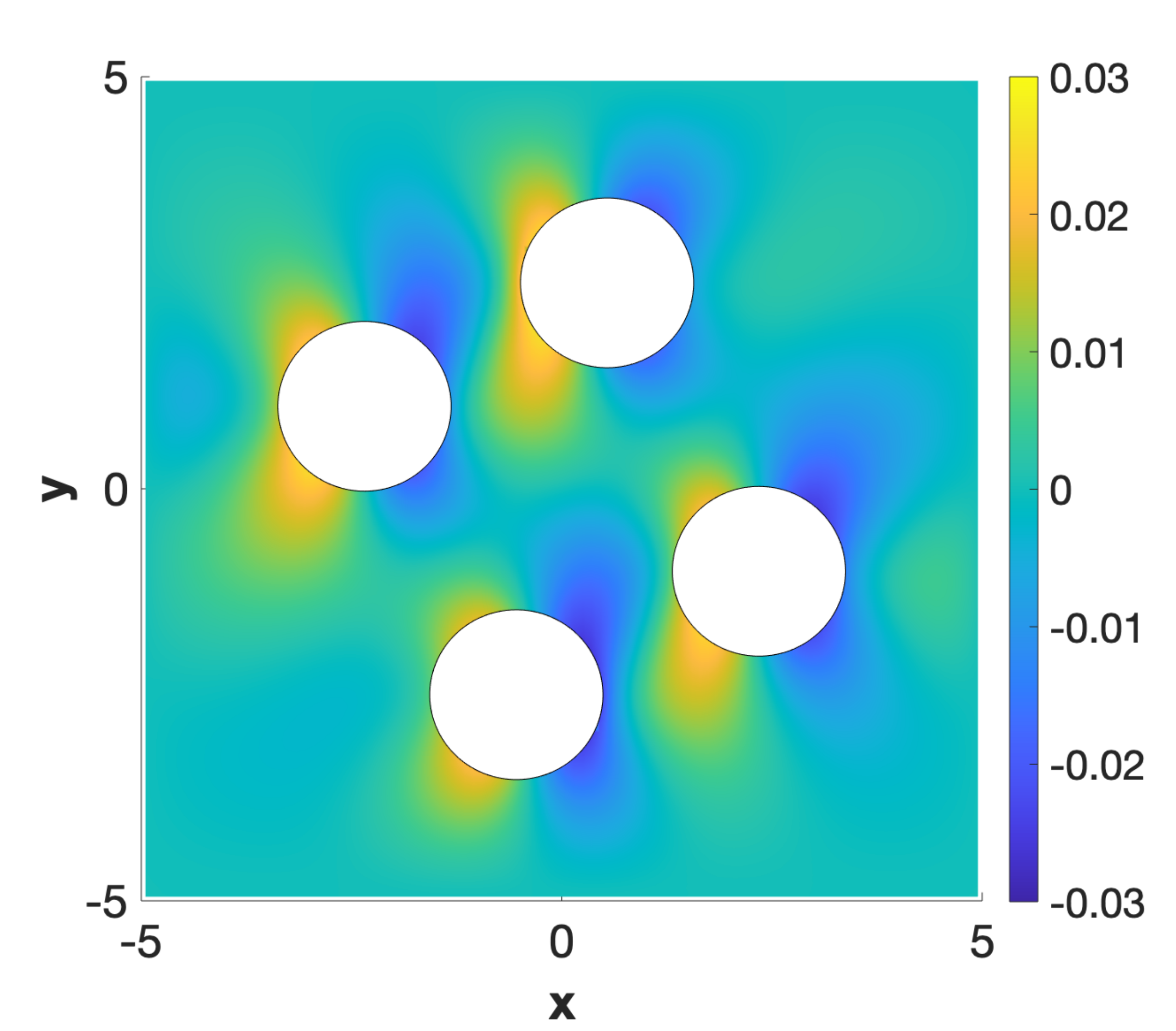}
}
\quad
\subfigure[Velocity in $y$ predicted by the ROM without correction]{
\includegraphics[width=4.8cm]{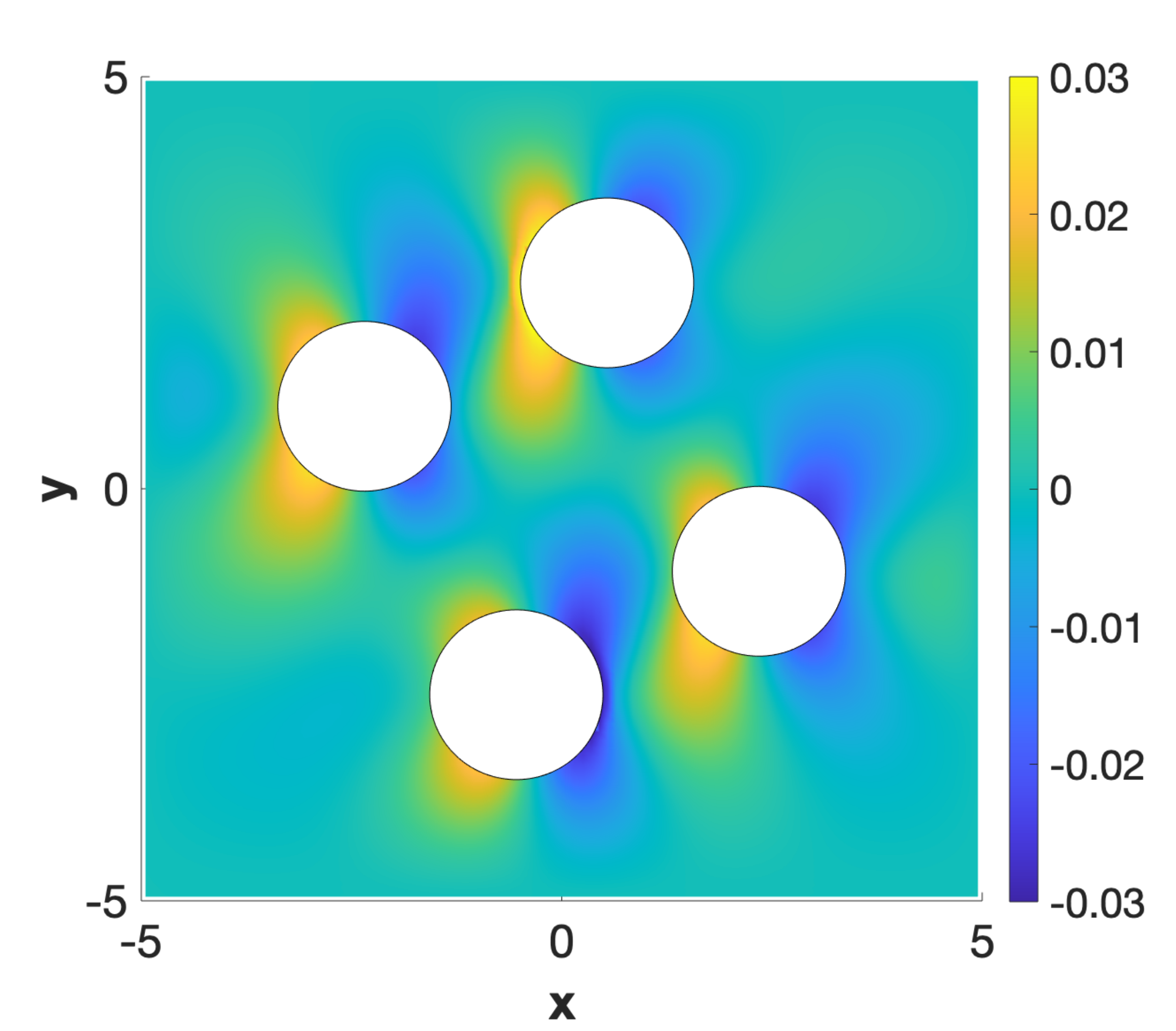}
}
\quad
\subfigure[Velocity in $y$ predicted by the ROM with correction]{
\includegraphics[width=4.8cm]{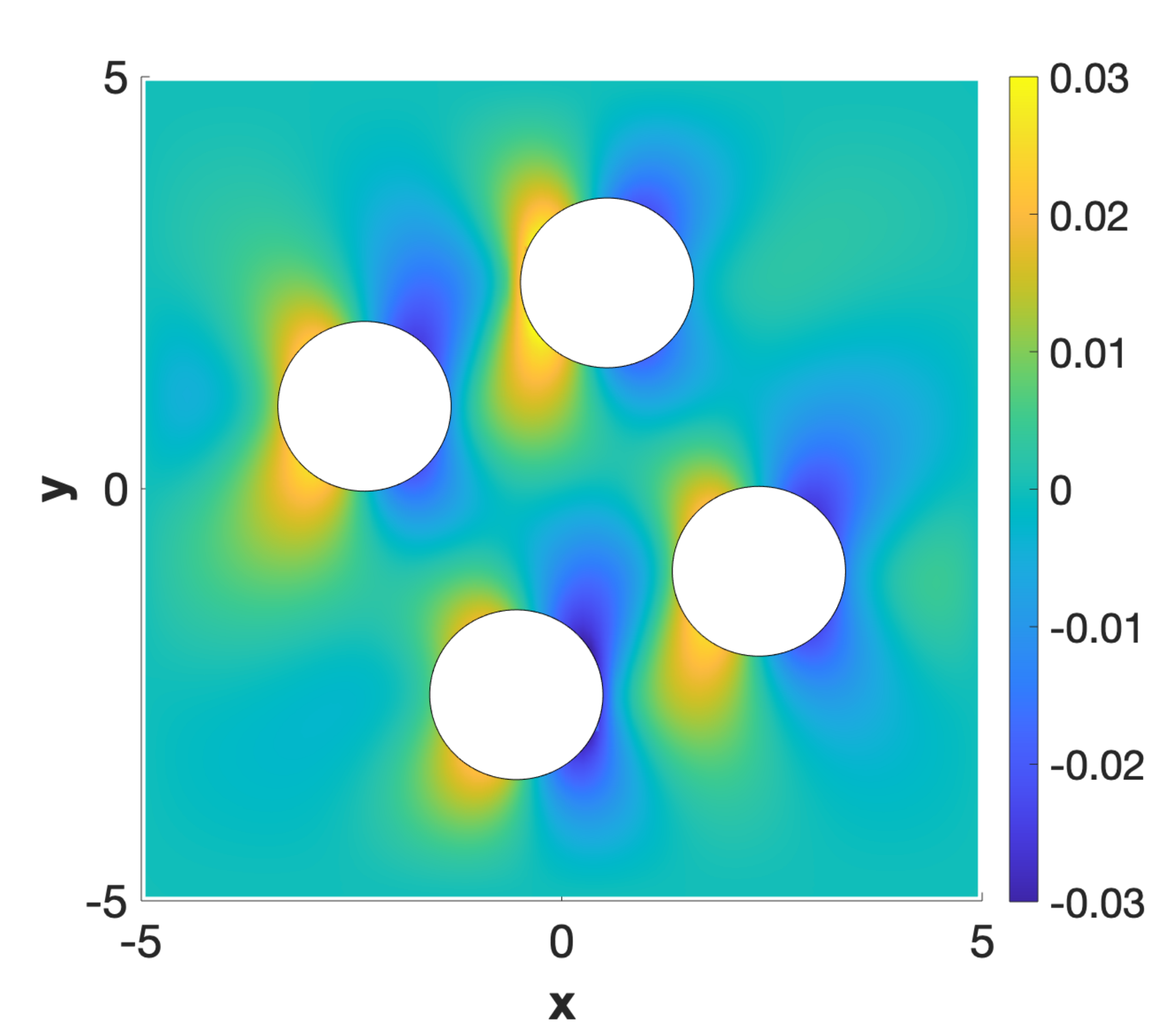}
}
\quad
\subfigure[Pressure field computed from numerical simulations]{
\includegraphics[width=4.8cm]{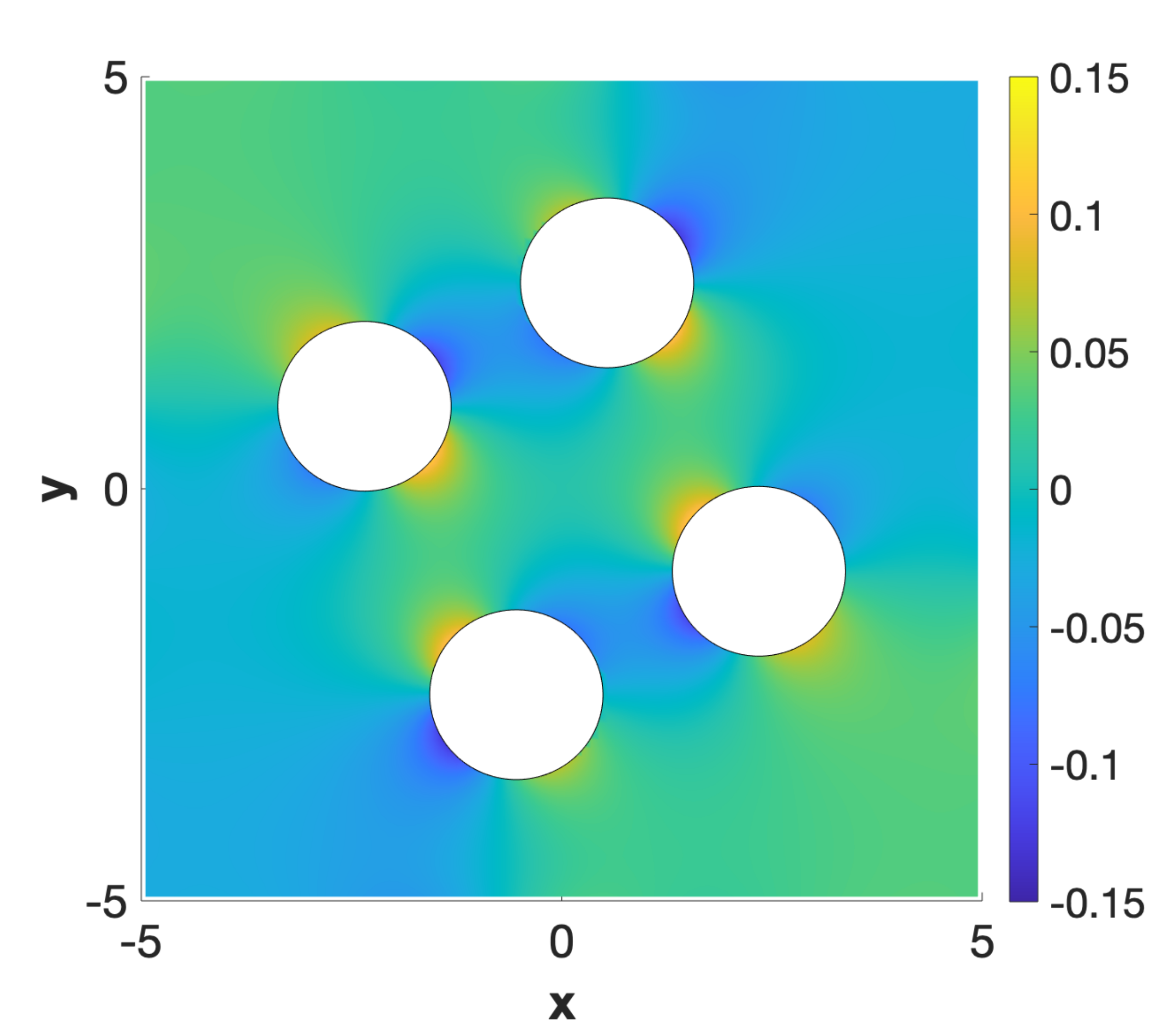}
}
\quad
\subfigure[Pressure field predicted by the ROM without correction]{
\includegraphics[width=4.8cm]{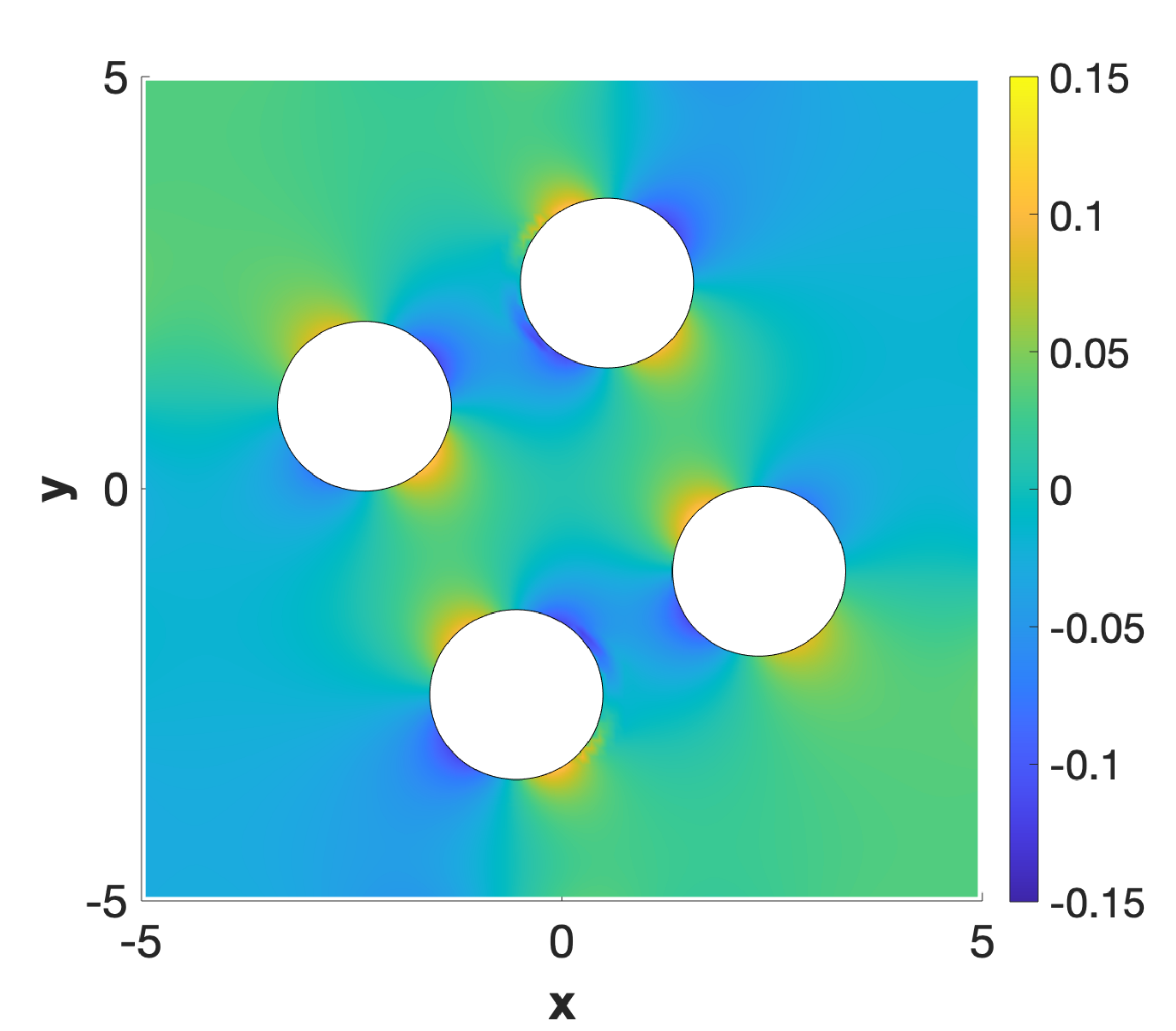}
}
\quad
\subfigure[Pressure field predicted by the ROM with correction]{
\includegraphics[width=4.8cm]{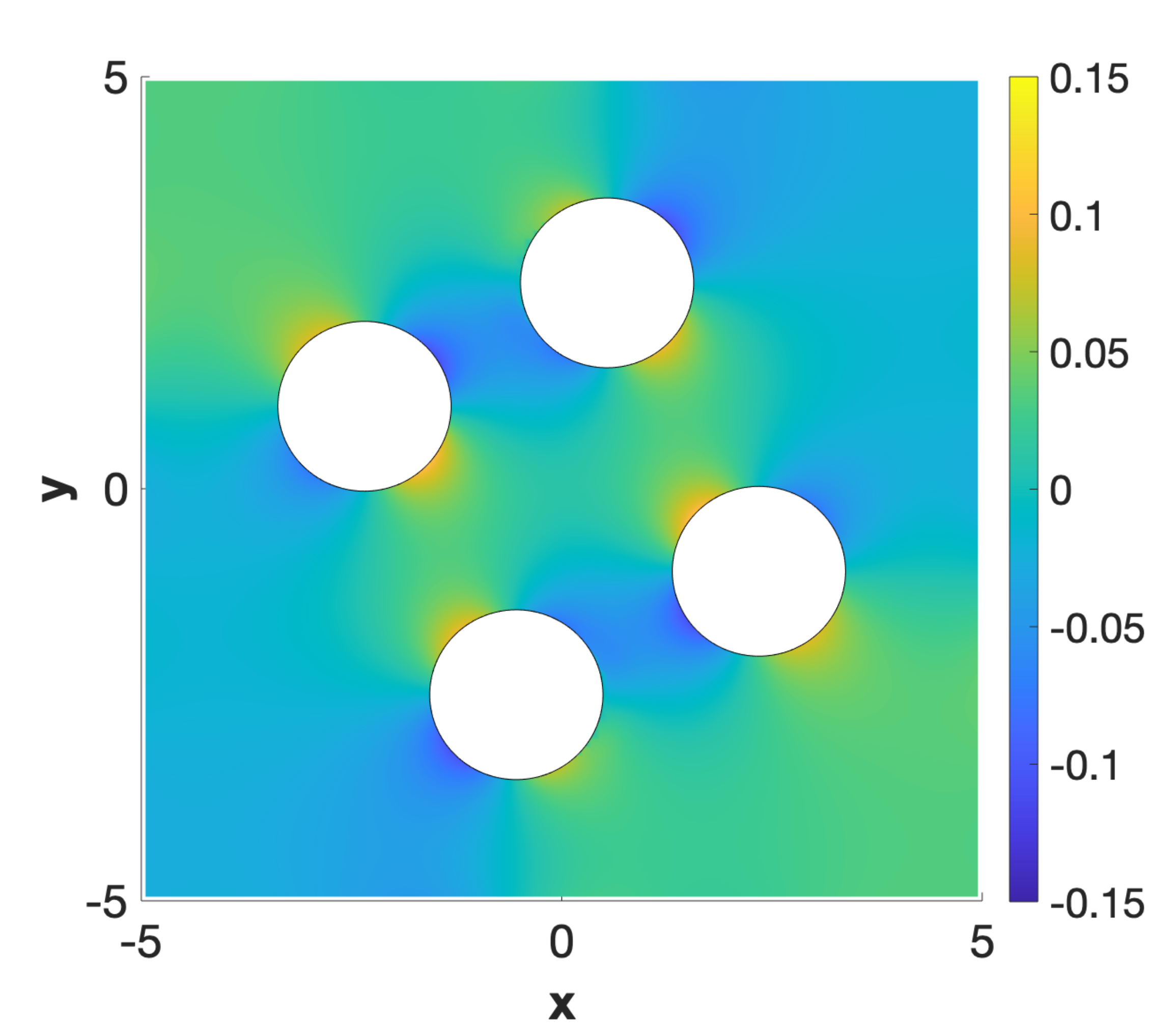}
}
\caption{Four cylinders under a shear flow: The velocity and pressure fields in the fluid predicted by the ROM at $t^*=3.0$, compared with the full-order solutions by numerical simulations.}
\label{fig:four_cylinders}
\end{figure}

\begin{figure}[htbp]
\centering
\subfigure[Velocity in $y$ at $x=0.5$]{
\includegraphics[width=6.5cm]{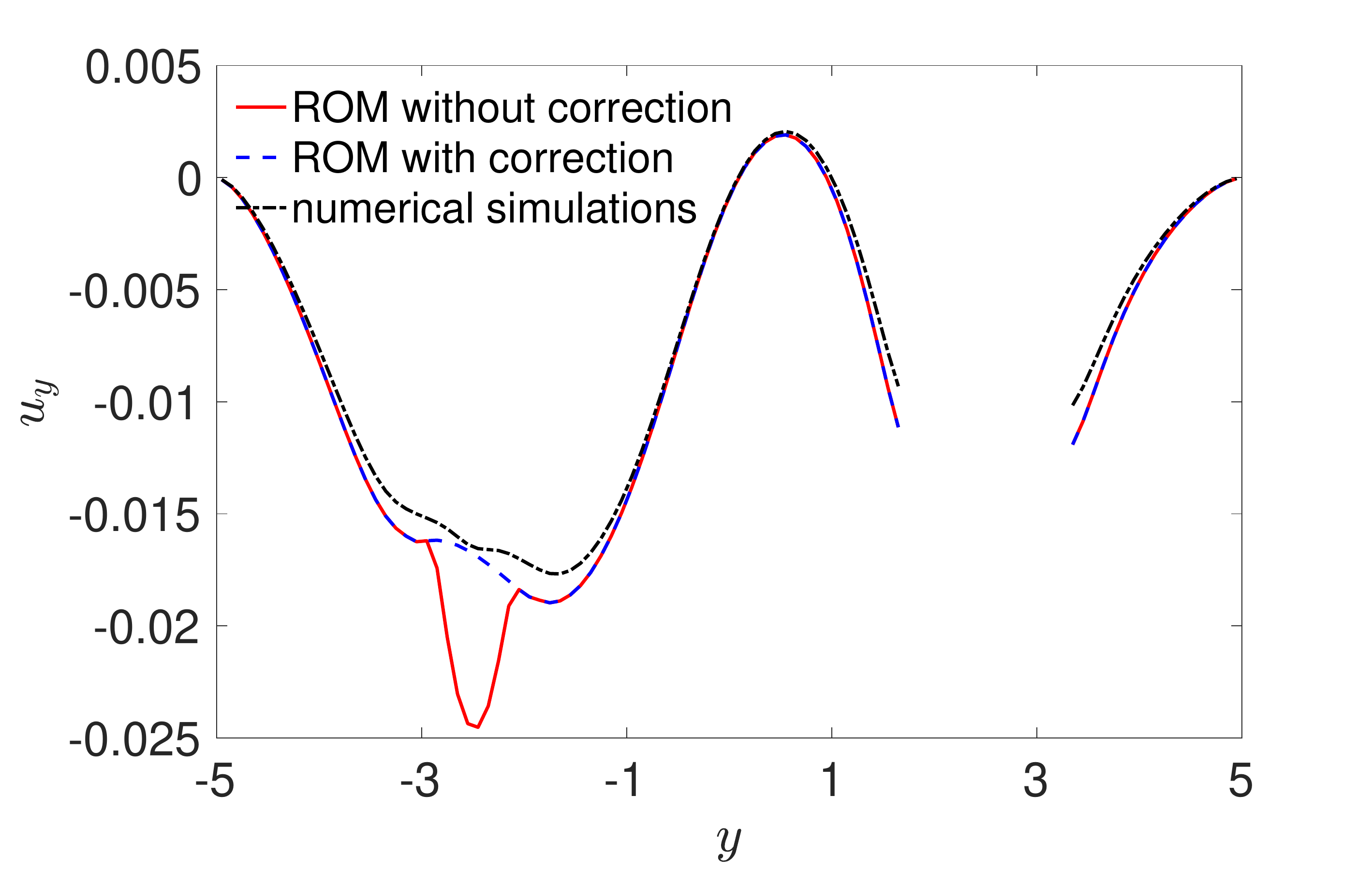}
}
\quad
\subfigure[Velocity in $y$ at $y=-2.5$]{
\includegraphics[width=6.5cm]{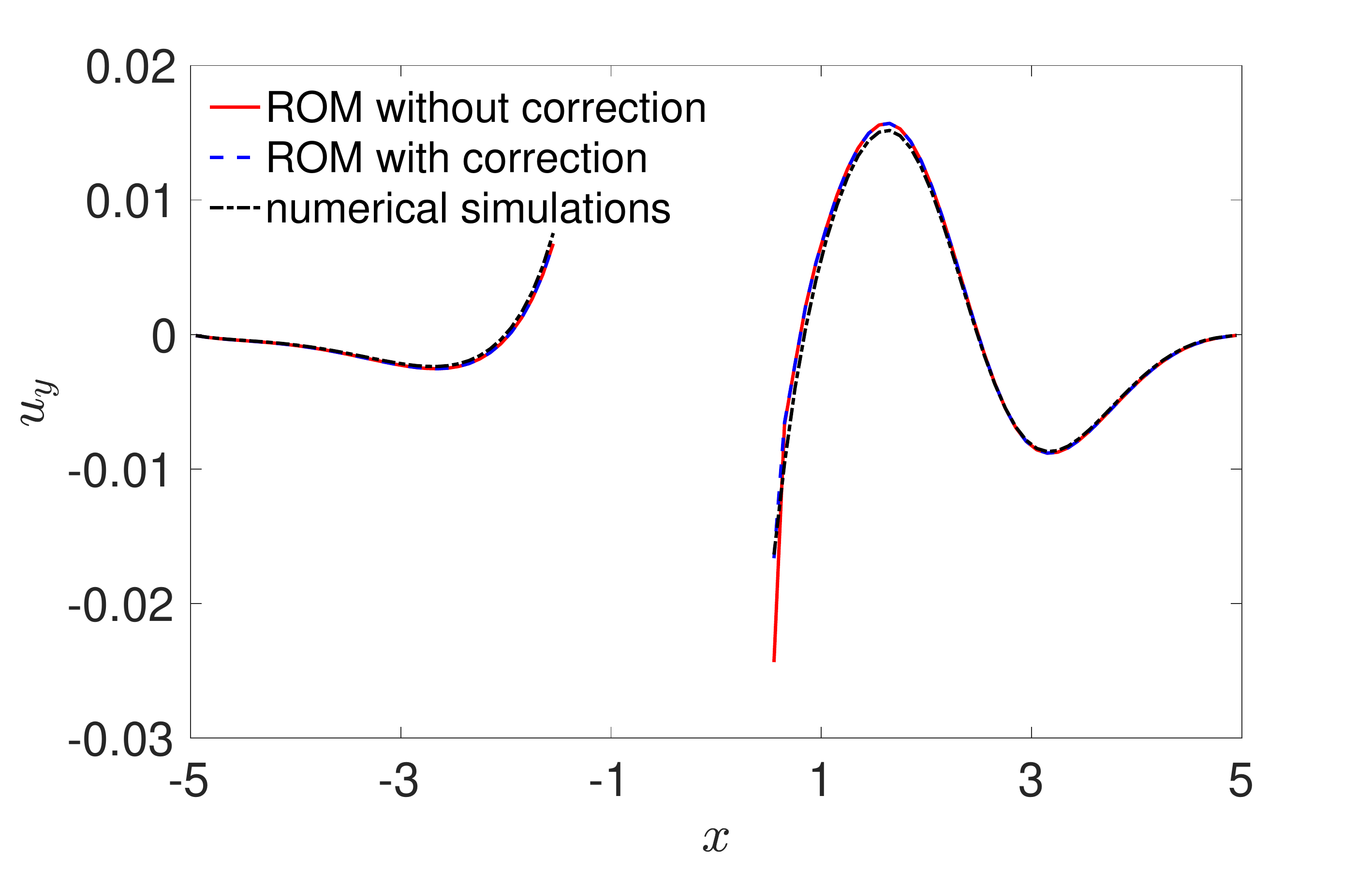}
}
\quad
\subfigure[Pressure at $x=0.5$]{
\includegraphics[width=6.5cm]{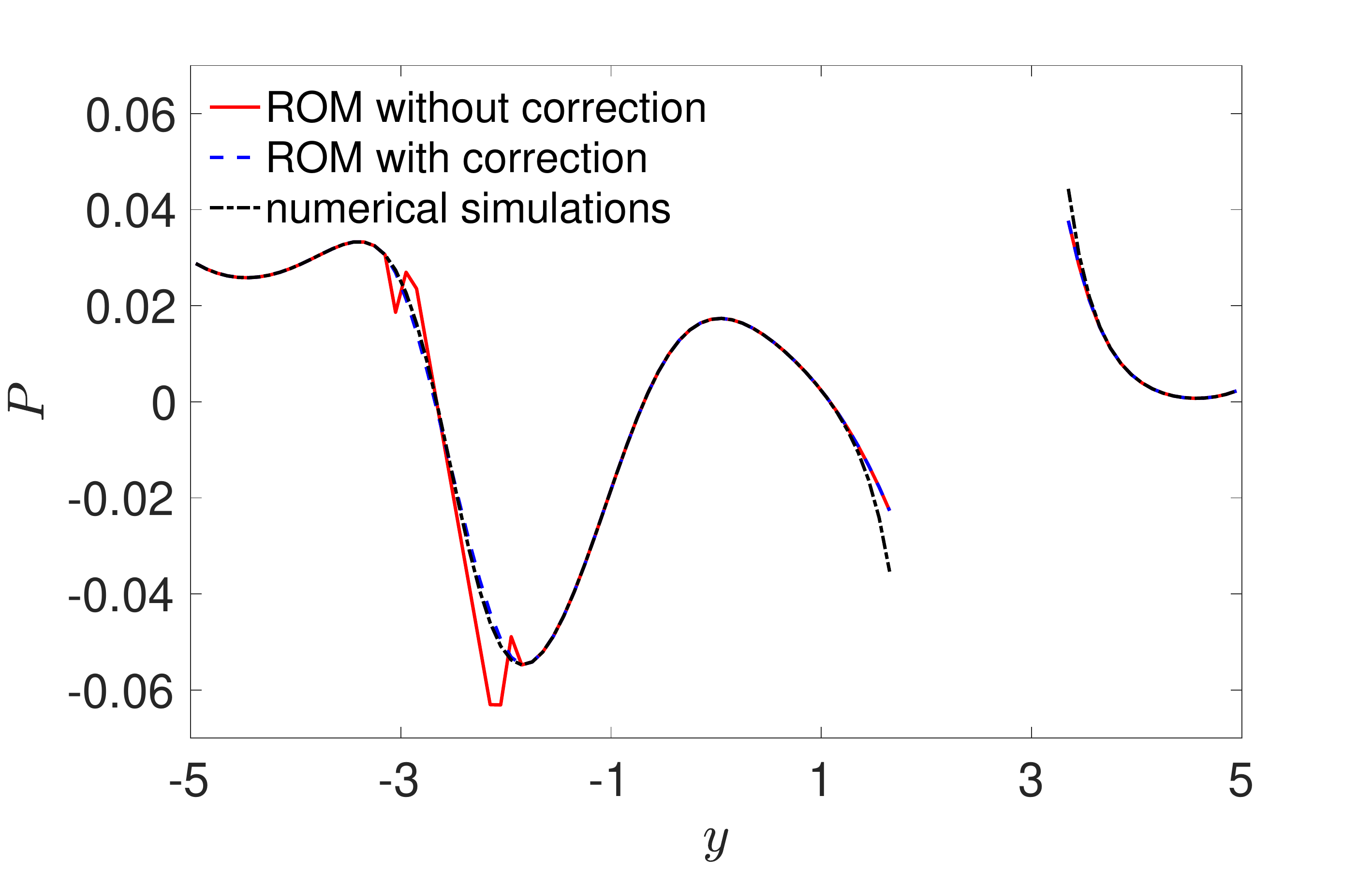}
}
\quad
\subfigure[Pressure at $y=-2.5$]{
\includegraphics[width=6.5cm]{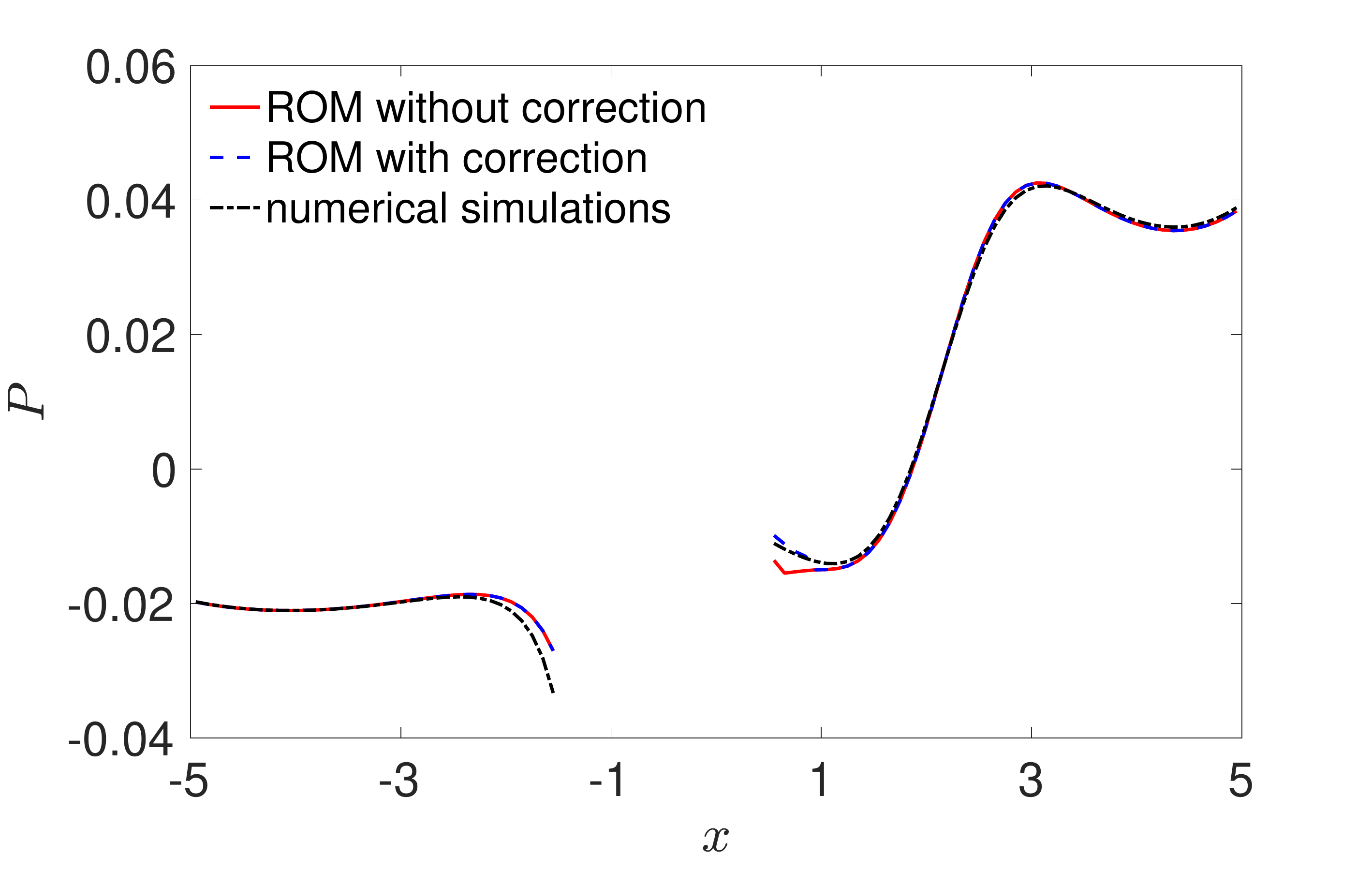}
}
\caption{Four cylinders under a shear flow: The velocity and pressure along the lines $x=0.5$ and $y=-2.5$ at $t^*=3.0$ predicted by the ROM with and without correction, compared with the full-order solutions by numerical simulations.}
\label{fig:four_cylinders_lines}
\end{figure}

\begin{figure}[htbp]
\centering
\subfigure[Velocity in $y$ computed from numerical simulations]{
\includegraphics[width=4.8cm]{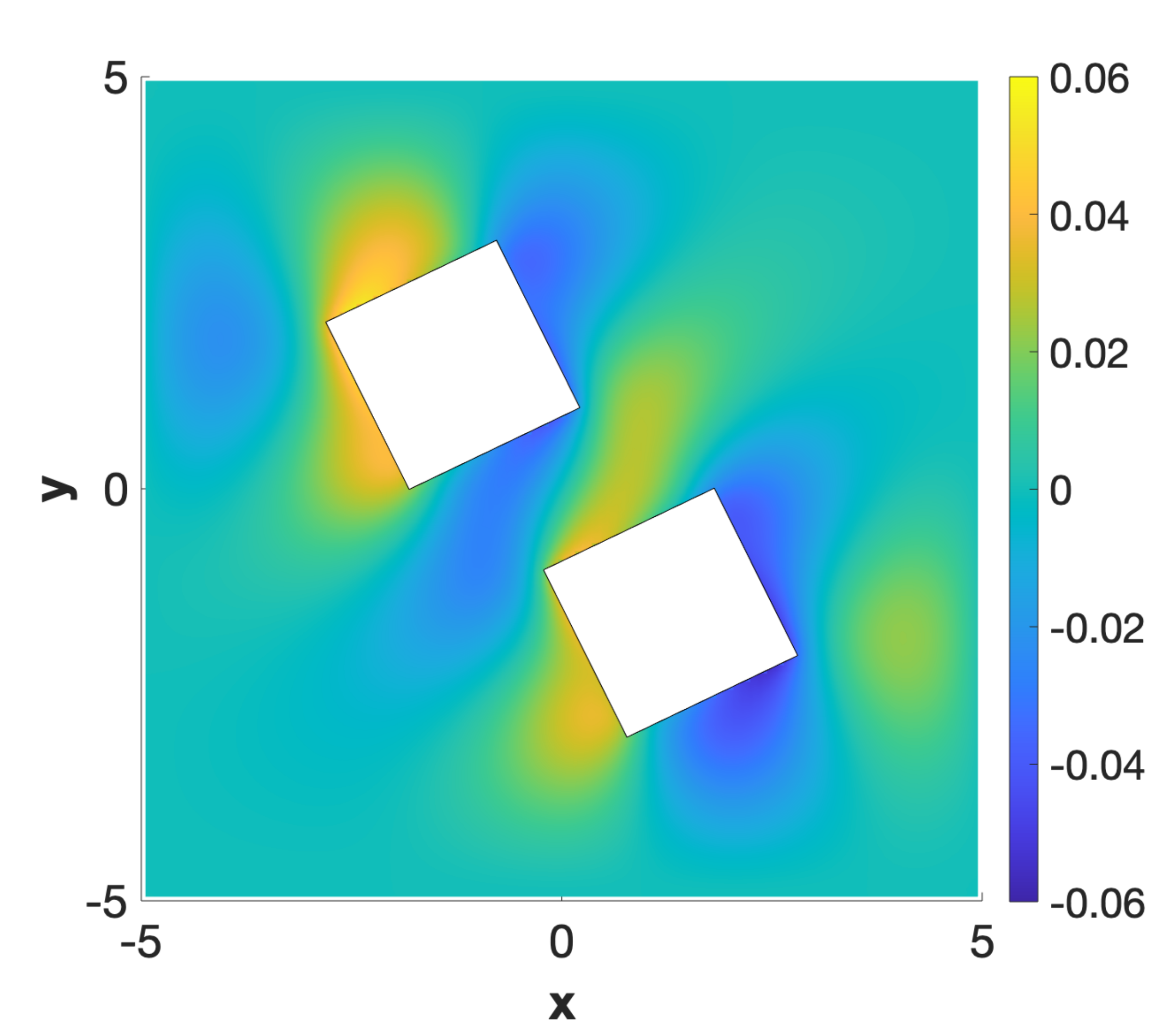}
}
\quad
\subfigure[Velocity in $y$ predicted by the ROM without correction]{
\includegraphics[width=4.8cm]{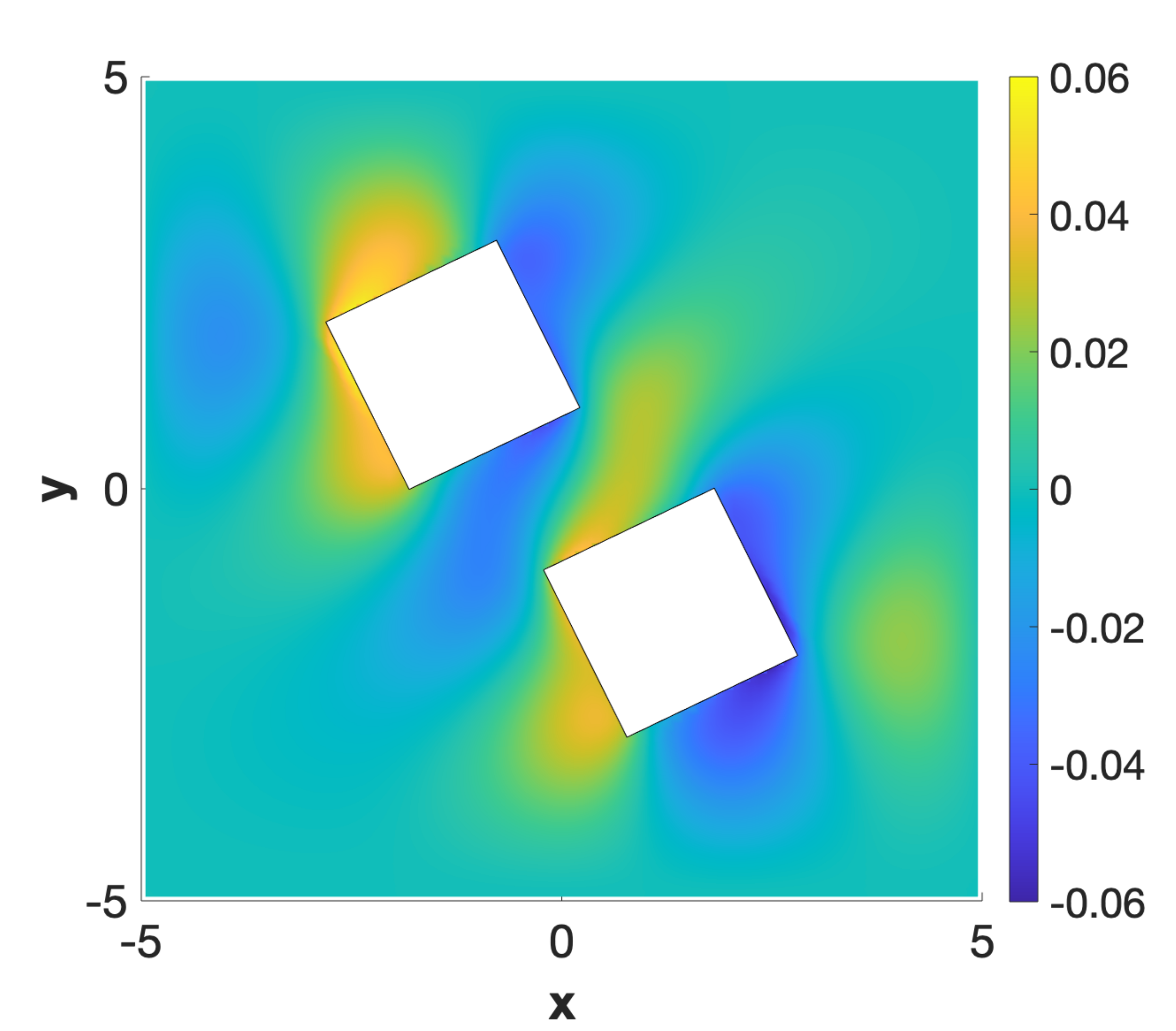}
}
\quad
\subfigure[Velocity in $y$ predicted by the ROM with correction]{
\includegraphics[width=4.8cm]{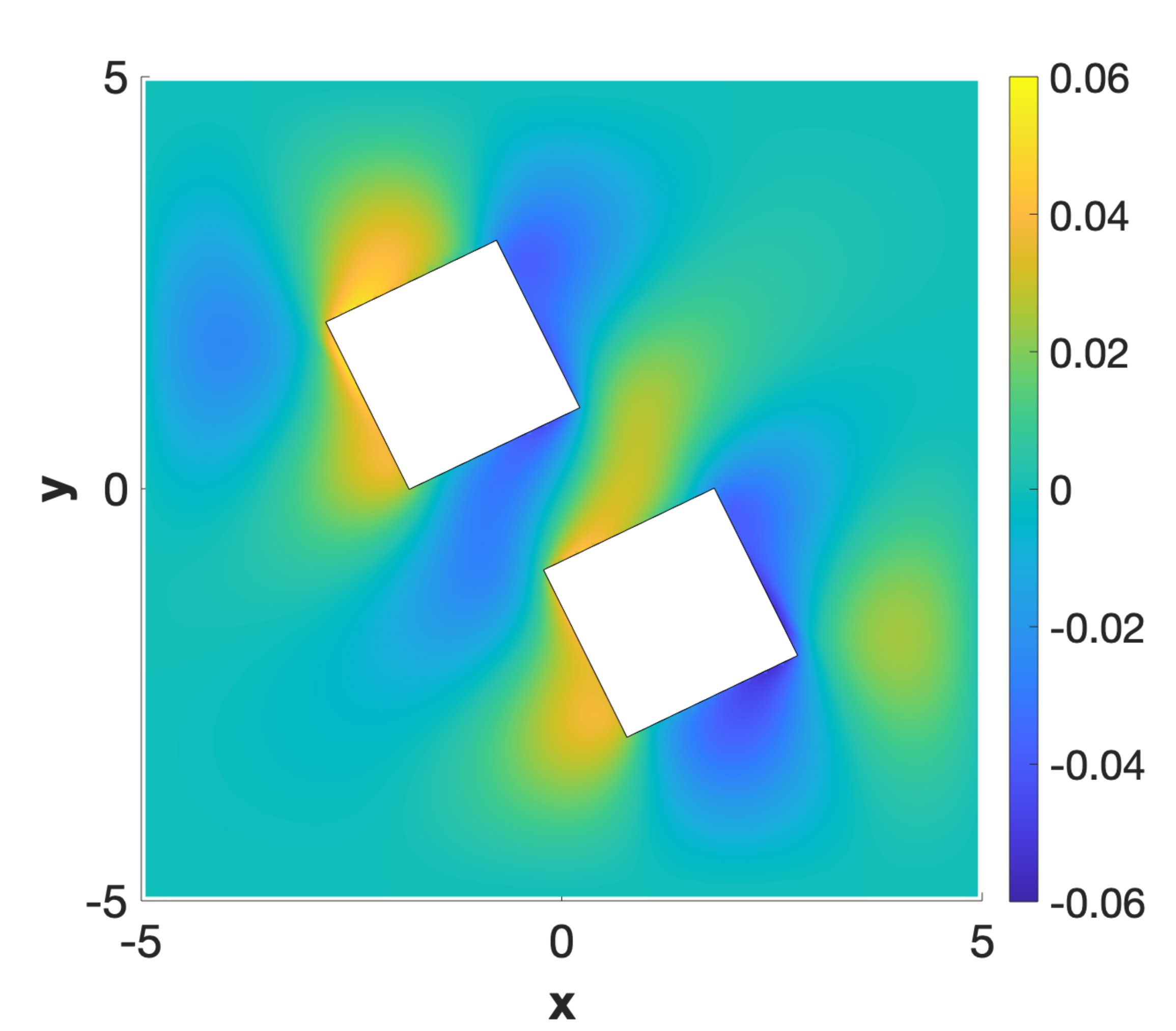}
}
\quad
\subfigure[Pressure field computed from numerical simulations]{
\includegraphics[width=4.8cm]{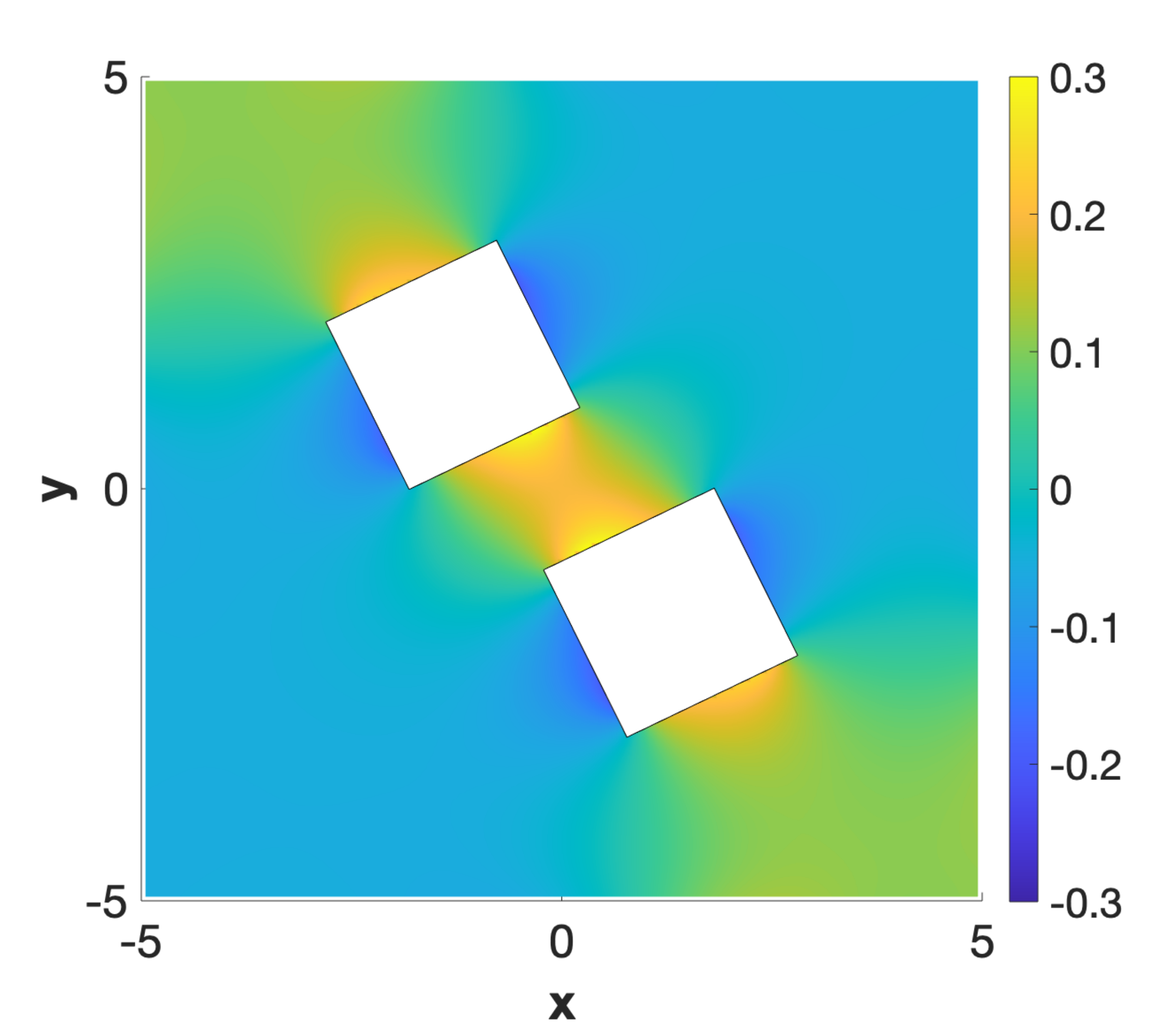}
}
\quad
\subfigure[Pressure field predicted by the ROM without correction]{
\includegraphics[width=4.8cm]{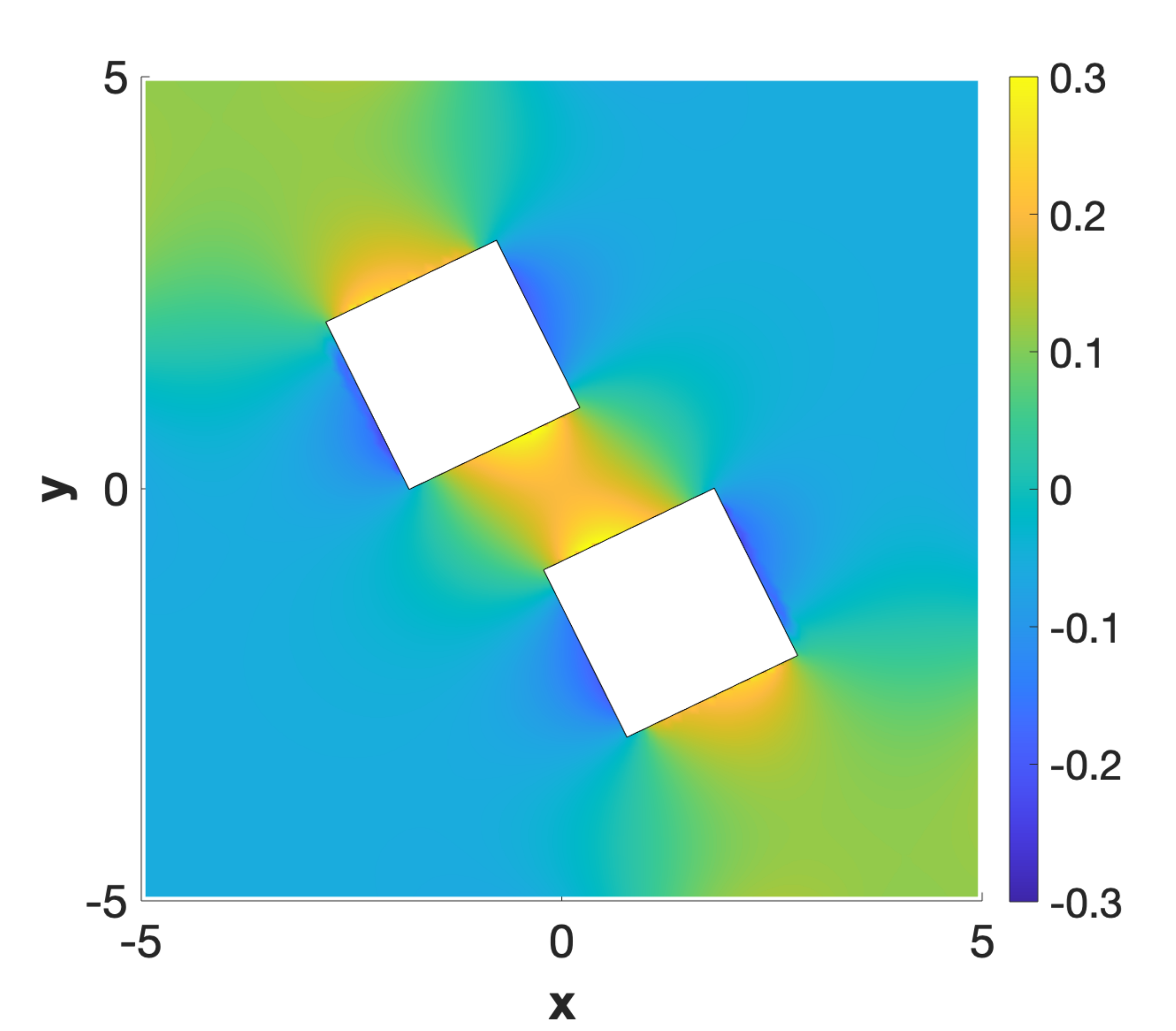}
}
\quad
\subfigure[Pressure field predicted by the ROM with correction]{
\includegraphics[width=4.8cm]{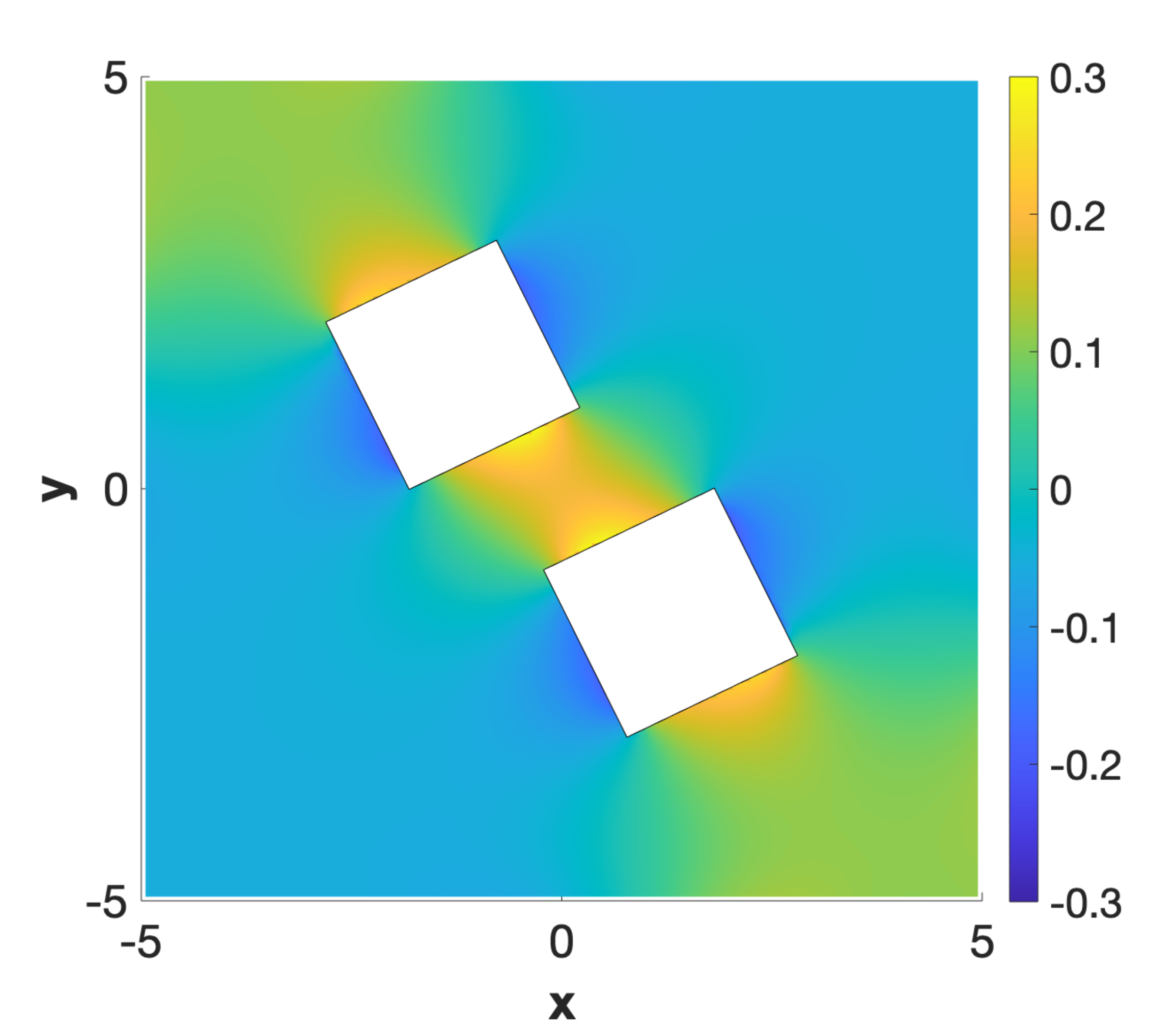}
}
\caption{Two squares under a shear flow: The velocity and pressure fields in the fluid predicted by the ROM at $t^*=3.0$, compared with the full-order solutions by numerical simulations.}
\label{fig:two_squares}
\end{figure}

\begin{figure}[htbp]
\centering
\subfigure[Velocity in $y$ at $x=-2.5$]{
\includegraphics[width=6.5cm]{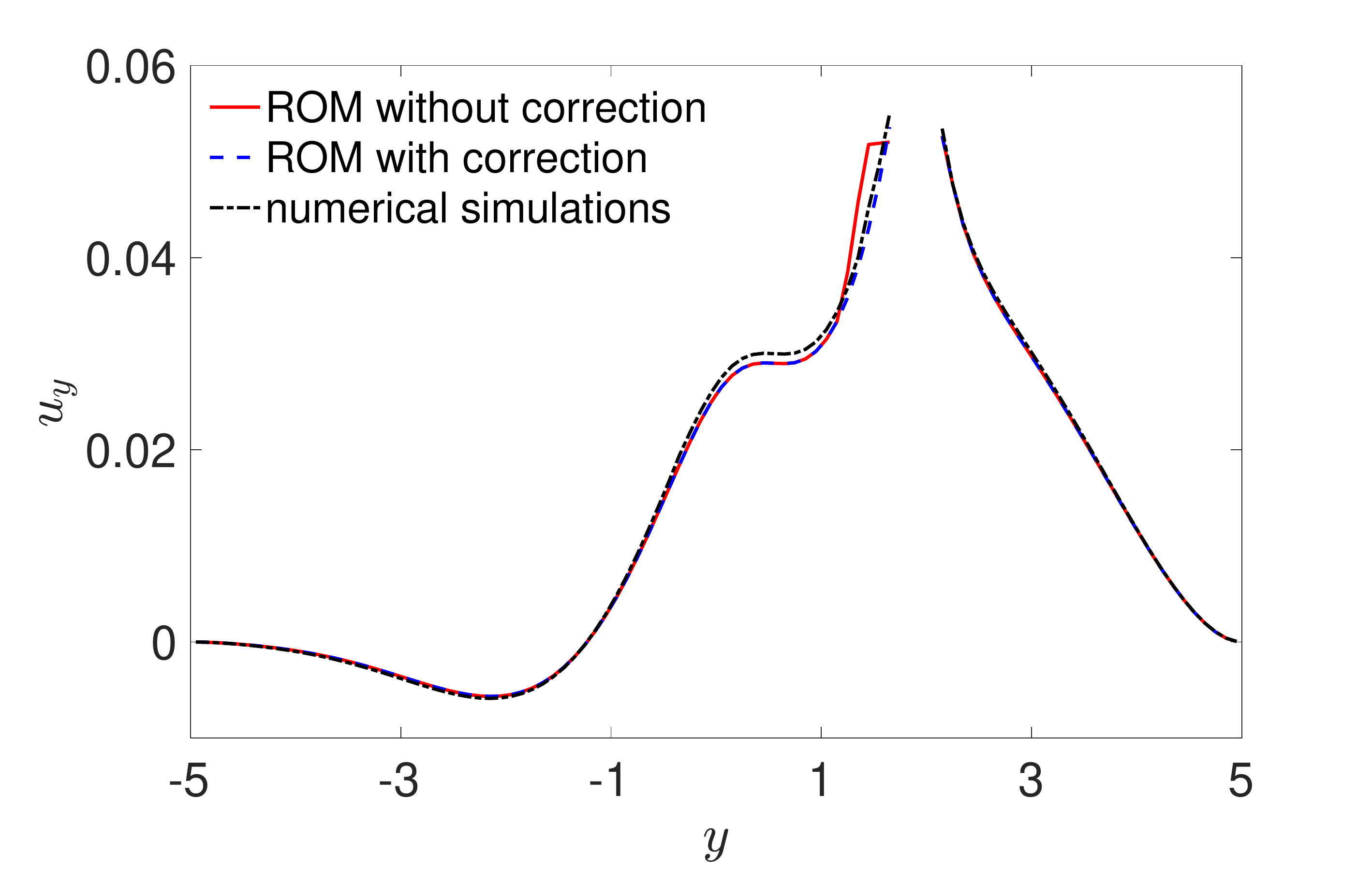}
}
\quad
\subfigure[Velocity in $y$ at $y=1.5$]{
\includegraphics[width=6.5cm]{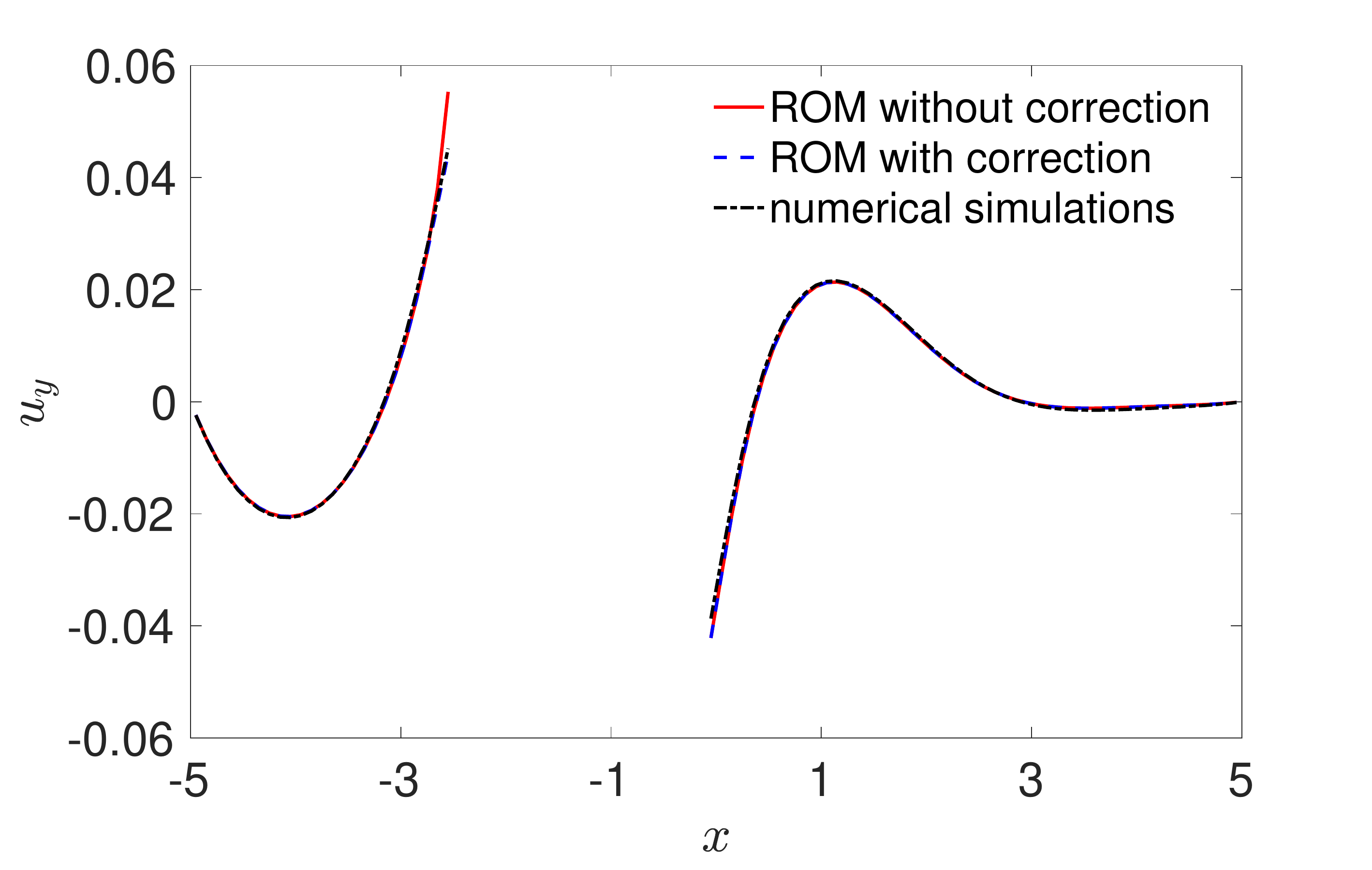}
}
\quad
\subfigure[Pressure at $x=-2.5$]{
\includegraphics[width=6.5cm]{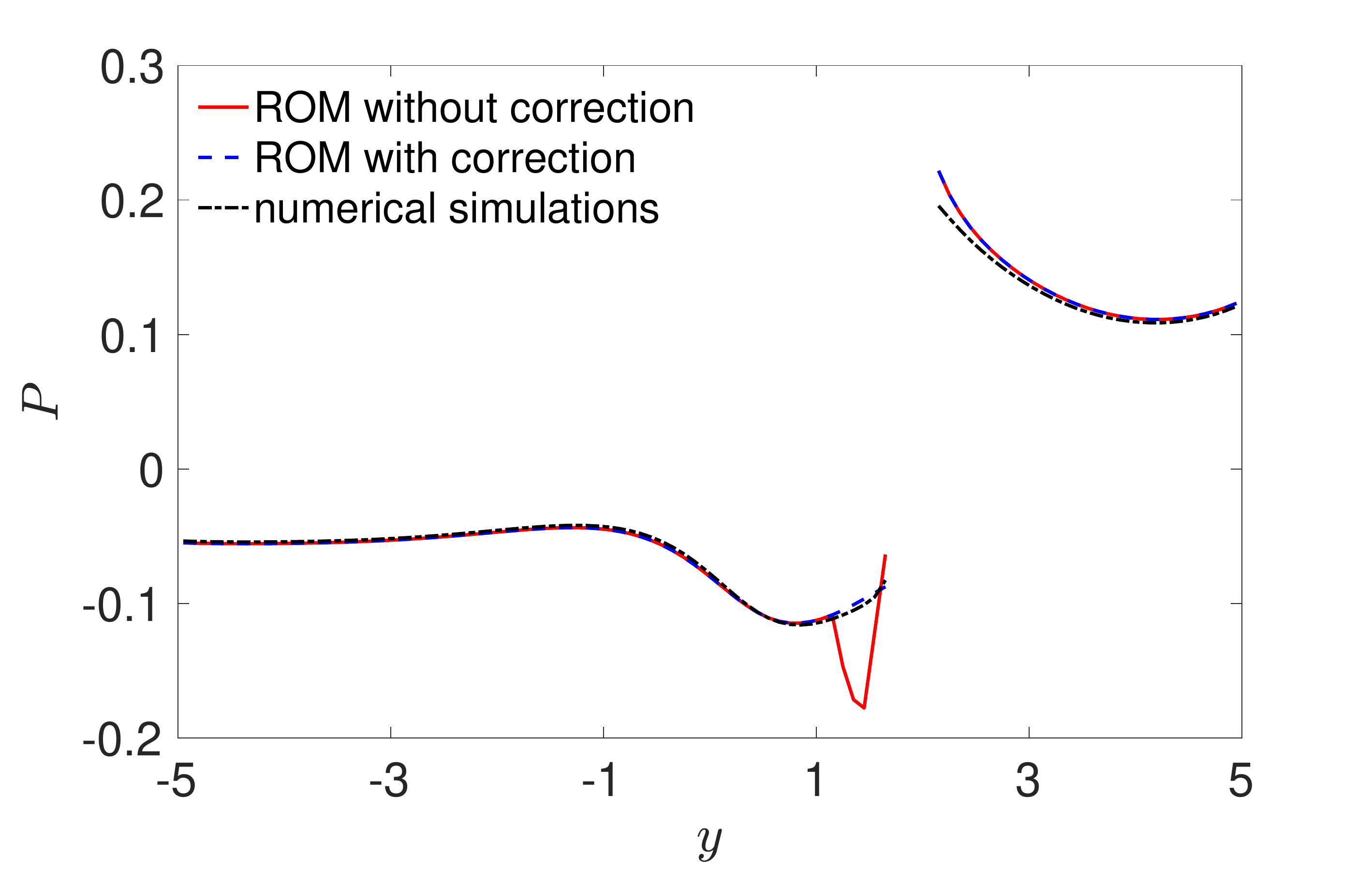}
}
\quad
\subfigure[Pressure at $y=1.5$]{
\includegraphics[width=6.5cm]{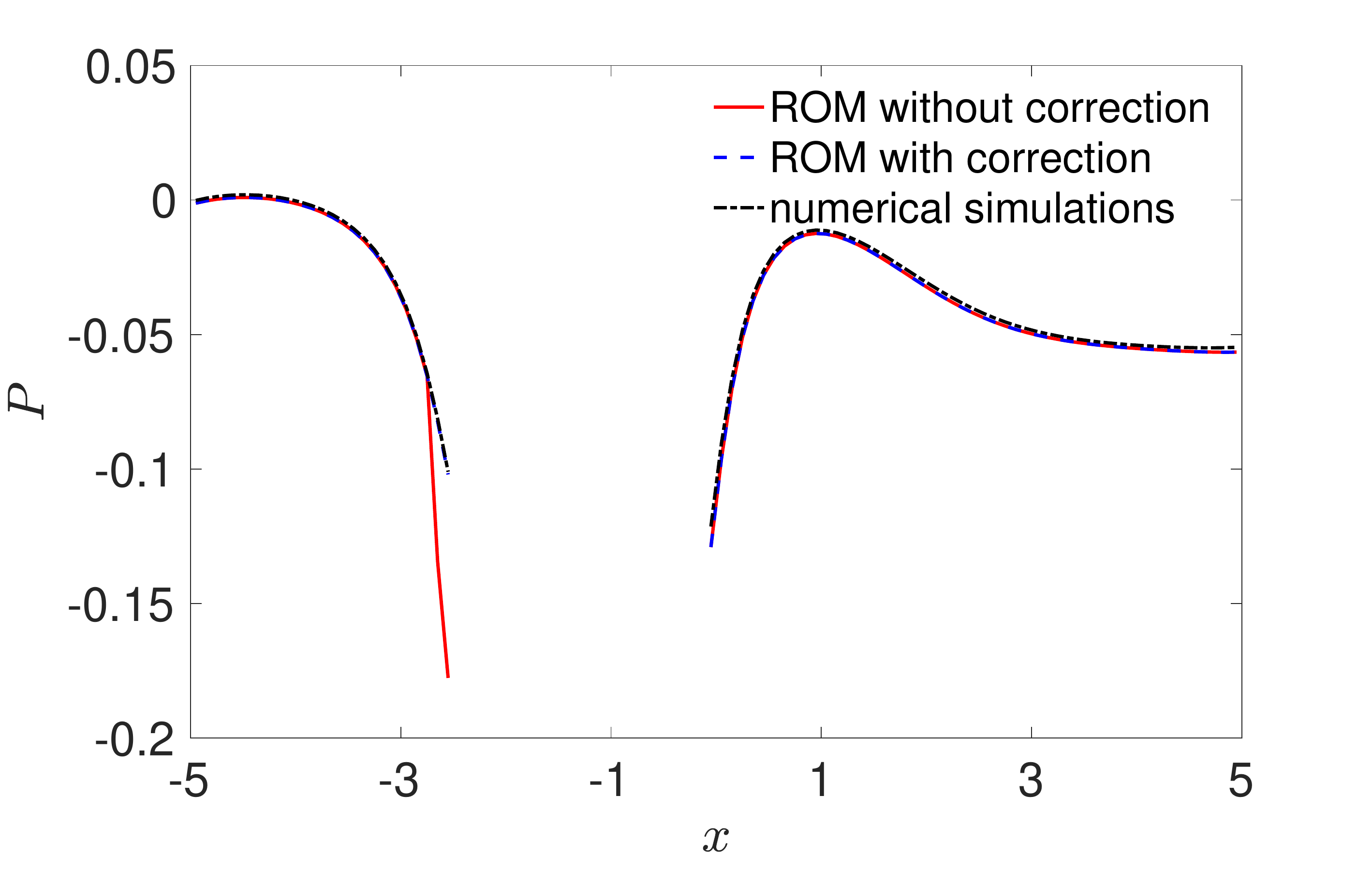}
}
\caption{Two squares under a shear flow: The velocity and pressure along the lines $x=-2.5$ and $y=1.5$ at $t^*=3.0$ predicted by the ROM with and without correction, compared with the full-order solutions by numerical simulations.}
\label{fig:two_squares_lines}
\end{figure}

To demonstrate the robustness of the proposed reduced order modeling method, we further examined a case of more cylinders in a shear flow. Figure \ref{fig:four_cylinders} and \ref{fig:four_cylinders_lines} summarize the velocity in $y$ direction and pressure predicted by the ROM with and without the correction near the moving solid boundaries, compared with the full-order solutions by numerical simulations.  
%%%%%%%%%%%%%%%%%%%%%%%%%%%%
%%%%%%%%%%%%%%%%%%%%%%%%%%%%%%%
%%%%%%%%%%%%%%%%%%%%%%%%%%%%
Finally, to demonstrate the applicability of the proposed reduced order modeling method to moving boundaries of arbitrary geometries, we also solved the fluid-solid interaction problem with two squares. The results are presented in Figure \ref{fig:two_squares} and \ref{fig:two_squares_lines}.
%%%%%%%%%%%%%%%%%%%%%%%%%%%%
%%%%%%%%%%%%%%%%%%%%%%%%%%%%
In either case, the predictions by the ROM with the correction achieve good accuracy. Figure \ref{fig:four_cylinders_lines} and \ref{fig:two_squares_lines} provide closer views for the difference between without and with the correction. We hence have demonstrated that the proposed reduced order modeling method is applicable to dynamical systems with general moving boundaries regardless of the number or geometry of the boundaries.

\section{Conclusion}\label{sec:conclu}

We have presented a model order reduction method for dynamical systems with moving boundaries, which draws on the POD, GPR, and MLS interpolation. The method is nonintrusive and applicable to experimental data. Given a set of snapshot data of state variables at discrete time instances, the reduced space is constructed via the POD. The time-evolution of the temporal POD coefficients and the parameters characterizing the moving boundaries is inferred from the data via the GPR. The cost of GPR would not increase due to the nonlinearity of the system. The errors in the ROM's prediction for the regions near the moving boundaries on the downstream side can be effectively reduced using the correction method based on the MLS interpolation. For a given set of snapshot data, the ROM constructed from the data can predict the full-order solution at a desired time inside or outside the dataset range. The forecast beyond the range of snapshot data is constrained by the POD and GPR as well as how many snapshot data are available. To avoid a trial-and-error approach, we have provided the criteria for \textit{a priori} determination of the furthest forecast time permitted in time extrapolation of the ROM. We have demonstrated the accuracy and efficiency of the proposed method in several benchmark problems, where the snapshot data used to construct and validate the ROMs were generated from analytical solutions, results of numerical simulations, and experimental data.  

%The proposed model reduction approach based on Gaussian process is applicable to both linear and nonlinear dynamical systems. The cost of GPR would not increase due to the nonlinearity of the system. Unlike the least squares method in \cite{PEHERSTORFER2016196}, if the full model has a nonlinear term,  the cost of operator inference grows exponentially in the order of the nonlinear term. 

When numerical simulations or experimental measurements are demanding or expensive, reduced order modeling provides an attractive alternative means to predict the full-order solutions. In practice, numerical simulations/experimental measurements and ROMs can be adaptively combined to achieve an efficient long-time prediction for a dynamical system. With the furthest forecast time of the ROM determined \textit{a priori} by the proposed criteria, the numerical simulations/experimental measurements and ROMs can be alternatively called in an automated process. We note that the ``equation-free" approach \cite{EquationFree_Kevrekidis,EquationFreeReview_Kevrekidis} proposed by Kevrekidis and collaborators for multiscale modeling also couples two levels of predictions ``on-the-fly": microscopic simulations and macroscopic models. In particular, it extrapolates ensemble-averaged macroscale quantities obtained from the microscopic simulations. The extrapolation is through a projective integrator that advances the macro variable over a macro time step with the time derivative of the macro variable computed from the results of the last few steps of the microscale simulations using small time steps. The ``equation-free" approach requires time-scale separation, i.e., the local relaxation time for the microscopic process is much smaller than the time scale for the macroscopic evolution of the system. The approach proposed in the present work uses Gaussian process regression for time extrapolation and does not require time-scale separation.

%\textcolor{blue}{In a future follow-up work, the GPR may be replaced with the LSTM recurrent neural network to infer the time-evolution of the POD temporal coefficients. According to \cite{ROM_LSTM_Vlachas2018}, the LSTM method's training cost scales linearly with the number of training samples, and it exhibits an $\mathcal{O}(1)$ inference computational cost. In contrast, the training cost of GPR exhibits a cubic scaling with respect to the number of training samples, and the inference cost scales quadratically with the number of inputs for inference.}

The slowly decaying Kolmogorov $n$-width is a major factor limiting the furthest forecast time of the ROM, which results from the fact that POD restricts the state to evolve in a linear subspace. To address this $n$-width limitation of linear subspaces, Lee and Carlberg proposed to construct ROMs in nonlinear manifolds that are computed based on convolutional autoencoders from deep learning \cite{Nonlinearbasis_LEE2020}. The ROM constructed by this method was shown significantly outperforming the linear-subspace ROM when forecast the full-order solution at a future time for advection-dominated Burgers equation. Thus, it may be worthwhile in a future work to consider replacing the linear spatial bases by nonlinear manifolds in our method to achieve a longer forecast time for the ROM.

Our method can be potentially extended to construct ROMs to predict the full-order solutions for different parameters/inputs. Suppose $u(\mathbf{x},t;\boldsymbol{\eta})$ is the full-order solution for a parameterized dynamical system, where %the time $t \in \mathcal{T}$, 
$\boldsymbol{\eta} \in \mathcal{P}$ denotes the parameter, and $\mathcal{P} \subset \mathbb{R}^d$ is the parameter space. The full-order solutions $u(\mathbf{x},t_i;\boldsymbol\eta_j)$ at sampled parameters $\boldsymbol\eta_j \in \mathcal{P}_S \subset \mathcal{P}$ ($j=1, 2, \dots, M_{\boldsymbol \eta}$) and time instances $t_i$ ($i=1, 2, \dots, M_t$) are available to form the snapshot dataset. We aim to predict the solution $u(\mathbf{x},t';\boldsymbol \eta^*)$ at $t'$ for a new parameter $\boldsymbol \eta^* \notin \mathcal{P}_S$. For this type of problems, the coefficients of POD modes become $a_k(t,\boldsymbol\eta)\in \mathbb{R}$. Thus, a multivariate Gaussian process model for each $a_k(t,\boldsymbol\eta)$ needs to be constructed from the snapshot data with the training input $\mathbf{z} = (t,\boldsymbol\eta) \in \mathbb{R}^{d+1}$. The multivariate Gaussian process model constructed is then used to predict the POD coefficient $a_k(t',\boldsymbol \eta^*)$ at $t'$ and $\boldsymbol \eta^*$. Finally, the full-order solution can be predicted by: $u(\mathbf{x},t';\boldsymbol \eta^*)\approx \bar u ({{\mathbf x}}) + \sum\limits_{k = 1}^R a_k(t',\boldsymbol \eta^*)\phi_k({{\mathbf x}})$ with $\bar u ({{\mathbf x}}) = \frac{1}{M_t M_{\boldsymbol \eta}} \sum\limits_{j=1}^{M_{\boldsymbol \eta}} \sum\limits_{i=1}^{M_t} u(\mathbf{x},t_i;\boldsymbol \eta_j)$. Detailed error analysis, criteria, and application to parameterized dynamical systems with moving boundaries will be investigated in our future work.

\section*{Acknowledgements}
This material is based upon work supported by the National Science Foundation under Grant No. CMMI-1761068. 
We gratefully acknowledge Dr. Christian Franck and Dr. Jin Yang in the Department of Mechanical Engineering at the University of Wisconsin-Madison for insightful discussions and providing us the experimental data in \S\ref{subsec:2DExp} and \S\ref{subsec:buble_cavity}. The authors also thank the two anonymous reviewers for their insightful comments and suggestions that helped improve the manuscript.

\bibliographystyle{elsarticle-num}
\bibliography{ref}
\end{document}